\documentclass[12pt]{article}
\usepackage{amsmath,amsfonts,amssymb}
\usepackage{hyperref}
\usepackage{graphicx}
\usepackage{feynmp}
\usepackage[dvips]{color}
\usepackage{epsfig}
\usepackage{psfrag}

\makeatletter\@addtoreset{equation}{section}\makeatother

\DeclareMathOperator{\Tr}{Tr}

\renewcommand{\Re}{\operatorname{Re}}
\renewcommand{\Im}{\operatorname{Im}}

\def\bR {\mathbb{R}}
\def\bT {\mathbb{T}}

\def\bN {\mathbb{N}}

\newcommand{\beq}{\begin{equation}}
\newcommand{\eeq}{\end{equation}}

\newcommand{\vev}[1]{{\left< {#1} \right>}}

\newcommand{\address}[1]{\vbox{\center\em#1}}
\renewcommand{\title}[1]{\vbox{\center\LARGE{#1}}\vspace{5mm}}

\newcommand{\Fcal}{\mathcal{F}}

\newcommand{\Ncal}{\mathcal{N}}
\newcommand{\Ocal}{\mathcal{O}}

\newcommand{\cF}{{\mathcal F}}

\newcommand{\cO}{{\mathcal O}}

\newcommand{\al}{\alpha}

\newcommand{\sign}{\operatorname{sign}}

\newcommand{\id}{\mathrm{id}}
\newcommand{\pr}{\operatorname{pr}}

\newcommand{\wD}{\widetilde{D}}

\begin{document}
\bibliographystyle{utphys}
\begin{fmffile}{graphs}

\begin{titlepage}
  \hfill{ }\\
  \hfill{ }\\
  \hfill{ }\\
  
  \hspace{-10mm} 
  \centerline{\LARGE{'t Hooft Operators in Gauge Theory from Toda CFT}}
  \vspace{10mm}
  
  \begin{center}
    \renewcommand{\thefootnote}{$\alph{footnote}$}
    Jaume Gomis\footnote{\href{mailto:jgomis@perimeterinstitute.ca}
      {\tt jgomis@perimeterinstitute.ca}}
    and Bruno Le Floch\footnote{\href{mailto:bruno@le-floch.fr}
      {\tt bruno@le-floch.fr}}
    \hfill{ }\\
    \hfill{ }\\
    \address{${}^{a,b}$Perimeter Institute for Theoretical Physics,\\
      Waterloo, Ontario, N2L 2Y5, Canada}
    \address{${}^{b}$\'Ecole Normale Sup\'erieure,\\
      Paris, 75005, France}
  \end{center}
  
  \renewcommand{\thefootnote}{\arabic{footnote}}
  \setcounter{footnote}{0}

  \vspace{15mm}

  \abstract{
  \medskip\medskip
  \normalsize{
  \noindent
  We construct loop operators in two dimensional Toda CFT and  calculate with them the  exact expectation value of certain supersymmetric 't Hooft and dyonic loop operators in four dimensional ${\cal N}=2$ gauge theories with $SU(N)$ gauge group. Explicit formulae for 't Hooft and dyonic operators in ${\cal N}=2^*$ and   ${\cal N}=2$ conformal SQCD with $SU(N)$ gauge group are presented. We also  briefly speculate on the Toda CFT realization of arbitrary loop operators in these gauge theories in terms of topological web operators in Toda CFT.
  }}
  
  \vspace{10mm}
  
  \noindent
  \today
  \vfill

\end{titlepage}

\tableofcontents

%===================================================
\section{Introduction}

 Loop operators are valuable observables with which to probe the quantum dynamics of  gauge theories.   't Hooft loop operators were introduced 
in~\cite{'tHooft:1977hy} as novel order parameters for the phases of gauge theories. The various phases are associated to different responses experienced by the external magnetic monopole which is inserted by the 't Hooft operator. These magnetic operators yield  complementary  insights beyond those provided by Wilson loop operators~\cite{Wilson:1974sk}, which insert electrically charged probes. It is therefore of interest to develop techniques to compute the expectation value of loop operators in gauge theories.

The perturbative computation of the expectation value of supersymmetric 't Hooft loop operators in ${\cal N}=4$ super Yang-Mills with arbitrary gauge group $G$ was performed in~\cite{Gomis:2009ir} using the refined description of 't Hooft operators introduced by Kapustin~\cite{Kapustin:2005py} (see also~\cite{Kapustin:2006pk}).  This computation demonstrates that the expectation value of an 't Hooft loop operator maps under the action of S-duality to that of a Wilson loop operator (see also~\cite{Gomis:2009xg}), which is captured by the matrix model proposed in \cite{Erickson,Dru-Gross}, as shown by Pestun~\cite{Pestun:2007rz}. This   reinforces that the main physical idea behind S-duality is a non-abelian generalization of electric-magnetic duality.

Recently, a new  powerful tool has emerged to study the expectation value of  supersymmetric observables in   four dimensional ${\cal N}=2$ gauge theories on $S^4$ arising from a punctured Riemann surface~\cite{Gaiotto:2009we}, starting with the work of Alday, Gaiotto and Tachikawa~\cite{AGT}. Correlation functions of various operators in two dimensional non-rational CFTs (Liouville and Toda)  capture the partition function \cite{AGT,Wyllard:2009hg} (see also \cite{Bonelli:2009zp}),  Wilson-'t Hooft loop operators \cite{Alday:2009fs,Drukker:2009id,Drukker:2010jp}, surface operators~\cite{Alday:2009fs} and domain wall operators~\cite{Drukker:2010jp} of these four dimensional  ${\cal N}=2$ gauge theories. The elegant subject of two dimensional CFT thus provides a novel framework with which to study four dimensional supersymmetric gauge theories. This approach to gauge theory has reproduced the results    for the partition function and Wilson loop operators obtained by Pestun~\cite{Pestun:2007rz},  but  perhaps   more interestingly has led to novel  computations of   gauge theory observables.

In this paper we present the {\it exact} expectation value of  't Hooft loop operators and Wilson-'t Hooft operators in the   four dimensional ${\cal N}=2$ superconformal  gauge theories with  a single $SU(N)$ vector multiplet. These theories are:

\begin{itemize}
\item ${\cal N}=2^*$:  vector multiplet coupled to a massive adjoint hypermultiplet
\item Conformal SQCD: vector multiplet coupled to $2N$ massive fundamental  hypermultiplets
\end{itemize}

We arrive at these results  from the computation of   correlation functions in two dimensional $A_{N-1}$ Toda CFT in the presence of Verlinde loop operators. Despite that Toda CFT  has not been solved ---unlike Liouville theory\footnote{$A_1$ Toda theory reduces to Liouville CFT. See~\cite{Teschner:2001rv} for an reference of this much more tractable theory.}--- we amass the ingredients needed to find explicit formulae for the expectation value of 't Hooft operators in ${\cal N}=2$ superconformal  $SU(N)$ gauge theories.
Even though we concentrate on the results for ${\cal N}=2^*$ and ${\cal N}=2$ conformal SQCD,  the techniques and formulae in this paper can be used to compute      't Hooft operators in arbitrary ${\cal N}=2$ $SU(N)$ superconformal quiver gauge theories when all gauge group factors are $SU(N)$.

This   extends to an arbitrary $SU(N)$ gauge group the computations of \cite{Alday:2009fs,Drukker:2009id}, where the expectation value of 't Hooft operators  in these   theories  with  $SU(2)$  gauge group     was mapped to the computation of loop operators in   Liouville CFT. Furthermore,  this provides the magnetic version of the  identification put forward in  \cite{Drukker:2010jp}, where Wilson loop operators in arbitrary representations of $SU(N)$ were given a Toda CFT representation (see also \cite{Passerini:2010pr}).\footnote{Further work on gauge theory loop operators from the perspective of two dimensional CFT include \cite{Wu:2009tq,Petkova:2009pe}.}  When the gauge group is $SU(2)$ the identification of gauge theory loop operators with Liouville loop operators relied on    
 the elegant characterization of the charges of gauge theory loop operators in terms of non-selfintersecting curves on the Riemann surface~\cite{Drukker:2009tz}. The corresponding description of arbitrary loop operators   in terms of data on the Riemann surface when  the gauge group is  $SU(N)$  is not known, but we nevertheless identify Toda CFT loop operators capturing specific 't Hooft operators in these gauge theories. We also speculate on the possibility that topological web operators  in Toda CFT   ---which are   supported on webs on the Riemann surface constructed  from trivalent graphs---  describe arbitrary gauge theory loop operators in these four dimensional ${\cal N}=2$ gauge theories when the gauge group is $SU(N)$.

The plan of the rest of the paper is as follows. In Section~\ref{sec:background} we set up the background. Specifically, we recall in Section~\ref{sec:partition} the equivalence~\cite{AGT,Wyllard:2009hg} between the partition function of  ${\cal N}=2^*$ and ${\cal N}=2$ conformal SQCD with $SU(N)$ gauge group and the correlation functions of two dimensional $A_{N-1}$ Toda CFT on the once punctured torus and four punctured sphere with suitable vertex operator insertions. We then discuss the admissible spectrum of 't Hooft loop operators in $\Ncal=2^*$ and $\Ncal=2$ conformal SQCD with $SU(N)$ gauge group   and introduce the basic notion of loop operators and topological defects necessary to perform our analysis and computations. Section~\ref{sec:movesloops} constitutes the bulk of the paper: here, we compute the matrix elements of Verlinde loop operators which encapsulate the expectation value of 't Hooft and dyonic operators in $\Ncal=2^*$, as well as 't Hooft and dyonic operators in $\Ncal=2$ conformal SQCD. We conclude in Section \ref{sec:discussion} with a discussion of our results and of how topological webs may describe 't Hooft loop operators in arbitrary representations.

The appendices contain many of the technical details and derivations. 
In Appendix~\ref{sec:Toda} we give a  brief review of the most salient features of $A_{N-1}$ Toda CFT required for our analysis. Appendix~\ref{sec:fusion} contains the derivation of the  fusion and braiding matrices of $A_{N-1}$ Toda CFT  used in the main text to compute the Verlinde loop operator corresponding to 't Hooft and dyonic loops in the $\Ncal=2^*$ gauge theory. Further ingredients needed for the case of $\Ncal=2$ conformal SQCD are presented in Appendix~\ref{sec:SQCDmonodromies}. Finally, Appendix~\ref{sec:Wilson} is devoted to the computation of the Verlinde loop operators corresponding in gauge theory to Wilson loop operators.

%===================================================
\section{Gauge theory and Toda CFT}
\label{sec:background}

\subsection{Partition function}
\label{sec:partition}

A useful starting point in the computation of the expectation value of 't Hooft operators in ${\cal N}=2^*$ and ${\cal N}=2$ conformal SQCD with $SU(N)$ gauge group is to engineer ${\cal N}=2$ gauge theories as the low energy limit of the  six dimensional $(0,2)$$A_{N-1}$  theory on $N$ M5-branes wrapped   on a genus $g$ Riemann surface with $n$ punctures $C_{g,n}$ \cite{Witten:1997sc,Gaiotto:2009we}. In this section we briefly summarize some ingredients that   enter in our computation of 't Hooft loop operators in Section~\ref{sec:movesloops}.

In the general construction of four dimensional ${\cal N}=2$ gauge theories from $N$ M5-branes on $C_{g,n}$ proposed in~\cite{Gaiotto:2009we}, each of the punctures on the Riemann surface is  characterized by a partition of $N$, which labels a supersymmetric codimension two singularity of the   $(0,2)$  $A_{N-1}$ theory.\footnote{Codimension two singularities for the general $(0,2)$ ADE theory are expected to be labeled by homomorphisms of $SU(2)$ into the corresponding Lie algebra, or equivalently by nilpotent orbits of the complexified Lie algebra up to conjugacy. The partition $N=n_1+\ldots n_N$ captures the decomposition of the fundamental representation of $SU(N)$ into a sum of $SU(2)$ representations of dimension $n_i$.}   
Two distinct types  of punctures appear in the construction of  ${\cal N}=2^*$ and ${\cal N}=2$   conformal SQCD with $SU(N)$ gauge group, known as ``full" and ``simple" punctures. A full puncture, which  will be denoted by $\odot$, corresponds to the partition $N= 1+\ldots+1$, while a simple puncture,  denoted by $\bullet$, corresponds to the partition $N=(N-1)+1$.

\begin{figure}[t]
\begin{center}
\begin{tabular}{ccc}
\includegraphics[scale=1.0]{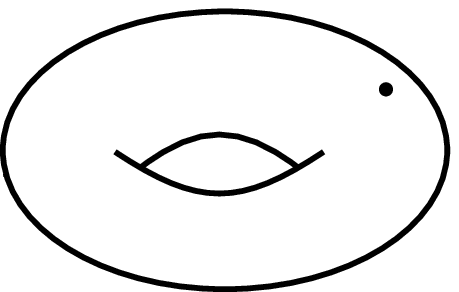}
&\hspace{30mm}&
\includegraphics[scale=1.0]{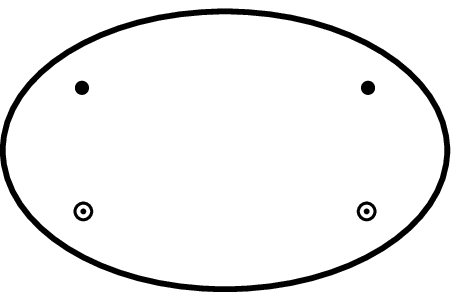}
\\
\\
(a) $\Ncal=2^*$ &&(b) $\Ncal=2$ conformal SQCD 
\end{tabular}
\parbox{5in}{
\caption{The punctured Riemann surfaces $C_{1,1}$ and $C_{0,4}$ associated to ${\cal N}=2^*$ and ${\cal N}=2$ conformal SQCD.}
\label{fig:punctures}
}
\end{center}
\end{figure}

The   mass parameters and flavour symmetries of the four dimensional gauge theory are encoded in the choice of punctures on the Riemann surface. The flavour symmetries associated to a full puncture and simple puncture are $SU(N)$ and $U(1)$ respectively. For ${\cal N}=2^*$ the relevant Riemann surface is the once-punctured torus $C_{1,1}$, with a simple puncture, while for ${\cal N}=2$ conformal SQCD it is $C_{0,4}$, the sphere with four punctures, two of which are simple and two full (see Figure~\ref{fig:punctures}). This makes manifest an $SU(N)^2\times U(1)^2$ subgroup of the $U(2N)$ flavour symmetry of ${\cal N}=2$ conformal SQCD and a $U(1)$  subgroup of the $SU(2)$ flavour symmetry  of ${\cal N}=2^*$. 

The partition function on $S^4$ of ${\cal N}=2^*$ and ${\cal N}=2$ conformal SQCD with an $SU(N)$ vector multiplet and massive hypermultiplets is captured by a correlation function in the two dimensional $A_{N-1}$ Toda CFT on the associated Riemann surfaces  $C_{1,1}$ and $C_{0,4}$, respectively~\cite{AGT,Wyllard:2009hg}. The mass parameters and flavour symmetries of the four dimensional gauge theory are encoded in the choice of Toda CFT vertex operators inserted at the punctures on the Riemann surface $C_{g,n}$. The matching between the allowed punctures ---which are labeled by a partition of $N$--- and Toda CFT representations was discussed in \cite{Kanno:2009ga,Drukker:2010jp}.

Vertex operators creating  primary states in the  $A_{N-1}$  Toda CFT  can be  constructed from $N-1$ scalar fields $\phi=\sum \phi_i e_i$, and are given semiclassically by (see Appendix~\ref{sec:Toda} for a summary of some properties of Toda CFT)
\begin{equation}
V_\alpha = e^{\langle \alpha, \phi\rangle}\,,
\label{vertex}
\end{equation}
where $e_i$ are the simple roots and $\langle~, ~\rangle$ is the bilinear form on the weight lattice of the $A_{N-1}$ Lie algebra.  Representations of the $W_N$-symmetry algebra of $A_{N-1}$ Toda CFT are then labeled, modulo the Weyl group action $\alpha\rightarrow  Q+\sigma(\alpha-Q)$ $\forall \sigma \in S_N$, by the momentum $\alpha=\sum_i\alpha_i e_i$ of the vertex operator creating their highest weight state.

A full puncture  in the M5-brane construction corresponds to inserting the primary of a non-degenerate representation of the $W_N$-algebra, created by a vertex operator $V_{\rm m}$ with momentum
\begin{equation*}
{\rm m}=Q+i m\,,
\end{equation*}
where $Q=(b+1/b)\rho\equiv q \rho$ is the background charge for the scalars $\phi$ and $\rho$ is the Weyl vector of the $A_{N-1}$ algebra.\footnote{It can be written as  $\rho=\sum\omega_l$, 
with $\omega_l$ being the fundamental weights of the $A_{N-1}$  Lie algebra, which are dual to the $A_{N-1}$   simple roots $e_k$, that is $\langle e_k,\omega_l\rangle=\delta_{kl}$.} The $N-1$ real component vector $m$ encodes  the masses   of hypermultiplets in the gauge theory.  The flavour symmetry group of the corresponding gauge theory hypermultiplets is the commutant in $SU(N)$ of the stability group of the  momentum vector ${\rm m}$. Since $m$ is an arbitrary $N-1$ component vector, an $SU(N)$ flavour symmetry is attached to a full puncture.

A simple puncture corresponds to inserting a vertex operator $V_{\rm \hat m}$ which generates a semi-degenerate representation of the $W_N$-algebra with $N-2$ level one null vectors (it reduces to a full puncture when $N=2$). The Toda momentum for a simple puncture is\footnote{We could equivalently use ${\rm \hat m}^*= N({q/2}+i\hat m)\, \omega_{1}$.}
\begin{equation*}
{\rm \hat m}= N\left({q\over 2}+i\hat m\right)\, \omega_{N-1}\,,
\end{equation*}
where $\omega_{N-1}$ is the   fundamental weight of the $A_{N-1}$ Lie algebra associated to the the last node of the $A_{N-1}$ Dynkin diagram. A simple puncture is thus labeled by a single real parameter $\hat m$, which encodes the mass of a hypermultiplet in the gauge theory.\footnote{This agrees  with the identification in~\cite{Kanno:2010kj}.} The momentum vector ${\rm \hat m}$ is left invariant by an $SU(N-1)$ subalgebra of the $A_{N-1}$ Lie algebra. Therefore, a $U(1)$ flavour symmetry is attached to a  simple puncture.

\begin{figure}[t]
\begin{center}
\begin{tabular}{cccccc}
\includegraphics[scale=1.0]{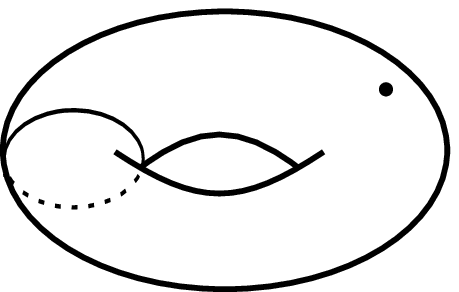}
&\hspace{30mm}&
\includegraphics[scale=1.0]{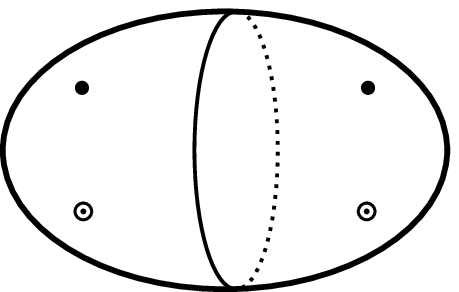}
\\
\\
\begin{fmfgraph*}(30,15)
  \fmfstraight
  \fmfleft{1}
  \fmfright{3}
  \fmf{plain,width=2,left}{1,2}
  \fmf{plain,width=2,right}{1,2}
  \fmf{plain,width=2,tension=4}{2,3}
  \fmfv{label=$\alpha$,label.angle=180}{1}
  \fmfv{label=${\rm \hat m}$,label.angle=0}{3}
\end{fmfgraph*}
&&
\begin{fmfgraph*}(30,20)
  \fmfstraight
  \fmfleft{b1,t1}
  \fmfright{b2,t2}
  \fmf{plain,width=2}{t1,m1}
  \fmf{plain,width=2}{b1,m1}
  \fmf{plain,width=2}{t2,m2}
  \fmf{plain,width=2}{b2,m2}
  \fmf{plain,width=2,label=$\alpha$}{m1,m2}
  \fmfv{label=${\rm \hat m}^*_3$,label.angle=180}{t1}
  \fmfv{label=${\rm m}^*_4$,label.angle=180}{b1}
  \fmfv{label=${\rm \hat m}_2$,label.angle=0}{t2}
  \fmfv{label=${\rm m}_1$,label.angle=0}{b2}
\end{fmfgraph*}
\\
\\
(a) $\Ncal=2^*$ &&(b) $\Ncal=2$ conformal SQCD 
\end{tabular}
\parbox{5in}{
\caption{Decomposition of the punctured Riemann surfaces $C_{1,1}$ and $C_{0,4}$ into pairs of pants, and the corresponding trivalent graphs $\Gamma_\sigma$.  The momenta ${\rm \hat m}, {\rm \hat m}_2, {\rm \hat m}^*_3$ are semi-degenerate, and $\alpha, {\rm m}_1, {\rm m}^*_4$ are non-degenerate.}
\label{fig:pants}}
\end{center}
\end{figure}

For a choice of duality frame $\sigma$ (see Figure~\ref{fig:pants}), the explicit formulae for the gauge theory partition functions  of ${\cal N}=2^*$ and ${\cal N}=2$ conformal SQCD with $SU(N)$ gauge group are given in terms of a Toda CFT correlation function with the corresponding (simple and full) punctures\footnote{The conjugation operation is defined by $h_j^*=-h_{N+1-j}$. In particular, $Q^*=Q$ and $\omega_1^*=\omega_{N-1}$.} by~\cite{Wyllard:2009hg}\footnote{Upon setting $b=1$ and appropriately normalizing the vertex operators.}
\begin{equation}
Z_{{\cal N}=2^*}=\langle V_{\rm \hat m}\rangle_{C_{1,1}}=\int d\alpha\, C(\alpha,{\rm \hat m} , 2Q-\alpha)\,    {\overline
\cF^{(\sigma)}_{\alpha,E}}(\tau)\,\cF^{(\sigma)}_{\alpha,E}(\tau)\,,
\label{N=2*}
\end{equation}
\begin{equation}
Z_{SQCD}\hskip-3pt= \hskip-5pt\langle V_{{\rm m}_4^*}(\infty)\hskip-1ptV_{{\rm \hat m}_3^*}(1)\hskip-1ptV_{{\rm \hat m}_2}(\tau) \hskip-1pt V_{{\rm m}_1}(0)\hskip-1pt\rangle_{C_{0,4}} \hskip-5pt = \hskip-5pt\int \hskip-4pt d\alpha\, \hskip-1pt C({\rm m}^*_4,{\rm \hat m}^*_3,\alpha)C(2Q-\alpha,{\rm \hat m}_2,{\rm m}_1)
 {\overline
\cF^{(s)}_{\alpha,E}}(\tau)\, \hskip-2pt\cF^{(s)}_{\alpha,E}(\tau)\,,
\label{SQCD}
\end{equation}
where the momentum that is being integrated over is non-degenerate\footnote{The real   momentum vector $a$ can be represented as $a\equiv (a_1,\ldots, a_N)$ with the constraint $\sum_{l=1}^N a_l=0$.}
\begin{equation*}
\alpha=Q+ia\,.
\end{equation*}
Here, $C(\alpha_1,\alpha_2,\alpha_3)$ is the three-point function of primaries carrying momenta $\alpha_1,\alpha_2$ and $\alpha_3$, and $\cF^{(\sigma)}_{\alpha,E}(\tau) / \overline \cF^{(\sigma)}_{\alpha,E}(\tau)$ are the holomorphic/antiholomorphic $W_N$-conformal blocks. These capture the contribution of the $W_N$-symmetry descendants to  the  correlation functions in the choice of factorization $\sigma$ of the correlator.\footnote{We recall that a choice of factorization channel $\sigma$ in the CFT correlation function corresponds to a choice of duality frame $\sigma$ in the gauge theory. Associativity of the CFT OPE implies that the answers of \eqref{N=2*} and \eqref{SQCD} do not depend on $\sigma$.} This is characterized by a trivalent graph $\Gamma_\sigma$, which also represents a sewing $\sigma$ of the Riemann surface $C_{g,n}$ from pairs of pants (trinions) and tubes (see Figure~\ref{fig:pants}).  The internal edge is labeled by a momentum $\alpha$, $E$ denotes the momenta of the external edges in $\Gamma_\sigma$ and $\tau $ denotes the moduli of $C_{g,n}$. 

Even though the general three-point function  $C(\alpha_1,\alpha_2,\alpha_3)$    is not known in Toda CFT, only  three-point functions involving a semi-degenerate primary  $C({\rm \hat m},\cdot\,,\cdot)$ enter in the calculation of the partition function of ${\cal N}=2^*$ and ${\cal N}=2$ conformal SQCD in a suitable choice of pants decomposition. Precisely the formula for the three-point function with a semi-degenerate primary and two non-degenerate primaries was proposed by Fateev and Litvinov~\cite{Fateev:2007ab,Fateev:2008bm} (see Appendix~\ref{sec:Toda} for the explicit formula). 
  
Conformal blocks in theories with $W_N$-symmetry  are generically not fixed by the $W_N$-algebra and therefore not explicitly computable~\cite{Bowcock:1993wq}. However,  the conformal blocks relevant for the computation of the partition function of ${\cal N}=2^*$ and ${\cal N}=2$ conformal SQCD with $SU(N)$   gauge group, which are constructed by sewing three-point conformal blocks with a semi-degenerate field for each three-punctured sphere, are fixed by the $W_N$-algebra and can be computed explicitly~\cite{Bowcock:1993wq}.\footnote{This holds when the Toda correlation function on $C_{0,4}$ is written in the factorization channel where each simple puncture collides with a full puncture. The duality frame of  ${\cal N}=2$ conformal SQCD  corresponding to the factorization channel where the two full punctures (and respectively two simple punctures) collide   involves more exotic building blocks, discussed  in~\cite{Gaiotto:2009we}.}  
%These conformal blocks depend on the moduli of the punctured Riemann surface, the momenta of the primary operators  and the momentum vector  $\alpha$ labeling the sum over delta function normalizable primary states in the intermediate channel, which  correspond to non-degenerate representations with $\alpha=Q+ia$. 

The identifications \eqref{N=2*} and \eqref{SQCD} of gauge theory partition functions with expectation values of Toda CFT vertex operators are obtained by matching the different factors in the CFT correlators with gauge theory objects, which combine to reproduce the result 
\begin{equation}
{\cal Z}=\int [da]\, \overline{Z_\text{Nekrasov}(ia,m, \tau)}\, Z_\text{Nekrasov}(ia,m,\tau)
\label{Pestun}
\end{equation}
derived by Pestun \cite{Pestun:2007rz} using localization. Here, $a\equiv (a_1,\ldots, a_N)\in \bR^N$ with $\sum_{l=1}^N a_l=0$, and $[da]$  is the integration measure over the zero mode on $S^4$ of a scalar field\footnote{Which takes values on Hermitean Lie algebra generators. After integrating over the angular variables one gets the Vandermonde determinant $\prod_{e} e\cdot a=(-1)^{N(N-1)/2}\prod_{i<j} (a_i-a_j)^2$.} in the ${\cal N}=2$ vector multiplet. The expression\footnote{To avoid cluttered formulae, we let $m$ denote the mass parameters associated to full and simple punctures.}
\begin{equation}
Z_\text{Nekrasov}(ia,m, \tau)= Z_\text{cl}(ia, \tau) Z_\text{1-loop}(ia,m) Z_\text{inst}(ia,m, \tau)\,.
\label{nikita}
\end{equation}
is the equivariant instanton partition function on $\bR^4$ \cite{Nekrasov:2002qd,Nekrasov:2003rj} with equivariant parameters ${\epsilon_1=\epsilon_2=1}$,   and  $\tau$ is the complexified gauge theory coupling constant. 
The   first factor  $Z_\text{cl}(ia, \tau)$ in \eqref{nikita} describes the classical contribution
\beq
Z_\text{cl}(ia, \tau)= \exp\left(2\pi i \tau \vev{a,a}\right)\,,
\eeq
 and the other two factors are the  one loop and instanton contribution to the localization computation respectively (see~\cite{Nekrasov:2002qd,Nekrasov:2003rj} for explicit formulae).

The conformal blocks $\cF^{(\sigma)}_{\alpha,E}(\tau)$ / $\overline \cF^{(\sigma)}_{\alpha,E}(\tau)$ reproduce the products $Z_{\text{cl}}\times Z_{\text{inst}}$ and $\overline Z_{\text{cl}} \times \overline Z_{\text{inst}}$, and three-point functions combine to give the measure of integration multiplied by the one-loop contribution $|Z_\text{1-loop}|^2$, and one obtains the identifications \eqref{N=2*} and \eqref{SQCD}. More generally,   the partition function of a ${\cal N}=2$ $SU(N)$ superconformal quiver gauge theory where all gauge group factors are $SU(N)$ is captured by Toda CFT correlators with two full punctures and $n-2$ simple puntures in $C_{0,n}$ and $n$ simple punctures in $C_{1,n}$.

\subsection{Loop operators}
\label{sec:defect}

The goal of this paper is to calculate the expectation value of 't Hooft loop operators in four dimensional ${\cal N}=2$ gauge theories.
 As we shall explain, the proposal is that a Verlinde loop operator~\cite{Verlinde:1988sn} labeled by a completely degenerate representation of the $W_N$-algebra, and supported on a closed curve  traversing the tube in the pants decomposition of $C_{1,1}$ and $C_{0,4}$ describes an 't Hooft operator in  ${\cal N}=2^*$ and ${\cal N}=2$ conformal SQCD with $SU(N)$ gauge group. Before delving into the details of the computation in Section \ref{sec:movesloops} we introduce the required ingredients.

\subsubsection{Spectrum of 't Hooft operators in gauge theory}
\label{sec:loopy}

An 't Hooft loop operator inserts a Dirac monopole propagating around a closed (or noncompact) worldline in spacetime. It creates a non-trivial magnetic flux through the $S^2$  that surrounds the magnetic monopole. Locally around any point in the loop, the magnetic flux inserted by an 't Hooft operator  is
\begin{equation}
F=\frac{B}{2}d\Omega_2\,,
\label{mono}
\end{equation}
where $d\Omega_2$ is the volume form of $S^2$.  $B$ determines an embedding of a $U(1)$ Dirac monopole into the gauge group, or by conjugation an embedding of $U(1)$ into the maximal torus $\bT$ of the gauge group. This implies that $B$ is a coweight (or magnetic weight), that is, a vector in the coweight lattice $\Lambda_{cw}$ of the Lie algebra of the gauge group~\cite{Goddard:1976qe}.\footnote{More precisely, it is an element of the cocharacter lattice $\Lambda_{cochar}$, which is a sublattice of the coweight lattice $\Lambda_{cochar}\subset \Lambda_{cw}$.} Therefore, for gauge theories with an $SU(N)$ gauge group an 't Hooft operator is characterized by the diagonal matrix
\begin{align*}
  B=\left(\begin{array} {cccc} 
    n_1-\frac{|n|}{N} &0                &  \dots & 0                \cr
    0                 &n_2-\frac{|n|}{N}&  \dots & 0                \cr
    \vdots            & \vdots          & \ddots & \vdots           \cr
    0                 &0                &  \dots & n_N-\frac{|n|}{N}\cr
  \end{array}
  \right) \,,
\end{align*}
where $(n_1,\ldots,n_N)$ is a collection of integers satisfying $n_1\geq n_2\geq \ldots \geq n_N \geq 0$  and $|n|=\sum_i n_i$. The charge vector $(n_1,\ldots,n_N)$ is a   weight vector\footnote{The coweight lattice and weight lattice  of $SU(N)$ are isomorphic.} of $SU(N)$, which can be associated an irreducible representation labeled by  a Young tableau $Y$ with $n_i$ boxes in the $i$-th row.

In general, not all choices of $(n_1,\ldots,n_N)$ give rise to a consistent 't Hooft operator in a given gauge theory. The gauge field configuration created by the 't Hooft operator has a Dirac string singularity, and   consistent 't Hooft operators are those for which the Dirac string is invisible. If the gauge theory has a matter field  transforming in a representation $R$ of the gauge group, then upon encircling the Dirac string, the matter field acquires a phase, given by an element of the center $Z(G)$ of the gauge group $G$
\begin{equation*}
\exp\left( 2\pi i\rho_R(B)\right)\in Z(G)\,,
\end{equation*}
where $\rho_R(B)$ is the matrix $B$ in the representation $R$.  Therefore, consistent 't Hooft operators are labeled by coweights $B$ satisfying $\exp\left( 2\pi i\rho_R(B)\right)=1$ for all representations $R$ for which there are matter fields present in the gauge theory.

For the theories of interest in this paper, we have matter fields in the adjoint representation of $SU(N)$ for ${\cal N}=2^*$ while ${\cal N}=2$ conformal SQCD has matter fields in the fundamental representation of $SU(N)$. This implies that the spectrum of 't Hooft operators in these theories are labeled by:
\begin{itemize}
\item ${\cal N}=2^*$: arbitrary highest weight of $SU(N)$; arbitrary Young tableau $Y$.
\item ${\cal N}=2$ conformal SQCD: highest weight  of $SU(N)$ neutral under the action of the center $\mathbb{Z}_N$; Young tableau $Y$ with $|n|$  a multiple of $N$.
\end{itemize}
%We will compute in both theories the expectation value of the simplest 't Hooft operators.

\subsubsection{Loop  operators and topological defects in 2d  CFT}  
 \label{sec:moves}

%Computation of exact expectation values of  't Hooft operators in ${\cal N}=2^*$ and ${\cal N}=2$ conformal SQCD (via Toda CFT): generalizes the proposal of~\cite{Alday:2009fs,Drukker:2009id} to an arbitrary $SU(N)$ gauge group.

A Verlinde loop operator~\cite{Verlinde:1988sn} in a two dimensional CFT with a chiral algebra ${\cal A}$ is labeled by a representation of ${\cal A}$ and is supported  on  an oriented  curve $p$ on a  Riemann surface $C_{g,n}$.

Physically, a loop operator  in $C_{g,n}$ ---which we denote by ${\cal O}_\mu(p)$--- encodes the action of quantum monodromy on the space of conformal blocks   in $C_{g,n}$ induced by  transporting the {\it chiral} vertex operator $V_{\mu}(z)$ around a closed curve $p$ on the Riemann surface $C_{g,n}$. The loop operator ${\cal O}_\mu(p)$ acting on the space of conformal blocks is defined by the following procedure~\cite{Verlinde:1988sn}:
\begin{enumerate}
\item Insert the identity operator $1$ into the trivalent graph $\Gamma_\sigma$ labeling the conformal blocks  ${\cal F}_{\alpha,E}^{(\sigma)}$ in  a choice of pants decomposition $\sigma$ of $C_{g,n}$  

\item Resolve the $1$ operator as the fusion of two conjugate chiral operators $V_{\mu}(z)$ and $V_{\mu^*}(z)$

\item Transport the chiral operator $V_{\mu}(z)$ along the closed curve $p$ in $C_{g,n}$

\item Fuse the two conjugate chiral operators $V_{\mu}(z)$ and $V_{\mu^*}(z)$ back to the identity operator
 \end{enumerate}
The action of the loop operator ${\cal O}_\mu(p)$  on   conformal blocks then takes the form
\begin{equation*}
{\cal F}_{\alpha,E}^{(\sigma)}\mapsto \left[{\cal O}_\mu(p)\, \cdot {\cal F}^{(\sigma)}\right]_{\alpha,E}=\int  d\nu(\alpha')  \,{\cal O}(\alpha,\alpha')\,{\cal F}_{\alpha',E}^{(\sigma)}(\tau)\,.
\end{equation*}
Therefore, a CFT  correlation function in $C_{g,n}$ in the presence of the loop operator
 ${\cal O}_\mu(p)$ is given by
\begin{equation*}
\vev{{\cal O}_\mu(p)}_{C_{g,n}}=\int d\nu(\alpha')d\nu(\alpha) \overline
{\cal F}_{\alpha,E}^{(\sigma)}(\tau)\,{\cal O}(\alpha,\alpha')\,{\cal F}_{\alpha',E}^{(\sigma)}(\tau)\,,
\end{equation*}
where the measure factor $\nu(\alpha)$ includes the product of the three point functions that appear in the sewing of $C_{g,n}$ from pairs of pants, that is, one for each vertex in $\Gamma_\sigma$.

This construction yields   an operator which encodes the monodromy action    obtained by   analytically continuing     conformal blocks.
The analytic continuation of conformal blocks is generated by three elementary transformations~\cite{Moore:1988qv} (see~\cite{Moore:1989vd} for a review)
\begin{itemize}
\item
Fusion move
\item
Braiding move
\item 
S-move. 
\end{itemize}
which generate the Moore-Seiberg groupoid ${\cal G}$. For each of these moves there is a matrix that implements the corresponding analytic continuation. The total monodromy induced by ${\cal O}_\mu(p)$ is obtained by multiplying together the matrices associated to the sequence of elementary moves that are generated by transporting the operator $V_{\mu}(z)$ around $p$. Since   monodromy is left invariant under smooth deformations of the curve $p$, the operator  ${\cal O}_\mu(p)$ depends only on the homotopy class of $p$.

Verlinde loop operators in CFT admit an    alternative description~\cite{Drukker:2010jp}  in terms of   topological defect operators  in CFT.\footnote{See e.g. \cite{Petkova:2000ip,Petkova:2001ag,Fuchs:2002cm,Bachas:2004sy,Sarkissian:2009aa}. Topological defect operators   have provided insights into symmetries, duality transformations, RG flows and boundary states in CFTs: \cite{Quella:2002ct,Graham:2003nc,Frohlich:2004ef,Frohlich:2006ch,Quella:2006de,Runkel:2007wd,Bachas:2007td}.}
These operators are  supported on oriented curves on a Riemann surface $C_{g,n}$ and commute with two copies of the chiral algebra ${\cal A}$, and therefore are labeled by a representation of  ${\cal A}$. Furthermore, since they are invariant under conformal transformations, the curve $p$ can be smoothly deformed without changing the operator, which therefore depends only on the homotopy class of the curve. Changing the orientation of the curve yields a topological defect operator labeled by the conjugate representation, that is $\cO_\mu(p^{-1})=\cO_{\mu^*}(p)$.

This formulation of loop operators in CFT  yields an elementary way of computing the expectation of a loop operator ${\cal O}_\mu(p)$    supported on a curve $p$ that wraps a tube in a pants decomposition $\sigma$ of $C_{g,n}$. For this choice of curve and pants decomposition,   conformal blocks are eigenstates of the operator ${\cal O}_\mu(p)$  and the correlation function is   given by\footnote{The 1 in $S_{1\alpha}$ denotes the trivial representation.}
\begin{equation}
\vev{{\cal O}_\mu(p)}_{C_{g,n}}=\int  d\nu(\alpha) \frac{S_{\mu\alpha}}{S_{1\alpha}}\,\overline
{\cal F}_{\alpha,E}^{(\sigma)}(\tau)\, {\cal F}_{\alpha,E}^{(\sigma)}(\tau)\,,
\label{Wilson}
\end{equation}
where $ S_{\mu\alpha}$ are the modular matrices of the CFT  relating the torus conformal blocks $\chi_\mu(\tau)=\sum_{\alpha} S_{\mu\alpha} \chi_\alpha(-1/\tau)$.

%===================================================
\section{'t Hooft loops in ${\cal N}=2^*$ and ${\cal N}=2$ conformal SQCD}
 \label{sec:movesloops}
 
In this section we compute the expectation value of 't Hooft   and Wilson-t' Hooft loop operators in  ${\cal N}=2^*$ and ${\cal N}=2$ conformal SQCD with $SU(N)$ gauge group by mapping the problem to correlation functions with loop operators in Toda CFT.
This extends to an arbitrary $SU(N)$ gauge group the proposal put forward in~\cite{Alday:2009fs,Drukker:2009id} identifying $SU(2)$ gauge theory 't Hooft loop operators with   loop operators in Liouville theory. Even though the analog of the classification~\cite{Drukker:2009tz} of loop operators in terms of data on the Riemann surface $C_{g,n}$ is not available for higher rank gauge theories, we obtain  explicit formulae for the expectation value of certain 't Hooft and Wilson-'t Hooft operators in ${\cal N}=2^*$ and ${\cal N}=2$ conformal SQCD with $SU(N)$ gauge group.
 
 Verlinde loop operators (and topological defects) in $A_{N-1}$ Toda CFT are labeled by a representation of the $W_N$-algebra (see e.g~\cite{Drukker:2010jp} for a recent description of the various representations). As mentioned in Section \ref{sec:partition},  representations of the $W_N$-algebra are encoded by the momentum vector $\alpha$ of the vertex operator \eqref{vertex} which creates the primary state in the representation. We denote by ${\cal O}_\mu(p)$ a Toda CFT loop operator supported on the curve $p$ in $C_{g,n}$ and labeled by the representation of the $W_N$-algebra corresponding to the Toda momentum vector $\alpha=\mu$.

The $A_{N-1}$ Toda CFT   operators ${\cal O}_\mu(p)$  representing  Wilson and 't Hooft loop operators in four dimensional ${\cal N}=2$ gauge theories are labeled by  completely degenerate representations of the $W_N$-algebra. These representations are characterized by primary operators  \eqref{vertex} with Toda momentum \cite{Fateev:1987zh}
\begin{equation*}
\mu=-b\lambda_1-\frac{1}{b} \lambda_2\,,
\end{equation*}
where $\lambda_1$ and $\lambda_2$ are   highest weight vectors of the $A_{N-1}$ Lie algebra corresponding to   representations $R_1$ and $R_2$ of $SU(N)$. 

Since Toda CFT captures the dynamics of gauge theories when $b=1$, the relevant Toda loop operators can be taken without loss of generality to carry momentum
\begin{equation}
\mu=-b\lambda_1\,,
\label{degen}
\end{equation}
where $\lambda_1$ is the highest weight vector of a representation $R$ of $SU(N)$.
The gauge theory interpretation of these  loop operators ${\cal O}_\mu(p)$ for curves $p$ wrapping a tube in the pants decomposition of $C_{g,n}$   was found in~\cite{Drukker:2010jp} by realizing  them in terms of topological defect operators in Toda CFT.

\begin{figure}[t]
\begin{center}
\begin{tabular}{cccccc}
\includegraphics[scale=1.0]{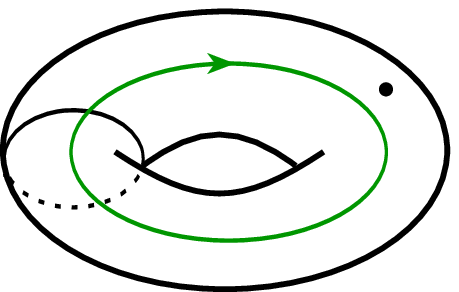}
&\hspace{30mm}&
\includegraphics[scale=1.0]{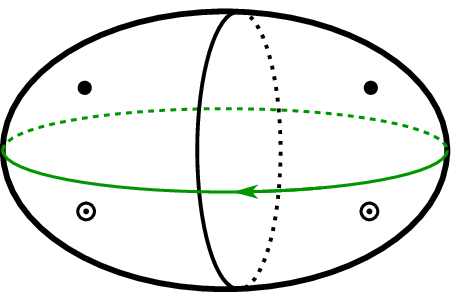}
&\hspace{30mm}&
\includegraphics[scale=1.0]{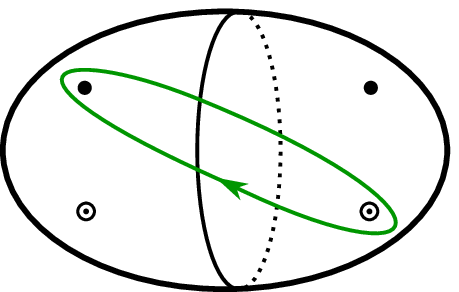}
\\
\\
(a)&&(b)&&(c)
\end{tabular}
\parbox{5in}{
\caption{Toda loop operators describing 't Hooft operators in   ${\cal N}=2^*$ and ${\cal N}=2$ conformal SQCD. The green curves denote the support of the loop operator and the black ones define the pants decomposition.}
\label{fig:tHooft}}
\end{center}
\end{figure}

The Toda CFT correlation function with the insertion of  a loop operator ${\cal O}_\mu(p)$ labeled by a   representation $R$ of $SU(N)$   and wrapping a tube in the pants decomposition\footnote{For example wrapping the black curves in Figure \ref{fig:tHooft}.} $\sigma$ of $C_{g,n}$   computes the expectation value of a Wilson loop operator in a   representation $R$ of 
 $SU(N)$ in the corresponding four dimensional ${\cal N}=2$ gauge theory~\cite{Drukker:2010jp}.\footnote{Topological defects labeled by a semi-degenerate representation of the $W_N$-algebra ---which are equivalent to Verlinde loop operators for quasi-rational theories and extend them for non quasi-rational theories--- were shown to describe domain walls in the four dimensional gauge theory~\cite{Drukker:2010jp}.} Using the   formula \eqref{Wilson} for   CFT correlators in the presence of such a topological defect operator  one gets\footnote{$a$ takes values in the Lie algebra with Hermitean generators.} 
\begin{equation*}
\vev{{\cal O}_\mu(p)}_{C_{g,n}}=\vev{W_R}=\int [da]\, \Tr_R(e^{2\pi a})\,\overline{Z_\text{Nekrasov}(ia,m,\tau)}\, Z_\text{Nekrasov}(ia,m, \tau)\,,
\end{equation*}
thus reproducing the gauge theory computation of Pestun~\cite{Pestun:2007rz}.
This was demonstrated in~\cite{Drukker:2010jp} by explicit computation of the characters  and   the modular matrices of degenerate representations in Toda CFT. This extended the proposal put forward in~\cite{Alday:2009fs,Drukker:2009id} to Wilson loop operators in arbitrary representations $R$  and arbitrary gauge group.

 The simplest  't Hooft loop operators are described by Toda CFT loop operators ${\cal O}_\mu(p)$  which cross a tube in the pants decomposition $\sigma$ of $C_{g,n}$ (see the green curves in Figure \ref{fig:tHooft}). These 
 can be obtained by the action of elements in the Moore-Seiberg groupoid ${\cal G}$  (which includes the mapping class group of the Riemann surface $C_{g,n}$)
  on the pants decomposition $\sigma'$  of $C_{g,n}$  in which the curve  $p$ wraps a tube defining the pants decomposition. Since the Moore-Seiberg groupoid  of $C_{g,n}$ corresponds to the S-duality group(oid) of the gauge theory, 
   this implies  that the gauge theory loop operators that are described by a Toda CFT loop operator  ${\cal O}_\mu(p)$  supported on a non-selfintersecting curve on $C_{g,n}$ are those that are in the S-duality orbit of any purely electric loop operator, that is, in the orbit of any Wilson loop operator. For gauge theory loop operators not in the orbit in of Wilson loop operators, one must consider other observables in the Toda CFT (see the discussion Section for a natural set of operators that may accomplish this).

The CFT computation of 't Hooft operators is significantly more involved than the computation of Wilson loop operators, just as it is in gauge theory.
The chiral field $V_{\mu}(z)$ has non-trivial fusion with the primary operators labeling the internal edges of the trivalent graph $\Gamma_\sigma$ which are being traversed by the path $p$. 
Therefore, the conformal blocks for this choice of pants decomposition $\sigma$ are not   eigenfunctions of the loop operator ${\cal O}_\mu(p)$, as the quantum numbers on the trivalent graph $\Gamma_\sigma$ are shifted upon circumnavigating $V_{\mu}(z)$ along $p$. 
This differs from the case of a  loop operator ${\cal O}_\mu(p)$ wrapping a tube in the pants decomposition $\sigma'$,  since in this case $p$ does not traverse the trivalent graph $\Gamma_{\sigma'}$ and the operator does not change the quantum numbers of the conformal blocks, which are therefore eigenstates of ${\cal O}_\mu(p)$.  Consequently, we expect from this argument that the expectation value of an 't Hooft loop takes the form 
 \begin{equation*}
\vev{T}=\int d\nu(\alpha')d\nu(\alpha) \overline
{\cal F}_{\alpha,E}^{(\sigma)}(\tau)\,{\cal O}(\alpha,\alpha')\,{\cal F}_{\alpha',E}^{(\sigma)}(\tau)\,.
\end{equation*}
The goal is to compute the kernel ${\cal O}(\alpha,\alpha')$.
 
We consider the 't Hooft operators in   ${\cal N}=2^*$ and ${\cal N}=2$ conformal SQCD corresponding to  the $A_{N-1}$ Toda CFT loop operator ${\cal O}_\mu(p)$   labeled by the highest weight vector of the fundamental representation of $SU(N)$, that is   
\begin{equation*}
\mu=-b \omega_1=-b h_1\,,
\end{equation*}
 where $\omega_1=h_1$ is the fundamental weight of the $A_{N-1}$  Lie algebra corresponding to the first node of the  $A_{N-1}$  Dynkin diagram and 
 \begin{equation*}
 (h_1,h_2,\ldots, h_N)
 \end{equation*}
are the $N$ weights of the fundamental representation of $SU(N)$.

\subsection{Fusion and braiding matrices in Toda CFT}
In order to   construct   the Verlinde loop operator ${\cal O}_\mu(p)$ we need to first fuse the chiral fields $V_{\mu}(z)$ with $\mu= -b \omega_1$, and $V_{\mu^*}(z)$ with $\mu^*=-b\omega_{N-1}$, which create primary states of two conjugate degenerate representations of the $W_N$-algebra. The vector $\omega_{N-1}=-h_N$  is the fundamental weight of the $A_{N-1}$  Lie algebra associated to last node in the Dynkin diagram. The goal is to calculate the monodromy acquired by $V_{\mu}(z)$ as it encircles the curve $p$ traversing the tube in the pants decomposition of $C_{1,1}$ and $C_{0,4}$ (see   Figure \ref{fig:tHooft}).

 As explained in Section \ref{sec:moves} the monodromy action on conformal blocks can be computed by combining sequentially the elementary moves in the Moore-Seiberg groupoid ${\cal G}$. For the computation of 't Hooft loops in ${\cal N}=2$ $SU(N)$ superconformal quiver gauge theories we need to determine  certain entries of the  fusion and braiding matrices of the $A_{N-1}$ Toda CFT, which relate the conformal blocks on the four punctured sphere in the three different pants decomposition of the four punctured sphere, conventionally called by $s$, $t$ and $u$-channels:  
  \begin{align*}
  \cF^{(s)}_{\al}\big[\begin{smallmatrix} \al_3 & \al_2\\ \al_4 & \al_1\end{smallmatrix}\big]
  \equiv
  \begin{matrix}
  \begin{fmfgraph*}(30,15)
    \fmfstraight
    \fmftop{tl,t1,t2,t3,tr}
    \fmfbottom{bl,b1,b2,b3,br}
    \fmf{fermion,width=2}{br,b3,b1,bl}
    \fmf{fermion,width=2}{t1,b1}
    \fmf{fermion,width=2}{t3,b3}
    \fmfv{label=$\alpha_4$,label.angle=-60,label.dist=6}{bl}
    \fmf{phantom,label=$\alpha$,label.side=left}{b3,b1}
    \fmfv{label=$\alpha_1$,label.angle=-120,label.dist=6}{br}
    \fmfv{label=$\alpha_2$,label.angle=0,label.dist=4}{t3}
    \fmfv{label=$\alpha_3$,label.angle=180,label.dist=4}{t1}
  \end{fmfgraph*}\end{matrix}
&&
  \cF^{(t)}_{\al}\big[\begin{smallmatrix} \al_3 & \al_2\\ \al_4 & \al_1\end{smallmatrix}\big]\equiv
  \begin{matrix}
  \begin{fmfgraph*}(20,15)
    \fmfstraight
    \fmftop{tl,t1,t2,t3,tr}
    \fmfbottom{bl,b2,br}
    \fmf{fermion,width=2}{br,b2,bl}
    \fmf{fermion,width=2}{t1,v}
    \fmf{fermion,width=2}{t3,v}
    \fmf{fermion,width=2,label=$\alpha$}{v,b2} 
    \fmfv{label=$\alpha_4$,label.angle=-60,label.dist=8}{bl}
    \fmfv{label=$\alpha_1$,label.angle=-120,label.dist=8}{br}
    \fmfv{label=$\alpha_3$,label.angle=180,label.dist=4}{t1}
    \fmfv{label=$\alpha_2$,label.angle=0,label.dist=4}{t3}
  \end{fmfgraph*}
  \end{matrix}
&&
 \cF^{(u)}_{\al}\big[\begin{smallmatrix} \al_3 & \al_2\\ \al_4 & \al_1\end{smallmatrix}\big]\equiv
 \begin{matrix}
   \begin{fmfgraph*}(30,15)
     \fmfstraight
     \fmftop{tl,t1,t2,t3,t4,tr}
     \fmfbottom{bl,b1,b2,b3,b4,br}
     \fmf{fermion,width=2}{br,b4,b1,bl}
     \fmf{fermion,width=2,right=0.4}{t4,b1}
     \fmf{fermion,width=2,left=0.4,rubout}{t1,b4}
     \fmf{phantom,tension=0,label=$\alpha$,label.side=left}{b4,b1}
     \fmfv{label=$\alpha_1$,label.angle=120,label.dist=6}{br}
     \fmfv{label=$\alpha_2$,label.angle=0,label.dist=4}{t4}
     \fmfv{label=$\alpha_3$,label.angle=180,label.dist=4}{t1}
     \fmfv{label=$\alpha_4$,label.angle=-60,label.dist=6}{bl}
   \end{fmfgraph*}
 \end{matrix}
 \end{align*}
 \vskip+5pt
\noindent
 The fusion and braiding moves are:
\begin{align*}
  \begin{matrix}
  \fmfframe(0,3)(0,3){\begin{fmfgraph*}(30,15)
    \fmfstraight
    \fmftop{tl,t1,t2,t3,tr}
    \fmfbottom{bl,b1,b2,b3,br}
    \fmf{fermion,width=2}{br,b3,b1,bl}
    \fmf{fermion,width=2}{t1,b1}
    \fmf{fermion,width=2}{t3,b3}
    \fmfv{label=$\alpha_4$,label.angle=-60,label.dist=6}{bl}
    \fmf{phantom,label=$\alpha$,label.side=left}{b3,b1}
    \fmfv{label=$\alpha_1$,label.angle=-120,label.dist=6}{br}
    \fmfv{label=$\alpha_2$,label.angle=0,label.dist=4}{t3}
    \fmfv{label=$\alpha_3$,label.angle=180,label.dist=4}{t1}
  \end{fmfgraph*}}\end{matrix}
& \equiv
  \int d\al'\;
  F_{\al\al'}\big[\begin{smallmatrix} \al_3 & \al_2\\ \al_4 & \al_1\end{smallmatrix}\big]
  \begin{matrix}
  \fmfframe(0,3)(0,3){\begin{fmfgraph*}(20,15)
    \fmfstraight
    \fmftop{tl,t1,t2,t3,tr}
    \fmfbottom{bl,b2,br}
    \fmf{fermion,width=2}{br,b2,bl}
    \fmf{fermion,width=2}{t1,v}
    \fmf{fermion,width=2}{t3,v}
    \fmf{fermion,width=2,label=$\alpha'$}{v,b2} 
    \fmfv{label=$\alpha_4$,label.angle=-60,label.dist=8}{bl}
    \fmfv{label=$\alpha_1$,label.angle=-120,label.dist=8}{br}
    \fmfv{label=$\alpha_3$,label.angle=180,label.dist=4}{t1}
    \fmfv{label=$\alpha_2$,label.angle=0,label.dist=4}{t3}
  \end{fmfgraph*}}
  \end{matrix}
\\
  \begin{matrix}
  \fmfframe(0,3)(0,3){\begin{fmfgraph*}(30,15)
    \fmfstraight
    \fmftop{tl,t1,t2,t3,tr}
    \fmfbottom{bl,b1,b2,b3,br}
    \fmf{fermion,width=2}{br,b3,b1,bl}
    \fmf{fermion,width=2}{t1,b1}
    \fmf{fermion,width=2}{t3,b3}
    \fmfv{label=$\alpha_4$,label.angle=-60,label.dist=6}{bl}
    \fmf{phantom,label=$\alpha$,label.side=left}{b3,b1}
    \fmfv{label=$\alpha_1$,label.angle=-120,label.dist=6}{br}
    \fmfv{label=$\alpha_2$,label.angle=0,label.dist=4}{t3}
    \fmfv{label=$\alpha_3$,label.angle=180,label.dist=4}{t1}
  \end{fmfgraph*}}\end{matrix}
& \equiv
  \int d\al'\;
  B_{\al\al'}^{\epsilon}\big[\begin{smallmatrix} \al_3 & \al_2\\ \al_4 & \al_1\end{smallmatrix}\big]
 \begin{matrix}
   \fmfframe(0,3)(0,3){\begin{fmfgraph*}(30,15)
     \fmfstraight
     \fmftop{tl,t1,t2,t3,t4,tr}
     \fmfbottom{bl,b1,b2,b3,b4,br}
     \fmf{fermion,width=2}{br,b4,b1,bl}
     \fmf{fermion,width=2,right=0.4}{t4,b1}
     \fmf{fermion,width=2,left=0.4,rubout}{t1,b4}
     \fmf{phantom,tension=0,label=$\alpha'$,label.side=left}{b4,b1}
     \fmfv{label=$\alpha_1$,label.angle=-120,label.dist=6}{br}
     \fmfv{label=$\alpha_2$,label.angle=0,label.dist=4}{t4}
     \fmfv{label=$\alpha_3$,label.angle=180,label.dist=4}{t1}
     \fmfv{label=$\alpha_4$,label.angle=-60,label.dist=6}{bl}
   \end{fmfgraph*}}
 \end{matrix}
\end{align*}
  $F_{\al\al'}^{}\big[\begin{smallmatrix} \al_3 & \al_2\\ \al_4 & \al_1\end{smallmatrix}\big]$ and $B_{\al\al'}^{\epsilon}\big[\begin{smallmatrix} \al_3 & \al_2\\ \al_4 & \al_1\end{smallmatrix}\big]$
are the fusion and braiding matrices of the CFT, where  $\epsilon=\pm 1$ determines the direction of the braiding. 
  The explicit derivation of the relevant matrices for the computation of 't Hooft loops in ${\cal N}=2$ $SU(N)$ superconformal quiver gauge theories is performed in Appendix \ref{sec:fusion}, and it yields the following results:\footnote{We recall that $q=b+1/b$.}
 \vskip+1pt
\begin{equation}
\label{fusion1}
\begin{matrix}
  \begin{fmfgraph*}(20,25)
    \fmftop{tl,tr}
    \fmfbottom{bl,br}
    \fmf{fermion,width=2}{br,b2,bl}
    \fmffreeze
    \fmf{fermion,width=0.5}{tl,v}
    \fmf{fermion,width=0.5}{tr,v}
    \fmf{dots_arrow,width=2,label=id}{v,b2} 
    \fmfv{label=$\alpha$,label.angle=-60,label.dist=6}{bl}
    \fmfv{label=$\alpha$,label.angle=-120,label.dist=6}{br}
    \fmfv{label=$\mu^*$,label.angle=30,label.dist=4}{tl}
    \fmfv{label=$\mu$,label.angle=150,label.dist=4}{tr}
  \end{fmfgraph*}
\end{matrix}
=
\frac{\Gamma(Nbq)}{\Gamma(bq)}
\sum_{l=1}^N 
 \prod_{j\neq l}
  \frac{\Gamma(b\langle \alpha-Q,h_j - h_l\rangle)}
  {\Gamma(bq+ b\langle \alpha-Q,h_j-h_l\rangle)}
\begin{matrix}
  \begin{fmfgraph*}(30,15)
    \fmftop{tl,tr}
    \fmfbottom{bl,br}
    \fmf{fermion,width=2}{br,b1,b3,bl}
    \fmf{phantom}{br,b1} \fmf{phantom}{b3,bl}
    \fmf{phantom}{tr,t1,t2,t3,tl}
    \fmffreeze
    \fmf{fermion,width=0.5}{t1,b1}
    \fmf{fermion,width=0.5}{t3,b3}
    \fmfv{label=$\alpha$,label.angle=-60,label.dist=6}{bl}
    \fmf{phantom,label=$\alpha-bh_l$,label.side=left}{b1,b3}
    \fmfv{label=$\alpha$,label.angle=-120,label.dist=6}{br}
    \fmfv{label=$\mu$,label.angle=180,label.dist=4}{t1}
    \fmfv{label=$\mu^*$,label.angle=0,label.dist=4}{t3}
  \end{fmfgraph*}
\end{matrix}
\end{equation}

\begin{equation}
\label{fusion2}
\pr\left[
\begin{matrix}
  \begin{fmfgraph*}(30,15)
    \fmftop{tl,tr}
    \fmfbottom{bl,br}
    \fmf{fermion,width=2}{br,b1,b3,bl}
    \fmf{phantom}{br,b1} \fmf{phantom}{b3,bl}
    \fmf{phantom}{tr,t1,t2,t3,tl}
    \fmffreeze
    \fmf{fermion,width=0.5}{t1,b1}
    \fmf{fermion,width=0.5}{t3,b3}
    \fmfv{label=$\alpha'$,label.angle=-60,label.dist=6}{bl}
    \fmf{phantom,label=$\alpha-bh_l$,label.side=left}{b1,b3}
    \fmfv{label=$\alpha$,label.angle=-120,label.dist=6}{br}
    \fmfv{label=$\mu$,label.angle=180,label.dist=4}{t1}
    \fmfv{label=$\mu^*$,label.angle=0,label.dist=4}{t3}
  \end{fmfgraph*}
\end{matrix}
\right]
= %\delta_{\alpha\alpha'}
\frac{\Gamma(1-Nbq)}{\Gamma(1-bq)}
 \prod_{j\neq l}
  \frac{\Gamma(1-b\langle \alpha-Q, h_j-h_l\rangle)}
  {\Gamma(-b^2-b\langle \alpha-Q,h_j-h_l\rangle)}
\begin{matrix}
  \fmfframe(0,5)(0,5){
  \begin{fmfgraph*}(20,25)
    \fmftop{tl,tr}
    \fmfbottom{bl,br}
    \fmf{fermion,width=2}{br,b2,bl}
    \fmffreeze
    \fmf{fermion,width=0.5}{tl,v}
    \fmf{fermion,width=0.5}{tr,v}
    \fmf{dots_arrow,width=2,label=id}{v,b2} 
    \fmfv{label=$\alpha'$,label.angle=-60,label.dist=6}{bl}
    \fmfv{label=$\alpha$,label.angle=-120,label.dist=6}{br}
    \fmfv{label=$\mu^*$,label.angle=30,label.dist=4}{tl}
    \fmfv{label=$\mu$,label.angle=150,label.dist=4}{tr}
  \end{fmfgraph*}}
\end{matrix}
\end{equation}
In \eqref{fusion2}, we project onto the $t$-channel conformal block with the identity as an internal state. The fusion of $V_\alpha$ with the identity operator $\id$ in the diagram on the right-hand side of \eqref{fusion2} has only one term, $V_\alpha$. Hence, for $\alpha\neq \alpha'$, the right hand side is zero. 

The braiding matrix is given by
\begin{align}
\notag
&\begin{matrix}
  \begin{fmfgraph*}(35,15)
    \fmfstraight
    \fmftop{tl,t1,t2,t3,tr}
    \fmfbottom{bl,b1,b2,b3,br}
    \fmf{fermion,width=2}{br,b3,b1,bl}
    \fmf{fermion,width=2}{t3,b3}
    \fmf{fermion,width=0.5}{t1,b1}
    \fmfv{label=$\alpha_2$,label.angle=60,label.dist=6}{bl}
    \fmf{phantom,label=$\alpha_2+bh_l$,label.side=right}{b3,b1}
    \fmfv{label=$\alpha_1$,label.angle=120,label.dist=6}{br}
    \fmfv{label=$\mu$,label.angle=180,label.dist=4}{t1}
    \fmfv{label=${\rm \hat m}$,label.angle=0,label.dist=4}{t3}
  \end{fmfgraph*}
\end{matrix}
=
\sum_{k=1}^N 
e^{i\pi\epsilon \varphi} \prod_{j\neq l} 
\frac{\Gamma(1+ b\langle \alpha_2-Q, h_j - h_l\rangle)}
{\Gamma(1+b\langle \alpha_2-Q,h_j\rangle 
        - b\langle \alpha_1-Q,h_k\rangle
        - b\langle \mu+{\rm \hat m}, h_1\rangle)}  
\cdot
\\  
\label{braiding}
&\cdot
 \prod_{j\neq k}
\frac{\Gamma(b\langle \alpha_1-Q, h_j- h_k\rangle)}
{\Gamma(b\langle \alpha_1-Q, h_j\rangle
        - b\langle \alpha_2-Q, h_l\rangle  
        + b\langle \mu+{\rm \hat m}, h_1\rangle )}
\begin{matrix}
  \begin{fmfgraph*}(35,15)
    \fmfstraight
    \fmftop{tl,t1,t2,t3,tr}
    \fmfbottom{bl,b1,b2,b3,br}
    \fmf{fermion,width=2}{br,b3,b1,bl}
    \fmffreeze
    \fmf{fermion,width=2}{t1,b1}
    \fmf{fermion,width=0.5}{t3,b3}
    \fmfv{label=$\alpha_2$,label.angle=60,label.dist=5}{bl}
    \fmf{phantom,label=$\alpha_1-b h_k$,label.side=left}{b1,b3}
    \fmfv{label=$\alpha_1$,label.angle=120,label.dist=5}{br}
    \fmfv{label=$\mu$,label.angle=180,label.dist=4}{t3}
    \fmfv{label=${\rm \hat m}$,label.angle=180,label.dist=4}{t1}
  \end{fmfgraph*}
\end{matrix}
\end{align}
with $\varphi=b\langle \alpha_2+b h_l-Q, h_l\rangle - b \langle \alpha_1-Q, h_k\rangle$. 

In Appendix \ref{sec:Wilson} we use these formulae to prove that the $A_{N-1}$ Toda  CFT loop operator ${\cal O}_\mu(p^r)$ wrapping $r$ times a curve $p$ defining a pants decomposition   of $C_{g,n}$ captures the expectation value of the Wilson loop operator (the trace is in the fundamental representation of $SU(N)$)
\begin{equation*}
\frac{1}{N}\vev{\hbox{Tr}\left(e^{2\pi  r a}\right)}_{{\cal N}=2}
\end{equation*} 
in the corresponding ${\cal N}=2$ gauge theory with $SU(N)$ gauge group. This reproduces the results in~\cite{Drukker:2010jp}  and~\cite{Passerini:2010pr} obtained using different methods.

\subsection{'t Hooft loop in ${\cal N}=2^*$}

In this section, we compute the Toda CFT correlator on $C_{1,1}$ in the presence of the loop operator ${\cal O}_\mu(p_0)$ wrapping the curve $p_0$ which traverses the tube in the pants decomposition $\sigma$ of $C_{1,1}$ (see Figure \ref{fig:tHooft}~(a)). More generally, we consider ${\cal O}_\mu(p_n)$, where the curve $p_n$ circles once around the b-cycle and $n$ times around the a-cycle of the torus. The Toda momentum at the puncture is semi-degenerate, with ${\rm \hat m}= N\left({q/2}+i\hat m\right)\, \omega_{N-1}$. Since we are transporting a completely degenerate field $V_{\mu}(z)$, which upon fusion with any primary in the CFT yields a finite number of representations (see Appendix \ref{sec:Toda}), we have that 
\begin{equation}
{\cal O}_\mu (p_n)\cdot {\cal F}^{(\sigma)}_{\alpha,{\rm \hat m}}=\sum_{\alpha'}{\cal O}_\mu(p_n)_{\alpha,\alpha'}{\cal F}^{(\sigma)}_{\alpha',{\rm \hat m}}\,.
\label{matrixele}
\end{equation}

\begin{figure}[t]
\begin{center}
\begin{align*}
&\begin{matrix}
\fmfframe(0,-8)(0,-8){
  \begin{fmfgraph*}(40,40)
    \fmfcmd{style_def identity expr p = 
      cdraw (subpath (.05length(p),.15length(p)) of p);
      cdraw (subpath (.25length(p),.35length(p)) of p);
      cdraw (subpath (.45length(p),.55length(p)) of p);
      cdraw (subpath (.65length(p),.75length(p)) of p);
      cdraw (subpath (.85length(p),.95length(p)) of p);
      shrink(.5); cfill (marrow (p,.5)); endshrink; 
      enddef;}
    \fmfcurved
    \fmfsurroundn{o}{8}
    \fmf{phantom,tension=.5}{o1,v1}
    \fmf{phantom,tension=.5}{o2,v2}
    \fmf{phantom,tension=.5}{o3,v3}
    \fmf{phantom,tension=.5}{o4,v4}
    \fmf{phantom,tension=.5}{o5,v5}
    \fmf{phantom,tension=.5}{o6,v6}
    \fmf{phantom,tension=.5}{o7,v7}
    \fmf{phantom,tension=.5}{o8,v8}
    \fmfcyclen{phantom}{v}{8} 
    \fmffreeze
    \fmf{fermion,width=2,right,label=$\alpha$,label.side=left}{v1,v5}
    \fmf{fermion,width=2,right,label=$\alpha$,label.side=left}{v5,v1}
    \fmf{fermion,width=2}{o1,v1}
    \fmf{phantom}{o4,o45,o5,o55,o6}
    \fmffreeze
    \fmf{fermion,width=.5}{o45,oo}
    \fmf{fermion,width=.5}{o55,oo}
    \fmf{identity,width=.5,label=id}{oo,v5}
    \fmfv{label=${\rm \hat m}$,label.angle=120,label.dist=6}{o1}
    \fmfv{label=$\mu$,label.angle=180,label.dist=6}{o45}
    \fmfv{label=$\mu^*$,label.angle=180,label.dist=6}{o55}
  \end{fmfgraph*}
}
\end{matrix}
\qquad \longrightarrow \qquad
\begin{matrix}
\fmfframe(0,-8)(0,-8){
  \begin{fmfgraph*}(40,40)
    \fmfcmd{style_def crazy_arrow expr p = 
      cdraw (subpath (.03length(p),.07length(p)) of p);
      cdraw (subpath (.13length(p),.17length(p)) of p);
      cdraw (subpath (.23length(p),.27length(p)) of p);
      cdraw (subpath (.33length(p),.37length(p)) of p);
      cdraw (subpath (.43length(p),.47length(p)) of p);
      cdraw (subpath (.53length(p),.57length(p)) of p);
      cdraw (subpath (.63length(p),.67length(p)) of p);
      cdraw (subpath (.73length(p),.77length(p)) of p);
      shrink(.5); cfill (harrow (p,.77)); endshrink; 
      enddef;}
    \fmfcurved
    \fmfsurroundn{o}{8}
    \fmf{phantom,tension=.5}{o1,v1}
    \fmf{phantom,tension=.5}{o2,v2}
    \fmf{phantom,tension=.5}{o3,v3}
    \fmf{phantom,tension=.5}{o4,v4}
    \fmf{phantom,tension=.5}{o5,v5}
    \fmf{phantom,tension=.5}{o6,v6}
    \fmf{phantom,tension=.5}{o7,v7}
    \fmf{phantom,tension=.5}{o8,v8}
    \fmfcyclen{phantom}{v}{8} 
    \fmffreeze
    \fmf{fermion,width=2,right=0.65,label=$\alpha$,label.side=left}{v1,v4}
    \fmf{fermion,width=2,right=0.4,label=$\alpha'_l$,label.side=left}{v4,v6}    
    \fmf{fermion,width=2,right=0.65,label=$\alpha$,label.side=left}{v6,v1}    
    \fmf{fermion,width=2}{o1,v1}
    \fmf{phantom}{o4,o45,o5,o55,o6}
    \fmffreeze
    \fmf{fermion,width=.5}{o45,v4}
    \fmf{fermion,width=.5}{o55,v6}
    \fmfv{label=${\rm \hat m}$,label.angle=120}{o1}
    \fmfv{label=$\mu$,label.angle=-90}{o45}
    \fmfv{label=$\mu^*$,label.angle=90}{o55}
    \fmf{phantom}{o45,m4,v4}
    \fmf{phantom}{o1,m1,v1}
    \fmf{crazy_arrow,width=.5,left=0.8,tension=0}{m4,m1}
  \end{fmfgraph*}
}
\end{matrix}
\\
\longrightarrow\quad
&\begin{matrix}
\fmfframe(0,-8)(0,-8){
  \begin{fmfgraph*}(40,40)
    \fmfcurved
    \fmfsurroundn{o}{8}
    \fmf{phantom,tension=.5}{o1,v1}
    \fmf{phantom,tension=.5}{o2,v2}
    \fmf{phantom,tension=.5}{o3,v3}
    \fmf{phantom,tension=.5}{o4,v4}
    \fmf{phantom,tension=.5}{o5,v5}
    \fmf{phantom,tension=.5}{o6,v6}
    \fmf{phantom,tension=.5}{o7,v7}
    \fmf{phantom,tension=.5}{o8,v8}
    \fmfcyclen{phantom}{v}{8} 
    \fmffreeze
    \fmf{fermion,width=2,right=1.5,label=$\alpha'_l$,label.side=left}{v1,v6}
    \fmf{fermion,width=2,right=0.4,label=$\alpha$,label.side=left}{v6,v8}    
    \fmf{fermion,width=2,right=0.2,label=$\alpha'_k$,label.side=right}{v8,v1}
    \fmf{fermion,width=2}{o1,v1}
    \fmf{phantom}{o4,o45,o5,o55,o6}
    \fmffreeze
    \fmf{fermion,width=.5}{o8,v8}
    \fmf{fermion,width=.5}{o55,v6}
    \fmfv{label=${\rm \hat m}$,label.angle=120}{o1}
    \fmfv{label=$\mu$,label.angle=60}{o8}
    \fmfv{label=$\mu^*$,label.angle=90}{o55}
    \fmf{phantom}{o45,m4,v4}
    \fmf{phantom}{o1,m1,v1}
  \end{fmfgraph*}
}
\end{matrix}
\qquad\longrightarrow\qquad
\begin{matrix}
\fmfframe(0,-8)(0,-8){
  \begin{fmfgraph*}(40,40)
    \fmfcmd{style_def identity expr p = 
      cdraw (subpath (.05length(p),.15length(p)) of p);
      cdraw (subpath (.25length(p),.35length(p)) of p);
      cdraw (subpath (.45length(p),.55length(p)) of p);
      cdraw (subpath (.65length(p),.75length(p)) of p);
      cdraw (subpath (.85length(p),.95length(p)) of p);
      shrink(.5); cfill (marrow (p,.5)); endshrink; 
      enddef;}
    \fmfcurved
    \fmfsurroundn{o}{8}
    \fmf{phantom,tension=.5}{o1,v1}
    \fmf{phantom,tension=.5}{o2,v2}
    \fmf{phantom,tension=.5}{o3,v3}
    \fmf{phantom,tension=.5}{o4,v4}
    \fmf{phantom,tension=.5}{o5,v5}
    \fmf{phantom,tension=.5}{o6,v6}
    \fmf{phantom,tension=.5}{o7,v7}
    \fmf{phantom,tension=.5}{o8,v8}
    \fmfcyclen{phantom}{v}{8} 
    \fmffreeze
    \fmf{fermion,width=2,right,label=$\alpha'_l$,label.side=left}{v1,v5}
    \fmf{fermion,width=2,right,label=$\alpha'_k$,label.side=left}{v5,v1}
    \fmf{fermion,width=2}{o1,v1}
    \fmf{phantom}{o4,o45,o5,o55,o6}
    \fmffreeze
    \fmf{fermion,width=.5}{o45,oo}
    \fmf{fermion,width=.5}{o55,oo}
    \fmf{identity,width=.5,label=id}{oo,v5}
    \fmfv{label=${\rm \hat m}$,label.angle=120,label.dist=6}{o1}
    \fmfv{label=$\mu^*$,label.angle=180,label.dist=6}{o45}
    \fmfv{label=$\mu$,label.angle=180,label.dist=6}{o55}
  \end{fmfgraph*}
}
\end{matrix}
\quad\phantom{\longrightarrow}
\end{align*}
\parbox{5in}{
\caption{'t Hooft loop as the monodromy associated with moving   $V_\mu(z)$ along the b-cycle of  $C_{1,1}$. The Toda  momenta labeling the edges are $\mu=-bh_1$, $\mu^* = b h_N$, ${\rm \hat m}= N\left({q/2}+i\hat m\right)\, \omega_{N-1}$, $\alpha'_l = \alpha - b h_l$, and $\alpha'_k = \alpha-b h_k$ for   $l,k=1,\ldots,N$.
\label{fig:tHooft-torus}
}}
\end{center}
\end{figure}
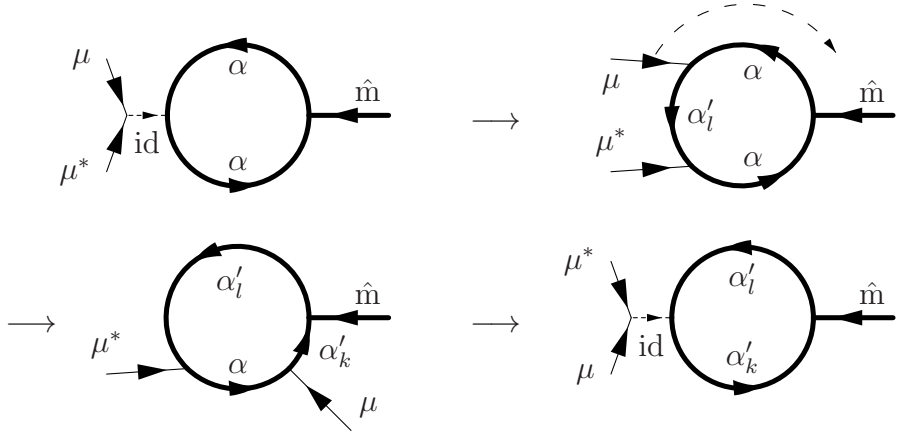

The operators ${\cal O}_\mu(p_n)$ are straightforward combinations of the building blocks \eqref{fusion1},  \eqref{fusion2}, and \eqref{braiding}  with particular momenta, shown in Figure~\ref{fig:tHooft-torus}.
We first concentrate on $\Ocal_\mu(p_0)$, which involves two fusion and one braiding move. 

After the first move \eqref{fusion1}, the internal momentum is
$\alpha'_l\equiv\alpha-bh_l$:
\begin{align*}
&
\begin{matrix}
\fmfframe(0,-5)(0,-5){
  \begin{fmfgraph*}(40,40)
    \fmfcmd{style_def identity expr p = 
      cdraw (subpath (.05length(p),.15length(p)) of p);
      cdraw (subpath (.25length(p),.35length(p)) of p);
      cdraw (subpath (.45length(p),.55length(p)) of p);
      cdraw (subpath (.65length(p),.75length(p)) of p);
      cdraw (subpath (.85length(p),.95length(p)) of p);
      shrink(.5); cfill (marrow (p,.5)); endshrink; 
      enddef;}
    \fmfcurved
    \fmfsurroundn{o}{8}
    \fmf{phantom,tension=.5}{o1,v1}
    \fmf{phantom,tension=.5}{o2,v2}
    \fmf{phantom,tension=.5}{o3,v3}
    \fmf{phantom,tension=.5}{o4,v4}
    \fmf{phantom,tension=.5}{o5,v5}
    \fmf{phantom,tension=.5}{o6,v6}
    \fmf{phantom,tension=.5}{o7,v7}
    \fmf{phantom,tension=.5}{o8,v8}
    \fmfcyclen{phantom}{v}{8} 
    \fmffreeze
    \fmf{fermion,width=2,right,label=$\alpha$}{v1,v5}
    \fmf{fermion,width=2,right,label=$\alpha$}{v5,v1}
    \fmf{fermion,width=2}{o1,v1}
    \fmf{phantom}{o4,o45,o5,o55,o6}
    \fmffreeze
    \fmf{fermion,width=.5}{o45,oo}
    \fmf{fermion,width=.5}{o55,oo}
    \fmf{identity,width=.5,label=id}{oo,v5}
    \fmfv{label=$\alpha$,label.angle=-90,label.dist=6}{v7}
    \fmfv{label=$\alpha$,label.angle=90,label.dist=6}{v3}
    \fmfv{label=${\rm \hat m}$,label.angle=120,label.dist=6}{o1}
    \fmfv{label=$\mu$,label.angle=90,label.dist=6}{o45}
    \fmfv{label=$\mu^*$,label.angle=-90,label.dist=6}{o55}
  \end{fmfgraph*}
}
\end{matrix}
=
\frac{\Gamma(Nbq)}{\Gamma(bq)}
\sum_{l=1}^N \left[
\left(\prod_{j\neq l}
  \frac{\Gamma(b\langle \alpha-Q,h_j - h_l\rangle)}
  {\Gamma(bq + b\langle \alpha-Q,h_j-h_l\rangle)}
\right)
\begin{matrix}
\fmfframe(0,-5)(0,-5){
  \begin{fmfgraph*}(40,40)
    \fmfcmd{style_def crazy_arrow expr p = 
      cdraw (subpath (.03length(p),.07length(p)) of p);
      cdraw (subpath (.13length(p),.17length(p)) of p);
      cdraw (subpath (.23length(p),.27length(p)) of p);
      cdraw (subpath (.33length(p),.37length(p)) of p);
      cdraw (subpath (.43length(p),.47length(p)) of p);
      cdraw (subpath (.53length(p),.57length(p)) of p);
      cdraw (subpath (.63length(p),.67length(p)) of p);
      cdraw (subpath (.73length(p),.77length(p)) of p);
      shrink(.5); cfill (harrow (p,.77)); endshrink; 
      enddef;}
    \fmfcurved
    \fmfsurroundn{o}{8}
    \fmf{phantom,tension=.5}{o1,v1}
    \fmf{phantom,tension=.5}{o2,v2}
    \fmf{phantom,tension=.5}{o3,v3}
    \fmf{phantom,tension=.5}{o4,v4}
    \fmf{phantom,tension=.5}{o5,v5}
    \fmf{phantom,tension=.5}{o6,v6}
    \fmf{phantom,tension=.5}{o7,v7}
    \fmf{phantom,tension=.5}{o8,v8}
    \fmfcyclen{phantom}{v}{8} 
    \fmffreeze
    \fmf{fermion,width=2,right=0.65,label=$\alpha$,label.side=left}{v1,v4}
    \fmf{fermion,width=2,right=0.4,label=$\alpha'_l$,label.side=left}{v4,v6}    
    \fmf{fermion,width=2,right=0.65,label=$\alpha$,label.side=right}{v6,v1}    
    \fmf{fermion,width=2}{o1,v1}
    \fmf{phantom}{o4,o45,o5,o55,o6}
    \fmffreeze
    \fmf{fermion,width=.5}{o45,v4}
    \fmf{fermion,width=.5}{o55,v6}
    \fmfv{label=${\rm \hat m}$,label.angle=120}{o1}
    \fmfv{label=$\mu$,label.angle=90}{o45}
    \fmfv{label=$\mu^*$,label.angle=-90}{o55}
    \fmf{phantom}{o45,m4,v4}
    \fmf{phantom}{o1,m1,v1}
    \fmf{crazy_arrow,width=.5,left=0.8,tension=0}{m4,m1}
  \end{fmfgraph*}
}
\end{matrix}
\right]
\end{align*}

The second move \eqref{braiding} is then applied, with $\alpha_1 = \alpha$, and $\alpha_2=\alpha'_l$.  Now an  internal momentum $\alpha'_k \equiv \alpha - b h_k$ appears. Using that $\alpha'_l = \alpha-b h_l$ and the formula for the scalar product of the weights $(h_1,\ldots,h_N)$ of the fundamental representation of $SU(N)$,  we get\footnote{The factor $\Gamma(bq + b\langle \alpha-Q, h_j-h_l\rangle)$ cancels between the fusion and braiding contribution.}
\begin{align*}
\frac{\Gamma(Nbq)}{\Gamma(bq)}
\sum_{l,k=1}^N   e^{i\pi\epsilon b\langle \alpha -Q, h_l - h_k\rangle}
&\left(\prod_{j\neq l}
  \frac{\Gamma(b\langle \alpha-Q,h_j - h_l\rangle)}
{\Gamma(bq + b\langle \alpha-Q, h_j-h_k\rangle 
- b \langle {\rm \hat m},h_1\rangle)}
\right)\cdot
\\
\cdot
&\left(\prod_{j\neq k}
\frac{\Gamma(b\langle \alpha-Q, h_j- h_k\rangle)}
{\Gamma(b\langle \alpha-Q, h_j - h_l\rangle + 
b\langle {\rm \hat m},h_1\rangle)}
\right)
\begin{matrix}
\fmfframe(0,-10)(0,-8){
  \begin{fmfgraph*}(40,40)
    \fmfcurved
    \fmfsurroundn{o}{8}
    \fmf{phantom,tension=.5}{o1,v1}
    \fmf{phantom,tension=.5}{o2,v2}
    \fmf{phantom,tension=.5}{o3,v3}
    \fmf{phantom,tension=.5}{o4,v4}
    \fmf{phantom,tension=.5}{o5,v5}
    \fmf{phantom,tension=.5}{o6,v6}
    \fmf{phantom,tension=.5}{o7,v7}
    \fmf{phantom,tension=.5}{o8,v8}
    \fmfcyclen{phantom}{v}{8} 
    \fmffreeze
    \fmf{fermion,width=2,right=1.5,label=$\alpha'_l$,label.side=left}{v1,v6}
    \fmf{fermion,width=2,right=0.4,label=$\alpha$,label.side=right}{v6,v8}    
    \fmf{fermion,width=2,right=0.2,label=$\alpha'_k$,label.side=right}{v8,v1}
    \fmf{fermion,width=2}{o1,v1}
    \fmf{phantom}{o4,o45,o5,o55,o6}
    \fmffreeze
    \fmf{fermion,width=.5}{o8,v8}
    \fmf{fermion,width=.5}{o55,v6}
    \fmfv{label=${\rm \hat m}$,label.angle=120}{o1}
    \fmfv{label=$\mu$,label.angle=60}{o8}
    \fmfv{label=$\mu^*$,label.angle=90}{o55}
    \fmf{phantom}{o45,m4,v4}
    \fmf{phantom}{o1,m1,v1}
  \end{fmfgraph*}
}
\end{matrix}
\end{align*}

We finally fuse back $\mu$ and $\mu^*$ using \eqref{fusion2}. As noted below \eqref{fusion2}, the projection onto the identity vanishes for $\alpha'_l\neq \alpha'_k$. In other words, only the terms with $l=k$ contribute, and in particular, the phase vanishes. The relevant fusion matrix  is  obtained by replacing $\alpha$ by $2Q-\alpha'_k$ in \eqref{fusion2}, yielding the result:\footnote{The factor $\Gamma(b\langle \alpha-Q, h_j- h_k\rangle)$ cancels between the fusion and braiding contribution. We also use Euler's reflection formula $\Gamma(x) \Gamma(1-x) =\pi/ \sin (\pi x)$.}
\begin{align*}
\hskip-30pt &
\frac{\sin\pi bq}{\sin\pi Nbq}
\sum_{k=1}^N  
\prod_{j\neq k}
  \frac{\Gamma(b\langle \alpha-Q,h_j - h_k\rangle)}
{\Gamma(bq + b\langle \alpha-Q, h_j-h_k\rangle 
- b \langle {\rm \hat m},h_1\rangle)}
\frac{\Gamma(bq + b\langle \alpha-Q, h_j - h_k\rangle)}
{\Gamma(b\langle \alpha-Q, h_j - h_k\rangle +
b\langle {\rm \hat m},h_1\rangle)}
\hskip-3pt \begin{matrix}
\fmfframe(0,-5)(0,-5){
  \begin{fmfgraph*}(30,30)
    \fmfcmd{style_def identity expr p = 
      cdraw (subpath (.05length(p),.15length(p)) of p);
      cdraw (subpath (.25length(p),.35length(p)) of p);
      cdraw (subpath (.45length(p),.55length(p)) of p);
      cdraw (subpath (.65length(p),.75length(p)) of p);
      cdraw (subpath (.85length(p),.95length(p)) of p);
      shrink(.5); cfill (marrow (p,.5)); endshrink; 
      enddef;}
    \fmfcurved
    \fmfsurroundn{o}{8}
    \fmf{phantom,tension=.5}{o1,v1}
    \fmf{phantom,tension=.5}{o2,v2}
    \fmf{phantom,tension=.5}{o3,v3}
    \fmf{phantom,tension=.5}{o4,v4}
    \fmf{phantom,tension=.5}{o5,v5}
    \fmf{phantom,tension=.5}{o6,v6}
    \fmf{phantom,tension=.5}{o7,v7}
    \fmf{phantom,tension=.5}{o8,v8}
    \fmfcyclen{phantom}{v}{8} 
    \fmffreeze
    \fmf{fermion,width=2,right,label=$\alpha'_k$,label.side=left}{v1,v5}
    \fmf{fermion,width=2,right,label=$\alpha'_k$,label.side=left}{v5,v1}
    \fmf{fermion,width=2}{o1,v1}
    \fmf{phantom}{o4,o45,o5,o55,o6}
    \fmffreeze
    \fmf{fermion,width=.5}{o45,oo}
    \fmf{fermion,width=.5}{o55,oo}
    \fmf{identity,width=.5,label=id}{oo,v5}
    \fmfv{label=${\rm \hat m}$,label.angle=120,label.dist=6}{o1}
    \fmfv{label=$\mu^*$,label.angle=90,label.dist=6}{o45}
    \fmfv{label=$\mu$,label.angle=-90,label.dist=6}{o55}
  \end{fmfgraph*}
}
\end{matrix}
\end{align*}
Setting ${\rm \hat m} = N(q/2+\hat m)\, \omega_{N-1}$, $\alpha=Q+i a$, and defining $a_j := \langle a, h_j\rangle\in\mathbb{R}$ (these obey $\sum_{j=1}^N a_j = 0$), the action  of the loop operator $\Ocal_{\mu}(p_0)$ on the conformal blocks of $C_{1,1}$ is given by \eqref{matrixele} with
\[
\Ocal_\mu(p_0)_{\alpha,\alpha-bh_k}
=
\frac{\sin \pi bq}{\sin \pi N bq}
\prod_{1\leq j\leq N}^{j\neq k}
  \frac{\Gamma(i b (a_j-a_k))}
{\Gamma(bq/2+i b( a_j-a_k)  - i b \hat m)}
\frac{\Gamma(bq + i b(a_j-a_k))}
{\Gamma(bq/2 + i b(a_j-a_k)  + i b \hat m)}\,.
\]
We note that the factor $\frac{\sin \pi N bq}{\sin \pi bq}$ is the quantum dimension of the fundamental representation of $q$-deformed $SL(N)$ with deformation parameter  $e^{i\pi bq}$.  

This computation, when evaluated at $b=1$,   yields   the exact expectation value of the supersymmetric 't Hooft loop operator in ${\cal N}=2^*$ super-Yang-Mills  labeled by the highest weight  of the fundamental representation of $SU(N)$.\footnote{This corresponds to an 't Hooft loop labeled by a Young tableau with $n_1=1$ and $n_2=\ldots=n_N=0$.} The answer, written in gauge theory variables, is given by
\begin{equation}
\label{vevTN2star}
\vev{T}_{{\cal N}=2^*}=\int [da] \left| Z_\text{1-loop}(ia,{\rm \hat m})\right|^2  \overline{Z_\text{cl}(ia,\tau)} \overline{Z_\text{inst}(ia,{\rm \hat m},\tau)}\sum_{k=1}^N T_k(a,\hat m)
Z_\text{cl}(ia-h_k,\tau)   Z_\text{inst}(ia-h_k,{\rm \hat m},\tau)\,,
\end{equation}
where 
\begin{equation*}
T_k(a,\hat m)=\frac{1}{N}\prod_{1\leq j\leq N}^{j\neq k}
  \frac{\Gamma(i (a_j-a_k))\Gamma(2+ i(a_j-a_k))}
{\Gamma(1+ i(a_j-a_k)  
-   i \hat m)\Gamma(1 + i(a_j-a_k)  + i \hat m)}\,,
\end{equation*}
and $\hat m$ is the mass of the adjoint hypermultiplet. This 't Hooft loop operator is the S-dual operator to the Wilson loop operator in the fundamental representation of $SU(N)$ in ${\cal N}=2^*$. Setting  $\hat m=0$  in \eqref{vevTN2star}  gives the result for the 't Hooft loop in ${\cal N}=4$ $SU(N)$ super-Yang-Mills\footnote{It is easy to show that the expression   $C(\alpha,{\rm \hat m},2Q-\alpha)$ in \eqref{3pt}  for ${\rm \hat m}=N(q/2)\,w_{N-1}$ when  $b=1$ just reduces to the Vandermonde determinant  
$\prod_{i<j} (a_i-a_j)^2$.} (see~\cite{Okuda:2010ke} for analogous statement when the gauge group is $SU(2)$).

We now calculate the expectation value of the dyonic operator obtained under the transformation $\theta\rightarrow \theta+2\pi n$, i.e. the Witten effect \cite{Witten:1979ey}. Under such a transformation the 't Hooft operator acquires electric charge, thus giving rise to a Wilson-'t Hooft operator. In the Toda CFT language this is obtained by encircling $V_\mu(z)$ around the curve $p_n$ wrapping $n$ times the a-cycle of the torus and once the b-cycle. Inserting $n$ rotations of $\mu$ around the a-cycle before the fusion of $\mu$ and $\mu^*$ introduces a phase factor 
\[
{\cal O}_\mu(p_n)_{\alpha,\alpha-bh_k}
=e^{-2\pi i n\left[\Delta(\alpha-bh_k)-\Delta(\mu)-\Delta(\alpha)\right]} 
\Ocal_\mu(p_0)_{\alpha,\alpha-bh_k}
=e^{- i\pi n bq (N-1)} e^{2\pi n ba_k} \Ocal_\mu(p_0)_{\alpha,\alpha-bh_k}\,,
\]
where $\Delta(\alpha)=\vev{Q,\alpha}-1/2\vev{\alpha,\alpha}$ is the conformal dimension of a primary with momentum $\alpha$.
The expectation value of the Wilson-'t Hooft loop is therefore given by (must 
set $b=1$)
\begin{equation}
\label{vevTN2stardyon}
\vev{W_nT}_{{\cal N}=2^*} = \int [da] \left| Z_\text{1-loop}(ia)\right|^2  \overline{Z_\text{cl}(ia)} \overline{Z_\text{inst}(ia)}\sum_{k=1}^N e^{2\pi na_k}T_k(a)
Z_\text{cl}(ia-h_k)   Z_\text{inst}(ia-h_k),
\vspace{-5 mm}
\end{equation}
where we suppressed the $\hat m$ and $\tau$ dependency.
%Explicitly
% \[
% \vev{W_nT}_{{\cal N}=2^*} 
% = \int [da] \left| Z_\text{1-loop}(ia,{\rm \hat m})\right|^2
%      \overline{Z_\text{cl}(ia,\tau)} 
%      \overline{Z_\text{inst}(ia,{\rm \hat m},\tau)}
%      \sum_{k=1}^N e^{2\pi n a_k} T_k(a,\hat m)
%         Z_\text{cl}(ia-h_k,\tau)   
%         Z_\text{inst}(ia-h_k,{\rm \hat m},\tau).
% \]

Weyl symmetry implies that each of the $N$ terms in the sums (\ref{vevTN2star})(\ref{vevTN2stardyon}) yields the same contribution. Therefore, the expectation value of this Wilson-'t Hooft loop is captured by $N$ times the contribution of the highest weight vector $h_1$,  coinciding with the following gauge theory expectation. The path integral of a   't Hooft loop labeled by a so-called miniscule representation\footnote{A representation for which all the weights are in  the Weyl orbit of the highest weight.} ---such as the fundamental representation of $SU(N)$---  does not receive subleading contributions,  besides the leading semiclassical saddle point, which is governed by the singularity created by the highest weight vector. 
In general, subleading saddle points ---with a  weaker effective monopole ``charge"--- are expected to contribute to the gauge theory path integral and are associated with saddle points where the singularity of the pointlike monopole labeled by the highest weight is screened by a regular monopole, by a mechanism known as monopole bubbling~\cite{Kapustin:2006pk} (these subleading saddle points appeared in the analysis of general 't Hooft loops in ${\cal N}=4$ super-Yang-Mills~\cite{Gomis:2009ir}).
 These subleading saddle points have a  milder singularity near the loop, due to the reduced monopole ``charge", but nevertheless should be included into the evaluation of the 't Hooft loop path integral.  
 These subleading saddle points will appear when discussing ${\cal N}=2$ conformal SQCD.

\subsection{'t Hooft loop in ${\cal N}=2$ conformal SQCD}
\label{sec:SQCD}

Let us first recall the Riemann surface description of ${\cal N}=2$ conformal SQCD, defined as an  ${\cal N}=2$  $SU(N)$ vector multiplet coupled to $2N$ fundamental hypermultiplets. It corresponds to the four punctured sphere $C_{0,4}$ with:\footnote{We recall that the conjugation operation is defined by $h_j^*=-h_{N+1-j}$. In particular, $Q^*=Q$ and $\omega_1^*=\omega_{N-1}$.}\begin{itemize}
\item two full punctures labeled by momenta ${\rm m}_1 = Q + i m_1$ and ${\rm m}_4^* = Q + i m_4^*$,
\item two semi-degenerate punctures labeled by momenta ${\rm \hat m}_2=N(q/2 + i \hat m_2)\,\omega_{N-1}$ and ${{\rm \hat m}_3^*= N(q/2+ i \hat m_3)\,\omega_1}$\,.
\end{itemize}
This description is in the pants decomposition $\sigma$ in which ${\rm m}_1$ and ${\rm \hat m}_2$ on the one hand, and ${\rm m}_4^*$ and ${\rm \hat m}_3^*$ collide, that is the $s$-channel (diagram $(a)$ below).
\[
\begin{matrix}
\fmfframe(5,5)(5,2){\begin{fmfgraph*}(24,9)
\fmfstraight
\fmfbottom{m4,bl,b,br,m1}
\fmftop{tl,m3,t,m2,tr}
\fmf{plain,width=2}{m3,bl}
\fmf{plain,width=2}{m2,br,bl}
\fmf{plain,width=2}{m4,bl}
\fmf{plain,width=2}{m1,br}
\fmfv{label=${\rm m}_1$}{m1}
\fmfv{label=${\rm \hat m}_2$}{m2}
\fmfv{label=${\rm \hat m}_3^*$}{m3}
\fmfv{label=${\rm m}_4^*$}{m4}
\end{fmfgraph*}}
&\qquad
&
\fmfframe(5,5)(5,2){\begin{fmfgraph*}(24,9)
\fmfstraight
\fmftop{00,10,20,30,40,50,60,70,80}
\fmfbottom{03,13,23,33,43,53,63,73,83}
\fmf{phantom}{10,11,12,13}
\fmf{phantom}{20,21,22,23}
\fmf{phantom}{60,61,62,63}
\fmf{phantom}{70,71,72,73}
\fmffreeze
\fmf{plain,width=2,rubout}{20,23}
\fmf{plain,width=2,rubout}{60,63}
\fmf{plain,width=2}{03,23,63,83}
\fmf{fermion,width=0.5,label=$p_t$,label.side=left}{11,71}
\fmf{plain,width=0.5,rubout=5}{12,72}
\fmf{plain,width=0.5,left}{12,11}
\fmf{plain,width=0.5,left}{71,72}
\fmfv{label=${\rm m}_1$}{83}
\fmfv{label=${\rm \hat m}_2$}{60}
\fmfv{label=${\rm \hat m}_3^*$}{20}
\fmfv{label=${\rm m}_4^*$}{03}
\end{fmfgraph*}}
&\qquad &
\fmfframe(5,5)(5,-1){\begin{fmfgraph*}(27,12)
\fmfstraight
\fmftop{00,10,20,30,40,50,60,70,80,90}
\fmfbottom{04,14,24,34,44,54,64,74,84,94}
\fmf{phantom}{00,01,02,03,04}
\fmf{phantom}{10,11,12,13,14}
\fmf{phantom}{20,21,22,23,24}
\fmf{phantom}{30,31,32,33,34}
\fmf{phantom}{50,51,52,53,54}
\fmf{phantom}{60,61,62,63,64}
\fmf{phantom}{70,71,72,73,74}
\fmf{phantom}{80,81,82,83,84}
\fmf{phantom}{90,91,92,93,94}
\fmffreeze
\fmf{phantom}{73,734,74}
\fmf{phantom}{83,834,84}
\fmffreeze
\fmf{plain,width=2,rubout}{20,23}
\fmf{plain,width=2}{63,60}
\fmf{plain,width=2}{03,63}
\fmf{plain,width=2,rubout}{63,93}
\fmf{plain,width=0.5,rubout=5}{72,12}
\fmf{plain,width=0.5,left}{12,11}
\fmf{plain,width=0.5}{11,31}
\fmf{fermion,width=0.5,label=$p_u$,label.side=left}{31,51}
\fmf{plain,width=0.5,rubout=5}{51,71}
\fmf{plain,width=0.5,left=0.4}{71,82}
\fmf{plain,width=0.5}{82,834}
\fmf{plain,width=0.5,left}{834,734}
\fmf{plain,width=0.5,rubout=5}{734,72}
\fmfv{label=${\rm m}_1$}{93}
\fmfv{label=${\rm \hat m}_2$}{60}
\fmfv{label=${\rm \hat m}_3^*$}{20}
\fmfv{label=${\rm m}_4^*$}{03}
\end{fmfgraph*}}
\\
(a)&&
(b)&&
(c)
\end{matrix}
\]

As shown in \cite{Drukker:2010jp}, Wilson loops in gauge theory are captured by Toda CFT loop operators winding around tubes of the   pants decomposition. After an S-duality transformation, in CFT terms, a change of pants decomposition, these loop operators wrap curves traversing tubes. Specifically, the   Wilson loops in the pants decompositions in which ${\rm m}_1$ collides  with ${\rm m}_4^*$ ($t$-channel), or ${\rm m}_1$ with ${\rm \hat m}_3^*$ ($u$-channel), are mapped under S-duality to Verlinde loops operators $\Ocal_\mu(p_t)$ and $\Ocal_\mu(p_u)$. The curves $p_t$ and $p_u$ in the $s$-channel are shown in the diagrams $(b)$ and $(c)$ above. 

In this section, we use results from Appendix~\ref{sec:SQCDmonodromies} to compute the matrix elements of $\Ocal_\mu(p_t)$ and $\Ocal_\mu(p_u)$. Just as for the case of the $\Ncal=2^*$ theory, the action of these loop operators on the $s$-channel conformal blocks has a finite number of terms: 
\begin{equation}
{\cal O}_\mu (p)\cdot {\cal F}^{(s)}_{\alpha,E}=\sum_{\alpha'}{\cal O}_{\alpha,\alpha'}{\cal F}^{(s)}_{\alpha',E}.
\label{shiftagain}
\end{equation}
In the equation above, $E$ stands for the external momenta $({\rm m}_1, {\rm \hat m}_2, {\rm \hat m}_3^*,{\rm m}_4^*)$.

\begin{figure}[ht]
\begin{center}
\begin{align*}
\begin{matrix}
  \fmfframe(0,5)(0,5){\begin{fmfgraph*}(50,15)
    \fmfcmd{style_def identity_arrow expr p = 
      cdraw (subpath (.03length(p),.07length(p)) of p);
      cdraw (subpath (.13length(p),.17length(p)) of p);
      cdraw (subpath (.23length(p),.27length(p)) of p);
      cdraw (subpath (.33length(p),.37length(p)) of p);
      cdraw (subpath (.43length(p),.47length(p)) of p);
      cdraw (subpath (.53length(p),.57length(p)) of p);
      cdraw (subpath (.63length(p),.67length(p)) of p);
      cdraw (subpath (.73length(p),.77length(p)) of p);
      cdraw (subpath (.83length(p),.87length(p)) of p);
      cdraw (subpath (.93length(p),.97length(p)) of p);
      shrink(.5); cfill (marrow (p,.5)); endshrink; 
      enddef;}
    \fmfstraight
    \fmftop{t1,t2,t3,t4,t5,t6}
    \fmfbottom{b1,b2,b3,b4,b5,b6}
    \fmf{phantom}{b4,b34,b3}
    \fmf{fermion,width=2}{b6,b5,b34,b2}
    \fmf{fermion,width=2}{b1,b2}
    \fmffreeze
    \fmf{identity_arrow,label.dist=2,label=$\id$,tension=2}{c,b34}
    \fmf{fermion,width=0.5}{t3,c}
    \fmf{fermion,width=0.5}{t4,c}
    \fmf{fermion,width=2}{t2,b2}
    \fmf{fermion,width=2}{t5,b5}
    \fmf{phantom,label=$\alpha$,label.side=left}{b5,b34}
    \fmf{phantom,label=$\alpha$,label.side=left}{b34,b2}
    \fmfv{label=${\rm m}_4^*$,label.angle=-60}{b1}
    \fmfv{label=${\rm \hat m}_3^*$,label.angle=90,label.dist=2}{t2}
    \fmfv{label=$\mu^*$,label.angle=90,label.dist=2}{t3}
    \fmfv{label=$\mu$,label.angle=90,label.dist=2}{t4}
    \fmfv{label=${\rm \hat m}_2$,label.angle=90,label.dist=2}{t5}
    \fmfv{label=${\rm m}_1$,label.angle=-120}{b6}
  \end{fmfgraph*}}
\end{matrix}
&\quad\mathop{\longrightarrow}\limits^{\text{step 1}}\quad
\begin{matrix}\fmfframe(0,5)(0,0){\begin{fmfgraph*}(50,20)
  \fmfcmd{style_def my_arrow expr p = 
    cdraw (subpath (length(p)/10,length(p)) of p);
    shrink(.7); cfill (marrow (p,.3)); endshrink; 
    enddef;}
  \fmfcmd{style_def my expr p = 
    cdraw (subpath (length(p)/10,length(p)) of p);
    enddef;}
  \fmfstraight
  \fmftop{s50,s40,s30,s20,s10,00,10,20,30,40,50}
  \fmfbottom{s58,s48,s38,s28,s18,08,18,28,38,48,58}
  \fmf{phantom}{00,02,04,06,08}
  \fmf{phantom}{s10,s11,s12,s13,s14,s15,s16,s17,s18}
  \fmf{phantom}{s20,s22,s24,s26,s28}
  \fmf{phantom}{s30,s32,s34,s36,s38}
  \fmf{phantom}{s40,s41,s42,s43,s44,s45,s46,s47,s48}
  \fmf{phantom}{s50,s52,s54,s56,s58}
  \fmf{phantom}{10,11,12,13,14,15,16,17,18}
  \fmf{phantom}{20,22,24,26,28}
  \fmf{phantom}{30,32,34,36,38}
  \fmf{phantom}{40,41,42,43,44,45,46,47,48}
  \fmf{phantom}{50,52,54,56,58}
  \fmf{phantom}{s47,s47a}
  \fmf{phantom}{s45,s45a}
  \fmf{phantom}{s45,s45b}
  \fmf{phantom}{s14,s14b}
  \fmf{phantom}{47,47a}
  \fmf{phantom}{45,45a}
  \fmf{phantom}{45,45b}
  \fmf{phantom}{14,14b}
  \fmffreeze
  \fmf{fermion,width=.5}{10,16}
  \fmf{fermion,width=2,rubout}{30,36}
%  \fmf{plain,right,width=.5}{47,44}
  \fmf{fermion,width=2,rubout}{56,36,16,s16,s36}
%  \fmf{plain,left,width=.5,rubout=5}{47a,45a}
  \fmf{plain,left,width=.5}{44,45}
  \fmf{my,width=.5,rubout=5}{14b,45b}
  \fmf{my_arrow,width=.5}{13,44}
  \fmf{fermion,width=.5}{s10,s16}
  \fmf{fermion,width=2,rubout}{s30,s36}
%  \fmf{plain,left,width=.5}{s47,s44}
  \fmf{fermion,width=2,rubout}{s56,s36}
%  \fmf{plain,right,width=.5,rubout=5}{s47a,s45a}
  \fmf{plain,right,width=.5}{s44,s45}
  \fmf{my,width=.5,rubout=5}{s14b,s45b}
  \fmf{my_arrow,width=.5}{s13,s44}
  \fmfv{label=$\mu^*$,label.angle=180}{s10}
  \fmfv{label=${\rm \hat m}_3^*$,label.angle=180}{s30}
  \fmfv{label=${\rm m}_4^*$,label.angle=-90}{s56}
  \fmfv{label=$\mu$,label.angle=0}{10}
  \fmfv{label=${\rm \hat m}_2$,label.angle=0}{30}
  \fmfv{label=${\rm m}_1$,label.angle=-90}{56}
  \fmf{phantom,label=$\alpha$,label.side=left}{36,16}
  \fmf{phantom,label=$\alpha'_l$,label.side=right}{s16,16}
  \fmf{phantom,label=$\alpha$,label.side=right}{s36,s16}
\end{fmfgraph*}}\end{matrix}
\\
\quad\mathop{\longrightarrow}\limits^{\text{step 2}}\quad
\begin{matrix}
  \fmfframe(0,5)(0,5){\begin{fmfgraph*}(50,15)
    \fmfstraight
    \fmftop{t1,t2,t3,t4,t5,t6}
    \fmfbottom{b1,b2,b3,b4,b5,b6}
    \fmf{fermion,width=2}{b6,b5,b4,b3,b2}
    \fmf{fermion,width=2}{b1,b2}
    \fmffreeze
    \fmf{phantom}{b1,mbl,b2}
    \fmf{phantom}{t1,mtl,t2}
    \fmf{phantom,tension=2}{mtl,mbl}
    \fmf{phantom}{b3,mb3,mt3,t3}
    \fmf{phantom}{b4,mb4,mt4,t4}
    \fmf{phantom,tension=2}{mtr,mbr}
    \fmf{phantom}{t5,mtr,t6}
    \fmf{phantom}{b5,mbr,b6}
    \fmffreeze
    \fmf{fermion,width=2,rubout}{t2,b2}
    \fmf{fermion,width=0.5,rubout}{t3,b3}
    \fmf{fermion,width=0.5,rubout}{t4,b4}
    \fmf{fermion,width=2,rubout}{t5,b5}
    \fmf{phantom,label=$\alpha''_{lr}$,label.side=right}{b2,b3}
    \fmf{phantom,label=$\alpha'_l$,label.side=right}{b3,b4}
    \fmf{phantom,label=$\alpha''_{lk}$,label.side=right}{b4,b5}
    \fmfv{label=${\rm m}_4^*$,label.angle=-60}{b1}
    \fmfv{label=${\rm \hat m}_3^*$,label.angle=90,label.dist=2}{t2}
    \fmfv{label=$\mu^*$,label.angle=90,label.dist=2}{t3}
    \fmfv{label=$\mu$,label.angle=90,label.dist=2}{t4}
    \fmfv{label=${\rm \hat m}_2$,label.angle=90,label.dist=2}{t5}
    \fmfv{label=${\rm m}_1$,label.angle=-120}{b6}
  \end{fmfgraph*}}
\end{matrix}
&\quad\mathop{\longrightarrow}\limits^{\text{step 3}}\quad
\begin{matrix}
  \fmfframe(0,5)(0,5){\begin{fmfgraph*}(50,15)
    \fmfcmd{style_def identity_arrow expr p = 
      cdraw (subpath (.03length(p),.07length(p)) of p);
      cdraw (subpath (.13length(p),.17length(p)) of p);
      cdraw (subpath (.23length(p),.27length(p)) of p);
      cdraw (subpath (.33length(p),.37length(p)) of p);
      cdraw (subpath (.43length(p),.47length(p)) of p);
      cdraw (subpath (.53length(p),.57length(p)) of p);
      cdraw (subpath (.63length(p),.67length(p)) of p);
      cdraw (subpath (.73length(p),.77length(p)) of p);
      cdraw (subpath (.83length(p),.87length(p)) of p);
      cdraw (subpath (.93length(p),.97length(p)) of p);
      shrink(.5); cfill (marrow (p,.5)); endshrink; 
      enddef;}
    \fmfstraight
    \fmftop{t1,t2,t3,t4,t5,t6}
    \fmfbottom{b1,b2,b3,b4,b5,b6}
    \fmf{phantom}{b4,b34,b3}
    \fmf{fermion,width=2}{b6,b5,b34,b2}
    \fmf{fermion,width=2}{b1,b2}
    \fmffreeze
    \fmf{identity_arrow,label.dist=2,label=$\id$,tension=2}{c,b34}
    \fmf{fermion,width=0.5}{t3,c}
    \fmf{fermion,width=0.5}{t4,c}
    \fmf{phantom,label=$\alpha''$,label.side=left}{b5,b34}
    \fmf{phantom,label=$\alpha''$,label.side=left}{b34,b2}
    \fmf{fermion,width=2}{t2,b2}
    \fmf{fermion,width=2}{t5,b5}
    \fmfv{label=${\rm m}_4^*$,label.angle=-60}{b1}
    \fmfv{label=${\rm \hat m}_3^*$,label.angle=90,label.dist=2}{t2}
    \fmfv{label=$\mu^*$,label.angle=90,label.dist=2}{t3}
    \fmfv{label=$\mu$,label.angle=90,label.dist=2}{t4}
    \fmfv{label=${\rm \hat m}_2$,label.angle=90,label.dist=2}{t5}
    \fmfv{label=${\rm m}_1$,label.angle=-120}{b6}
  \end{fmfgraph*}}
\end{matrix}
\end{align*}
\vskip-5mm
\parbox{5in}{
\caption{'t Hooft loop as the monodromy associated with moving $V_\mu(z)$ along $p_t$ on  $C_{0,4}$. The Toda  momenta labeling the edges are described in the main text: $\alpha'_l = \alpha - b h_l$, and $\alpha''_{l,k} = \alpha-b h_l + b h_k$ for $1\leq l,k\leq N$. For the curve $p_u$, the monodromies are replaced by:
\label{fig:tHooft-sphere-t}
}}
\[
\fmfframe(5,5)(5,3){\begin{fmfgraph*}(50,20)
  \fmfcmd{style_def my_arrow expr p = 
    cdraw (subpath (length(p)/10,length(p)) of p);
    shrink(.7); cfill (marrow (p,.3)); endshrink; 
    enddef;}
  \fmfcmd{style_def my expr p = 
    cdraw (subpath (length(p)/10,length(p)) of p);
    enddef;}
  \fmfstraight
  \fmftop{s50,s40,s30,s20,s10,00,10,20,30,40,50}
  \fmfbottom{s58,s48,s38,s28,s18,08,18,28,38,48,58}
  \fmf{phantom}{00,02,04,06,08}
  \fmf{phantom}{s10,s11,s12,s13,s14,s15,s16,s17,s18}
  \fmf{phantom}{s20,s22,s24,s26,s28}
  \fmf{phantom}{s30,s32,s34,s36,s38}
  \fmf{phantom}{s40,s41,s42,s43,s44,s45,s46,s47,s48}
  \fmf{phantom}{s50,s52,s54,s56,s58}
  \fmf{phantom}{10,11,12,13,14,15,16,17,18}
  \fmf{phantom}{20,22,24,26,28}
  \fmf{phantom}{30,32,34,36,38}
  \fmf{phantom}{40,41,42,43,44,45,46,47,48}
  \fmf{phantom}{50,52,54,56,58}
  \fmf{phantom}{s47,s47a}
  \fmf{phantom}{s45,s45a}
  \fmf{phantom}{s45,s45b}
  \fmf{phantom}{s14,s14b}
  \fmf{phantom}{47,47a}
  \fmf{phantom}{45,45a}
  \fmf{phantom}{45,45b}
  \fmf{phantom}{14,14b}
  \fmffreeze
  \fmf{fermion,width=.5}{10,16}
  \fmf{fermion,width=2,rubout}{30,36}
  \fmf{plain,right,width=.5}{47,44}
  \fmf{fermion,width=2,rubout}{56,36,16,s16,s36}
  \fmf{plain,left,width=.5,rubout=5}{47a,45a}
  \fmf{my,width=.5,rubout=5}{14b,45b}
  \fmf{my_arrow,width=.5,rubout=5}{13,44}
  \fmf{fermion,width=.5}{s10,s16}
  \fmf{fermion,width=2,rubout}{s30,s36}
%  \fmf{plain,left,width=.5}{s47,s44}
  \fmf{fermion,width=2,rubout}{s56,s36}
%  \fmf{plain,right,width=.5,rubout=5}{s47a,s45a}
  \fmf{plain,right,width=.5}{s44,s45}
  \fmf{my,width=.5,rubout=5}{s14b,s45b}
  \fmf{my_arrow,width=.5}{s13,s44}
  \fmfv{label=$\mu^*$,label.angle=180}{s10}
  \fmfv{label=${\rm \hat m}_3^*$,label.angle=180}{s30}
  \fmfv{label=${\rm m}_4^*$,label.angle=-90}{s56}
  \fmfv{label=$\mu$,label.angle=0}{10}
  \fmfv{label=${\rm \hat m}_2$,label.angle=0}{30}
  \fmfv{label=${\rm m}_1$,label.angle=-90}{56}
  \fmf{phantom,label=$\alpha$,label.side=left}{36,16}
  \fmf{phantom,label=$\alpha'$,label.side=left}{s16,16}
  \fmf{phantom,label=$\alpha$,label.side=right}{s36,s16}
\end{fmfgraph*}}
\]
\end{center}
\end{figure}
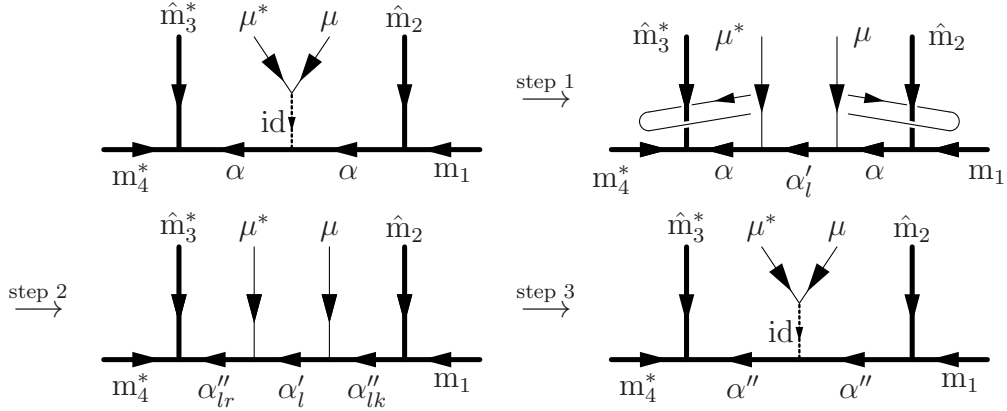

The sequences of moves for the paths $p_t$ and $p_u$ are given in Figure~\ref{fig:tHooft-sphere-t}. Both consist of the fusion move~\eqref{fusion1}, two monodromies, and the fusion of $\mu$ and $\mu^*$ given in~\eqref{fusion2}\footnote{With $\alpha$ changed to the new internal momentum $\alpha''$.}. At step 1, the fusion move introduces an internal momentum $\alpha'_l=\alpha-b h_l$. The monodromies of step~2 shift the internal momenta further: $\alpha''_{lr} = \alpha - b h_l + b h_r$ and $\alpha''_{lk}=\alpha - b h_l + b h_k$. The last fusion step forces $\alpha''_{lr}=\alpha''_{lk}=\alpha''$, hence $r=k$. All in all, the loops introduce a shift of the internal momentum from $\alpha$ to $\alpha'' = \alpha-bh_l + b h_k$, and we have to compute for both loops the coefficients $\Ocal_{\alpha,\alpha-bh_l+bh_k}$ (for $l\neq k$), and $\Ocal_{\alpha,\alpha}$ in (\ref{shiftagain}).

The monodromies appearing in step 2 of both calculations have a similar structure, and are computed in Appendix~\ref{sec:SQCDmonodromies}. They consist of two braiding moves (with orientations controlled by signs $\epsilon_2,\epsilon_2'=\pm 1$), and in some cases the insertion of a phase, controlled by $\epsilon_1=0,\pm 1$. A technical point, which we address in the same appendix, is to relate the signs appearing in the formulae for the braidings to the curve followed by $\mu$ (or $\mu^*$). 

We write the results using $\langle\alpha-Q,h_j\rangle = ia_j$, $\langle {\rm m}_1 -Q,h_j\rangle = i m_{1,j}$, and $\langle {\rm m}_4 -Q,h_j\rangle = i m_{4,j}$, as well as $\langle {\rm \hat m}_2,h_1\rangle = q/2 + i \hat m_2$ and $\langle {\rm \hat m}_3,h_1\rangle = q/2 + i \hat m_3$. 
The $2N$ mass parameters $m_f$ of the fundamental hypermultiplets, where $f=1,\ldots 2N$,  are
\[
m_f=\hat m_2 + m_{1,j} \quad \text{and} \quad 
m_f=\hat m_3 + m_{4,j} \quad \text{for} \quad j=1,\ldots N.
\]

\subsubsection{Operator $\Ocal_\mu(p_t)$}

In the notations of Appendix~\ref{sec:SQCDmonodromies}, the monodromies that enter the computation of the Verlinde loop operator $\Ocal_\mu(p_t)$ correspond to cases b* and b, and the signs controlling the different braiding moves ($\epsilon_1$ and $\epsilon_1^*$ do not play any role in the case of $p_t$) are $(\epsilon_2^*,\epsilon_1^*,{\epsilon_2^*}') = (-,0,-)$ and $(\epsilon_2,\epsilon_1,\epsilon_2') = (+,0,+)$. We compute separately the diagonal and off-diagonal terms of ${\cal O}_{\alpha,\alpha''}$.

The off-diagonal elements $\Ocal_\mu(p_t)_{\alpha,\alpha-bh_l+bh_k}$, $l\neq k$, are obtained by combining the fusion moves \eqref{fusion1} and \eqref{fusion2} with the matrix elements $M_{lk}^{*,-,0,-}$  and $M_{lk}^{+,0,+}$ given in \eqref{monolrb*} and \eqref{monolrb}.  One gets
\begin{align*}
&\Ocal_\mu(p_t)_{\alpha,\alpha-bh_l+bh_k} =
\, 4\pi^2 \frac{\sin\pi bq}{\sin \pi Nbq} e^{\pi b (N-2) (\hat m_2-\hat m_3)} \cdot
\\
\notag
&\cdot\frac{\prod_{j\neq l} \Gamma(ib(a_j-a_l)) \Gamma(bq+ ib(a_j-a_l))\prod_{j\neq k}\Gamma(b^2\delta_{jl} + ib(a_k-a_j))\Gamma(bq+b^2\delta_{lj} +i b (a_k -a_j))}{\prod_f \Gamma(bq/2 - iba_l + ibm_f)\Gamma(bq/2 + iba_k - ibm_f)}\,.
\notag
\end{align*}

All the diagonal coefficients $M_{ll}$ of the monodromies \eqref{monollb*} and \eqref{monollb} contribute to $\Ocal_{\alpha,\alpha}$, and we have to sum over $l$ to take into account all these intermediate shifts by $bh_l$:
\begin{align*}
\Ocal_\mu(p_t)_{\alpha,\alpha}=&
\frac{\sin\pi bq}{\sin \pi Nbq} e^{2b\pi [\hat m_3 - \hat m_2]}
\sum_l
\left[\prod_{j\neq l}\frac{\sin\pi(bq+ ib(a_j-a_l))}{\sin\pi(ib(a_j-a_l))}\right]\cdot
\\
& 
\cdot\left[1  
+ 2i e^{-i\pi (N-2) bq/2 - \pi N b\hat m_3}
\frac{\prod_j \sin\pi(bq/2 - iba_l +  ibm_{4,j} + ib\hat m_3)}
     {\prod_{j\neq l} \sin\pi(bq + ib(a_j-a_l))}\right]
\\
& 
\cdot\left[1
- 2i e^{i\pi (N-2) bq/2} e^{\pi b N\hat m_2} 
\frac{\prod_j \sin\pi(bq/2-iba_l+ib\hat m_2 + ibm_{1,j})}
     {\prod_{j\neq l} \sin\pi(bq + ib(a_j-a_l))} \right]\,.
\end{align*}
%\BLF{The flavour Wilson loops which constitute the singlet contribution correspond to the term 1 times 1 in the above product:
%\[
%\frac{\sin\pi bq}{\sin \pi Nbq} e^{2b\pi [\hat m_3 - \hat m_2]}
%\sum_l \left[\prod_{j\neq l}\frac{\sin\pi(bq+ ib(a_j-a_l))}{\sin\pi(ib(a_j-a_l))}\right]
%= e^{2b\pi [\hat m_3 - \hat m_2]}
%\]
%as we want [right signs!].}
We expand the product, and perform the sums of products of sines using a contour integral (this method is used in Appendices \ref{sec:SQCDmonodromies} and \ref{sec:Wilson} for similar sums) to get
\begin{align*}
&\Ocal_\mu(p_t)_{\alpha,\alpha}=
\frac{\sin(\pi (N-2) bq)}{\sin \pi Nbq} e^{2b\pi [\hat m_3 - \hat m_2]}
+ \frac{\sin\pi bq}{\sin \pi Nbq} e^{(N-2)b\pi [\hat m_2 - \hat m_3]}\Bigg(
2\cos\pi\Big[bq+ib\sum_f m_f\Big]
\\
&+4\sum_l
\frac{\prod_f \sin\pi(bq/2-iba_l+ibm_f)}{\prod_{j\neq l} \sin\pi(ib(a_j-a_l)) \sin\pi(bq + ib(a_j-a_l))}
\Bigg)\,.
\end{align*}

\subsubsection{Operator $\Ocal_\mu(p_u)$}

The operator $\Ocal_\mu(p_u)$ involves the same monodromy of $\mu^*$, but a different monodromy on the $\mu$ side. The monodromies that enter this computation correspond to the cases b* and c of the Appendix~\ref{sec:SQCDmonodromies}, with signs $(\epsilon_2^*,\epsilon_1^*,{\epsilon_2^*}') = (-,0,-)$ and 
$(\epsilon_2,\epsilon_1,\epsilon_2') = (-,+,+)$. Again, we compute separately the diagonal and off-diagonal terms  of ${\cal O}_{\alpha,\alpha''}$. 

The diagonal contribution $\Ocal_\mu(p_u)_{\alpha,\alpha}$ is a sum over all intermediate shifts $-bh_l$ of the product of the matrix elements $M_{ll}^{*,-,0,-}$  and $M_{ll}^{-,+,+}$ given in \eqref{monollb*} and \eqref{monollc}, together with the contribution of the initial and final fusion moves in Figure \ref{fig:tHooft-sphere-t}:
\begin{align*}
\Ocal_\mu(p_u)_{\alpha,\alpha}
&= e^{2\pi b(\hat m_2+\hat m_3)} e^{i\pi (N-3) bq} 
\frac{\sin\pi bq}{\sin \pi Nbq} 
\sum_l e^{-2 \pi b a_l} 
\left[\prod_{j\neq l} \frac{\sin\pi(bq+ ib(a_j-a_l))}{\sin\pi(i b (a_j - a_l))}\right]\cdot 
\\
&\quad \cdot\left[1 + 2i e^{-\pi N b \hat m_3} e^{i\pi (N-2)bq/2} \frac{\prod_j \sin\pi(bq/2 - iba_l+ibm_{4,j} + ib\hat m_3)}{\prod_{j\neq l} \sin\pi(bq+ib(a_j-a_l))}\right]
\\
&\quad \cdot\left[1 + 2i e^{-\pi N b \hat m_2} e^{i\pi (N-2)bq/2} \frac{\prod_j \sin\pi(bq/2-iba_l+ibm_{1,j} + ib\hat m_2)}{\prod_{j\neq l} \sin\pi(bq+ib(a_j-a_l))}\right]\,.
\end{align*}
This implies that
\begin{align*}
%\\%%%remove this
\Ocal_\mu(p_u)_{\alpha,\alpha}
&= 
e^{2\pi b(\hat m_2+\hat m_3)-i\pi bq} \frac{\sin\pi bq}{\sin\pi Nbq}
\Bigg( - e^{-i\pi bq} \sum_l [e^{-2\pi b a_l}]
+ \sum_f [e^{-2\pi b m_f}]
\\
&\quad - 4 e^{-\pi b \sum_f m_f} 
\sum_l \left[e^{-2 \pi b a_l} \frac{\prod_f  \sin\pi (bq/2 -iba_l+ibm_f)}{\prod_{j\neq l} \sin\pi(ib(a_j-a_l)) \sin\pi(bq+ib(a_j-a_l))}\right]\Bigg)\,.
\end{align*}
%\BLF{The $\sum_f [e^{-2\pi b m_f}]$ term, together with its prefactors, is 
%\begin{align*}
%&\frac{\sin\pi bq}{\sin\pi Nbq} e^{2\pi b(\hat m_2+\hat m_3)-i\pi bq} 
%\sum_j [e^{-2\pi b (\hat m_2+m_{1,j})} + e^{-2\pi b (\hat m_3+m_{4,j})}]
%\\
%&=
%e^{2\pi b \hat m_3-i\pi bq} \frac{\sin\pi bq}{\sin\pi Nbq} \sum_j [e^{-2\pi b m_{1,j}}]
%+ e^{2\pi b \hat m_2-i\pi bq} \frac{\sin\pi bq}{\sin\pi Nbq} \sum_j[e^{-2\pi b m_{4,j}}].
%\end{align*}
%The first term is the flavour Wilson loop that we want [even right signs]. Note that $2\pi b \hat m_3 - i\pi bq = - 2\pi ib\langle {\rm \hat m_3},h_1\rangle$.}

The off-diagonal matrix elements $\Ocal_{\alpha,\alpha-bh_l+bh_k}$, $l\neq k$, are obtained by combining the fusion moves \eqref{fusion1} and \eqref{fusion2} with the matrix elements $M_{lk}^{*,-,0,-}$  and $M_{lk}^{-,+,+}$, given in \eqref{monolrb*} and \eqref{monolrc}:
\begin{align*}
&\Ocal_\mu(p_u)_{\alpha,\alpha - bh_l +b h_k}
=4\pi^2 \frac{\sin\pi bq}{\sin \pi Nbq} 
e^{i\pi b^2-\pi b (N-2) (\hat m_2+\hat m_3)}e^{-2\pi b a_k} 
\\
&\cdot \frac{\prod_{j\neq l} \Gamma(i b (a_j - a_l)) \Gamma(bq + ib(a_j-a_l))\prod_{j\neq k} \Gamma(b^2\delta_{lj}+ib(a_k-a_j)) \Gamma(bq+b^2\delta_{lj}+ib(a_k-a_j))}{\prod_f \Gamma(bq/2 - iba_l + ibm_f)\Gamma(bq/2 + iba_k - ibm_f)}.
\end{align*}

The following fact is noteworthy: if one positions the different punctures as in the diagram below, then the analog of $\Ocal_\mu(p_u)$ is the Verlinde loop operator corresponding to the curve
\[
\fmfframe(0,0)(0,0){\begin{fmfgraph*}(27,18)
\fmfstraight
\fmftop{00,10,20,30,40,50,60,70,80,90}
\fmfbottom{06,16,26,36,46,56,66,76,86,96}
\fmf{phantom}{00,03,06}
\fmf{phantom}{10,11,12,13,14,15,16}
\fmf{phantom}{20,21,22,23,24,25,26}
\fmf{phantom}{60,61,62,63,64,65,66}
\fmf{phantom}{70,71,72,73,74,75,76}
\fmf{phantom}{80,81,82,83,84,85,86}
\fmf{phantom}{90,93,96}
\fmffreeze
\fmf{plain,width=2,rubout}{20,23}
\fmf{plain,width=2}{63,66}
\fmf{plain,width=2}{03,63}
\fmf{plain,width=2,rubout}{63,93}
\fmf{plain,width=0.5,rubout=5}{72,12}
\fmf{plain,width=0.5,left}{12,11}
\fmf{fermion,width=0.5}{11,71}
\fmf{plain,width=0.5,left=0.4}{71,82}
\fmf{plain,width=0.5}{82,84}
\fmf{plain,width=0.5,left}{84,74}
\fmf{plain,width=0.5,rubout=5}{74,72}
\fmfv{label=${\rm m}_1$}{93}
\fmfv{label=${\rm \hat m}_2$,label.angle=150}{66}
\fmfv{label=${\rm \hat m}_3^*$,label.angle=0}{20}
\fmfv{label=${\rm m}_4^*$}{03}
\end{fmfgraph*}}
\]
This loop operator is S-dual to the Verlinde operator capturing the Wilson loop of the theory described by the pants decomposition in which ${\rm m}_1$ and ${\rm \hat m}^*_3$ collide. It can be computed as the product of the braiding of ${\rm \hat m}_2$ and ${\rm m}_1$; $\Ocal_\mu(p_u)$; and a second braiding of ${\rm \hat m}_2$ and ${\rm m}_1$. Hence, the matrix elements only differ from $\Ocal_\mu(p_u)$ by an exponential factor. Specifically, the diagonal matrix elements are unchanged, and for the off-diagonal coefficients, one has to change the exponential factor $e^{i\pi b^2-\pi b (N-2) (\hat m_2+\hat m_3)}e^{-2\pi b a_k}$ to the new factor $e^{-\pi b (N-2) (\hat m_2+\hat m_3)}e^{-\pi b (a_k+a_l)}$, which is more symmetrical, and vanishes for $N=2$: in this case, $k\neq l$ implies $a_k+a_l=0$.

\subsubsection{Gauge theory results}
The computations of $\Ocal_\mu(p_t)$ and $\Ocal_\mu(p_u)$, when evaluated at $b=1$, yield the expectation value of 't Hooft operators which are S-dual to Wilson loops in the $\Ncal=2$ theories described by the $t$- and $u$-channel pants decompositions of the four punctured sphere. It is useful to split the answers as sums over the roots of the $A_{N-1}$ Lie algebra, given by  $h_l-h_k~\hbox{for}~ 1\leq l\neq k\leq N$, and the zero weight $0$. The expectation value  for each of the two 't Hooft loop operators can be written as 
\begin{equation}
\label{vevTSQCD}
\vev{T}_{SQCD}=\int [da] \left| Z_\text{1-loop}(ia,\hat m)\right|^2 \overline{Z_\text{cl}(ia, \tau)} \overline{Z_\text{inst}(ia,\hat m, \tau)} \sum_e T_e(a,m) Z_\text{cl}(ia-e, \tau)   Z_\text{inst}(ia-e,\hat m, \tau)\,,
\end{equation}
where $e$ is $0$ or a root $h_l-h_k$ of $A_{N-1}$.  We recall that the index $f=1,\ldots,2N$ labels the $2N$ fundamental hypermultiplets and $m_f$ their mass.

In the case of $p_t$, the formula for the factors are
\begin{align*}
&T_0(a,m)=
\frac{N-2}{N} e^{-2\pi [\hat m_2 - \hat m_3]}
+ \frac{1}{N} e^{(N-2)\pi [\hat m_2 - \hat m_3]}\Bigg[
2\cosh\pi \Big[\sum_f m_f \Big] 
-4\sum_l
\frac{\prod_f \sinh\pi(a_l-m_f)}
{\prod_{j\neq l} [\sinh\pi(a_j-a_l)]^2}
\Bigg],
\end{align*}
and for $e=h_l-h_k$ a root,
\begin{align*}
T_e(a,m)
&=\frac{1}{N} 4\pi^2 e^{\pi (N-2) (\hat m_2-\hat m_3)} \cdot
\\
\notag
&\cdot\frac{\prod_{j\neq l} \Gamma(i(a_j-a_l)) \Gamma(2+ i(a_j-a_l))\prod_{j\neq k}\Gamma(\delta_{jl} + i(a_k-a_j))\Gamma(2+\delta_{lj} +i (a_k -a_j))}{\prod_f \Gamma(1 - ia_l + im_f)\Gamma(1+ ia_k -im_f)}.
\end{align*}

For the path $p_u$, we get
\begin{align*}
T_0(a,m) = \frac{1}{N} e^{2\pi (\hat m_2+\hat m_3)}\left[
- \sum_j [e^{-2\pi a_j}]+ \sum_f [e^{-2\pi  m_f}]
+4 e^{-\pi \sum_f m_f}
\sum_l e^{-2 \pi  a_l} \frac{\prod_f  \sinh\pi (a_l-m_f)}{\prod_{j\neq l} [\sinh\pi(a_j-a_l)]^2}\right]\,,
\end{align*}
and when $e=h_l-h_k$ with $k\neq l$
\begin{align*}
T_e(a,m) 
&= -\frac{1}{N} 4\pi^2 e^{-\pi  (N-2) (\hat m_2+\hat m_3)} \cdot
\\
&\cdot e^{-2\pi  a_k} \frac{\prod_{j\neq l} \Gamma(i (a_j - a_l)) \Gamma(2 + i(a_j-a_l))\prod_{j\neq k} \Gamma(\delta_{lj}+i(a_k-a_j)) \Gamma(2+\delta_{lj}+i(a_k-a_j))}{\prod_f \Gamma(1 - ia_l + im_f)\Gamma(1 + ia_k - im_f)}\,.
\end{align*}

We note that the first operator ---obtained from the path $p_t$--- describes a purely magnetic insertion while the second operator ---obtained from the path $p_u$--- describes a dyonic insertion, where the electric charge is in the fundamental representation of $SU(N)$. These describe, in turn, an 't Hooft operator   and a Wilson-'t Hooft operator in ${\cal N}=2$ conformal SQCD with $SU(N)$ gauge group.

The expression of the 't Hooft and Wilson-'t Hooft loop expectation value contains a sum over the weights of the adjoint representation of $SU(N)$,\footnote{This representation has a Young tableau with $n_1=2$ and $n_2=\ldots=n_{N-1}=1$, $n_N=0$.} thus automatically giving rise to an operator for which the Dirac string of the monopole singularity is invisible. Our incomplete knowledge of the basis of 't Hooft operators in gauge theories forbids a precise identification of the labels of the two 't Hooft operators,  as either an adjoint or adjoint plus singlet highest 't Hooft loop. Our calculations, however, are unambiguous in the characterization of these loop operators as S-duals of Wilson loop operators of the theory obtained from the four-punctured sphere.

 The calculation of these 't Hooft loop operators has a novel feature compared to the 't Hooft loop analysis in ${\cal N}=2^*$. All the weights of the adjoint representation of $SU(N)$ are not related by the action of the Weyl group. The $N(N-1)$ terms in $T_e(a,m)$ for the non-zero weights of the adjoint representation yield the same contribution to the 't Hooft loop expectation value while the $N-1$ copies of the zero weight of the adjoint give a different contribution. In the case of the contribution of the zero weight, the singularity is weaker  than that for the non-zero weights (actually the effective monopole ``charge" vanishes), signaling that the monopole singularity has been completely screened by a conventional monopole. Our computation makes an explicit prediction for the contribution of the subleading saddle point of the 't Hooft loop path integral, which is associated with non-trivial monopole bubbling.

%===================================================
\section{Discussion and conclusions}
\label{sec:discussion}

In this paper we have found expressions for the exact expectation value of supersymmetric Wilson-'t Hooft loop operators in ${\cal N}=2^*$ and ${\cal N}=2$ conformal SQCD  with $SU(N)$ gauge group on $S^4$.   The answer involves a sum over all the weights in the representation which labels the 't Hooft operator. Each weight determines the shift in the vector multiplet scalar appearing in the holomorphic Nekrasov partition function.  Furthermore, a non-trivial function multiplies each term in the sum. This function can be associated  to the one loop determinant of the ${\cal N}=2$ vector multiplet and ${\cal N}=2$ hypermultiplets in the monopole background, the numerator corresponding to the vector multiplet and the denominator to the hypermultiplets.

The results have been obtained by mapping the gauge theory problem to a computation   involving loop operators in two dimensional $A_{N-1}$ Toda CFT, which we have constructed. It is pleasing to note that the Toda CFT  loop operators automatically yield  consistent 't~Hooft operators in the corresponding gauge theory,  such that  the matter fields of the gauge theory are single valued around the corresponding monopole singularity.
The  't Hooft loop operators we have considered  include the S-dual operators to the Wilson loop operator in the fundamental representation of $SU(N)$ in ${\cal N}=2^*$ and ${\cal N}=2$ conformal SQCD. Our results also indicate that monopole bubbling needs to be taken into account in the full path integral definition of an 't Hooft loop (as discussed in \cite{Gomis:2009ir}). With the techniques developed in this paper it is possible to extend our computations to arbitrary ${\cal N}=2$ superconformal quiver gauge theories. 

There is an important qualitative difference in the properties of   loop operators   in gauge theories with  $SU(2)$ and $SU(N)$ gauge groups with $N>2$. In the theories that arise from a  Riemann surface $C_{g,n}$~\cite{Gaiotto:2009we},  the charges of the Wilson-'t Hooft operators in $SU(2)$ gauge theories      are encoded in the choice of  a non-selfintersecting closed curve $p$ on the Riemann surface $C_{g,n}$, up to homotopy~\cite{Drukker:2009tz}.   These charges   depend on the choice of pants decomposition of the Riemann surface or in gauge theory parlance, on the choice of duality frame.
Moreover, the action of the S-duality  group(oid) on gauge theory loop operators is induced by the geometrical action of the Moore-Seiberg groupoid of $C_{g,n}$, which relates different pants decompositions. This implies that \emph{any} loop operator in this class of theories   is in the duality orbit of a Wilson loop, as any non-selfintersecting curve can be used to define a pants decomposition, and such curves corresponding to purely electric operators \cite{Drukker:2010jp}.
   
 The situation in theories with higher rank is rather different. It is no longer true that all loop operators are in the   duality orbit of a Wilson loop operator. This implies that a general gauge theory loop operator cannot be encoded by a Toda CFT loop operator supported on a non-selfintersecting curve on the Riemann surface, unlike the case with $SU(2)$ gauge group. These Toda CFT loop operators can describe only the gauge theory loop operators in the duality orbits of Wilson loops.
 
 The example of ${\cal N}=2^*$ with $SU(N)$ gauge group is illustrative of the main features. A general dyonic operator in this theory is labeled by a pair of   vectors   $(\mu_e,\mu_m)$  taking values in the coweight lattice $\Lambda_{cw}$ and weight lattice $\Lambda_{w}$ respectively, modulo the simultaneous action of the Weyl group. In this case, all gauge theory loop operators with parallel charge vectors $\mu_e \parallel \mu_m$ (and only those) can be transformed into a purely Wilson loop operator $(\mu'_e,0)$ by the action of S-duality. Given that a Toda CFT loop operator supported on a non-self-intersecting curve describes a Wilson loop operator in some choice of duality frame, this implies that only gauge theory loop operators with charges $(\mu_e,\mu_m)$ such that $\mu_e \parallel \mu_m$ can be described by such Toda CFT loop operators.

There is, however,  a natural candidate class  of Toda CFT operators that could account for the gauge theory loop operators that are not in the duality orbit of Wilson loops. For ${\cal N}=2^*$ these are the loop operators labeled by the pair of vectors 
$(\mu_e,\mu_m)$ which are not parallel $\mu_e \nparallel \mu_m$.
These are constructed by joining  topological junction  operators in Toda CFT to define topological web operators.

\begin{figure}[h]
\centering
\begin{align*}
\epsfig{file=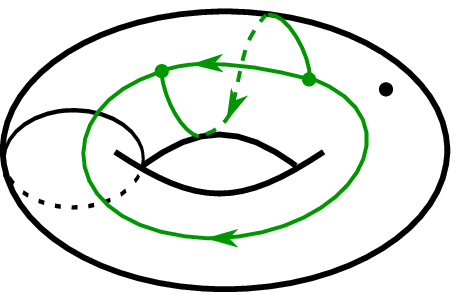,width=2in}
&&
\epsfig{file=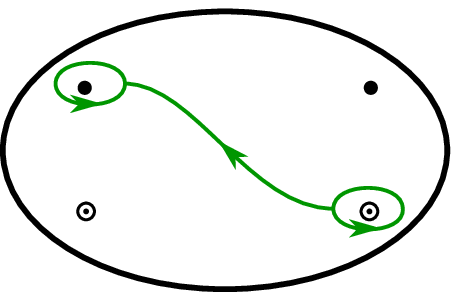,width=2in}
\end{align*}
\parbox{5in}{\caption{Topological defect webs  in $C_{1,1}$ and $C_{0,4}$.} 
\label{fig:web}}
\end{figure}

Topological web operators (see Figure \ref{fig:web}) are constructed from topological junction operators with  trivalent vertices\footnote{Topological junctions with higher valence can be resolved into trivalent junctions.} whose oriented edges correspond to topological line operators, and are therefore labeled by representations of the $W_N$-algebra. If all defects are labeled by a degenerate representation of $SU(N)$, then a topological junction operator is labeled by three representations $(R_1,R_2,R_3)$ of $SU(N)$.  If all lines are incoming, there are  $N_{R_1 R_2}^{R_3^*}$ topological junctions, where $N_{R_1 R_2}^{R_3^*}$  are the Littlewood-Richardson coefficients. By joining these junctions on the Riemann surface $C_{g,n}$ we can construct topological web operators.
 As discussed  in~\cite{Drukker:2010jp}, these operators together with topological defect operators in $C_{g,n}$ form an interesting algebra, defined by the way these operators act on the space of conformal blocks in $C_{g,n}$.  
 
In the case when the gauge group   is $SU(2)$, corresponding to Liouville theory, intersecting Liouville  loop operators can be detached by using the Skein relations~\cite{Drukker:2009id}. 
 This signifies that the Liouville loop operators generate a  complete basis of gauge theory loop operators, as any more complicated configuration can be resolved in terms of Liouville loop operators (which are supported on non-selfintersecting curves).
 As discussed in~\cite{Drukker:2010jp}, this is no longer the case in Toda CFT,  as intersecting Toda loop operators can at best be resolved in terms of Toda loop operators and Toda   topological webs. One may interpret this as a signal of   the necessity of having to consider   topological webs  in order  to describe general gauge theory loop operators, and in particular those that are not in the duality orbit of Wilson loops. This leads  to conjecture that gauge theory loop operators in theories with higher rank are naturally labeled by topological junction operators, and not just topological defect (loop) operators. 
 It would be interesting to find a basis of Toda CFT operators with which to describe an arbitrary gauge theory loop operator.
 
 It would be desirable to find an explicit description of the gauge theory loop operators in terms of Toda CFT topological webs and loop operators, and develop techniques  to calculate   Toda CFT correlators in the presence of    topological webs. In particular, by considering the algebra of Toda loop operators and topological webs one may be able to compute the operator product expansion of gauge theory loop operators, as was done in~\cite{Drukker:2009id} using Liouville loop operators and in~\cite{Kapustin:2007wm} using gauge theory techniques (see also~\cite{Gaiotto:2010be}). Such methods would yield new    tools with which to  calculate exactly arbitrary gauge theory loop operators in this class of gauge theories.

%===================================================
\section*{Acknowledgements}
We would like to thank N. Drukker, D. Gaiotto, T. Okuda, V. Pestun and N. Saulina for discussions. 
J.G. would like to thank the KITP for hospitality. B.L.F. is greatly indebted to the Perimeter Scholars International program of the Perimeter Institute.
Research at the Perimeter Institute is supported in part by the Government of Canada through NSERC and by the Province of Ontario through MRI. 
J.G. also acknowledges further support from an NSERC Discovery Grant and from an ERA grant by the Province of Ontario. This research was supported in part by DARPA under Grant No.
HR0011-09-1-0015 and by the National Science Foundation under Grant
No. PHY05-51164.
B.L.F. received additional support from the \'Ecole Normale Sup\'erieure.

\vfill\eject

\appendix

%===================================================
\section{Toda CFT formulae}
\label{sec:Toda} 
 
The local dynamics of the $A_{N-1}$ Toda CFT  is governed by the Lagrangian density
\begin{equation}
{\cal L}=\frac{1}{8\pi} \vev{\partial_ a \phi, \partial^ a \phi}+\mu\sum_{i=1}^{N-1}e^{b\vev{e_i,\phi}}\,,
\label{action}
\end{equation}
where   $e_i$ are the simple roots of the $A_{N-1}$ Lie algebra.
The scalar fields $\phi=\sum_i \phi_i e_i$ are subject to a background charge $Q=(b+1/b)\rho\equiv q \rho$, which makes the interaction terms in the Lagrangian \eqref{action}   exactly marginal and implies that the CFT has central charge
\begin{equation*}
c=N-1+12 \langle Q ,Q\rangle=(N-1)\left[1+N(N+1)q^2\right]\,.
\end{equation*}
 $\rho=\sum_i \omega_i$ is the Weyl vector and $\omega_i$ are the fundamental weights of the $A_{N-1}$ Lie algebra, which are dual to the simple roots
\begin{equation*}
\vev{\omega_i,e_j}=\delta_{ij}\,,
\end{equation*}
so that $\vev{\rho,e_i}=1$. We will also use the weights $(h_1,\ldots,h_N)$ of the fundamental representation, in terms of which $e_i=h_i-h_{i+1}$.

$A_{N-1}$ Toda CFT  admits the action of a holomorphic and anti-holomorphic  $W_N$-algebra. Each $W_N$-algebra is generated by $N-1$ conserved currents $\{W^{(2)},W^{(3)},\ldots,W^{(N-1)}\}$, where $W^{(l)}$ has spin $l$.
The current $W^{(2)}=T$ denotes the usual energy-momentum tensor of a CFT and generates the Virasoro subalgebra ${\it Vir}\subset W_N$. The vertex operators in Toda CFT
\begin{equation*}
V_\alpha=e^{\langle \alpha, \phi\rangle},
\end{equation*}
labelled by their momentum $\alpha=\sum_i \alpha_i e_i$, create the highest weight states of the representations of the $W_N$-algebra. Under the action of the $W_N$-algebra current $W^{(l)}$, the primary  $e^{\langle \alpha, \phi\rangle}$  carries weight  $\Delta^{(l)}(\alpha)$. This weight, which is a polynomial of degree $l$ in $\alpha$,    is left invariant under the following action of the permutation group $S_N$ (the $A_{N-1}$  Weyl group)
\begin{equation}
\alpha\longrightarrow \alpha_\sigma\equiv Q+\sigma(\alpha-Q) \qquad  \sigma\in S_N\,,
\label{weyl}
\end{equation}
such that $\Delta^{(l)}(\alpha)=\Delta^{(l)}(\alpha_\sigma)~ \forall \sigma\in S_N$.
This implies that a representation  of the $W_N$-algebra is labeled by a momentum vector  $\alpha$ modulo the action \eqref{weyl} of the Weyl group.

$W_N$-symmetry constrains the two and three-point functions in Toda CFT. In particular, it  requires that the momenta $\alpha_1$ and $\alpha_2$ of the primaries in the two-point function obey $\Delta^{(l)}(\alpha_1)=(-1)^l\Delta^{(l)}(\alpha_2)$, hence $\alpha_1=2Q-\alpha_2$. The basic two-point function of primary operators is 
\begin{equation}
\vev{V_{2Q-\alpha}(x)V_\alpha(0)}=\frac{1}{|x|^{4\Delta(\alpha)}}\,,
\label{2pt}
\end{equation}
where $\Delta(\alpha)=\Delta(2Q-\alpha)\equiv \Delta^{(2)}(\alpha)=\vev{2Q-\alpha,\alpha}/2$. The other non-zero two-point functions occur at values of the momenta related to those in \eqref{2pt} by the action of the Weyl group \eqref{weyl}, and are encoded in the reflection amplitudes computed in~\cite{Fateev:2001mj}. As usual, the two-point functions of $W_N$-descendants are determined in terms of the two-point function of the primaries.

The three-point functions of primaries are encoded in the coefficients $C(\alpha_1,\alpha_2,\alpha_3)$
\begin{equation*}
\vev{V_{\alpha_1}(x_1)V_{\alpha_2}(x_2)V_{\alpha_3}(x_3)}
=\frac{C(\alpha_1,\alpha_2,\alpha_3)} 
      {|x_1-x_2|^{2\Delta_{12}}|x_1-x_3|^{2\Delta_{13}}|x_2-x_3|^{2\Delta_{23}}}\,,
\end{equation*}
where $\Delta_{ij}=\Delta(\alpha_i)+\Delta(\alpha_j)-\Delta(\alpha_k)$. The general three-point function in Toda CFT is not known. 
Furthermore, the three-point functions of all $W_N$-descendants are not completely determined in terms of the three-point function of the $W_N$-primaries, unlike the case when the symmetry of the theory is just the Virasoro algebra. This means that without further input the higher point correlation functions in Toda CFT are not known, as the conformal blocks, which capture the contribution of $W_N$-descendants to correlators, are not constructed.

However, the expression for the three point function with an insertion of a semi-degenerate field with momentum $\alpha_3=\kappa \omega_{N-1}=-\kappa h_N$ and two non-degenerate fields with momenta $\alpha_1$ and $\alpha_2$ was presented in~\cite{Fateev:2005gs}:\footnote{For $\alpha_3=\kappa \omega_1=\kappa h_1$ we replace the products over weights of the fundamental representation in \eqref{3pt} with products of weights of the antifundamental representation, given by $\bar{h}_i=-h_{N+1-i}$.}
\begin{equation}
C(\alpha_1,\alpha_2,-\kappa h_N)
= A \frac{\prod_{e>0}\Upsilon(\vev{Q-\alpha_1,e})\Upsilon(\vev{Q-\alpha_2,e})}
{\prod_{i,j=1}^N \Upsilon(\kappa/N+\vev{Q-\alpha_1,h_i}+\vev{Q-\alpha_2,h_j})}\,,
  \label{3pt}
\end{equation}
where $e>0$ denote the positive roots of $SU(N)$, $e=h_i-h_j$, $i<j$ in terms of the weigths of the fundamental representation, and $A$ is a prefactor (see~\cite{Fateev:2005gs} for the prefactor and the definition of $\Upsilon(x)$). 

Furthermore,   for the choice of semi-degenerate momentum $\alpha_3=\kappa h_1$ or  $\alpha_3=-\kappa h_N$ and   arbitrary   $\alpha_1$ and $\alpha_2$, the three-point function of $W_N$-descendants   are  completely determined~\cite{Bowcock:1993wq}, as the   null vectors  of the semi-degenerate representation labeled by $\alpha_3$ can be used to solve for the unknown descendant correlators. Therefore, correlation functions which admit a factorization in which a 
semi-degenerate field appears in each of the three-punctured spheres  defining the factorization  can be explicitly computed.

In this paper an important role is played by the $W_N$-primary with momentum
\begin{equation*}
\mu=-bh_1\,,
\end{equation*}
 which is the lowest weight state of a completely degenerate representation of the $W_N$-algebra, and is labeled by the highest weight of the fundamental representation of $SU(N)$.  A  key property of such a degenerate primary is that fusing it with a  primary with generic momentum $\alpha$ yields a finite number of representations \cite{Fateev:2005gs} 
  \begin{equation}
 \left[V_{\mu}\right]\cdot \left[V_{\alpha}\right]=\sum_{l=1}^N \left[V_{\alpha-bh_l}\right]\,.
 \label{fuserep}
 \end{equation} 
 If the momentum $\alpha$ is that of a semi-degenerate representation with $\alpha=-\kappa h_N$ the fusion has only two terms 
 \begin{equation}
 \label{semidegeneratefuserep}
 \left[V_{\mu}\right]\cdot \left[V_{-\kappa h_N}\right]=  \left[V_{-\kappa h_N-bh_1}\right]+\left[V_{-\kappa h_N-bh_N}\right]\,.
 \end{equation}
 Some useful formulae for computations include:
 \begin{align*}
\langle h_i, h_j\rangle &= \delta_{ij} - 1/N
&\langle \rho, h_k \rangle &= \frac{N+1}{2}-k 
\\
\notag
\langle h_i, e_j\rangle &= \delta_{ij} - \delta_{i,j+1}
&\langle\rho,\rho\rangle &= \frac{1}{12} N (N-1)(N+1).
\end{align*}

%===================================================
\section{Fusion and braiding matrices in  $A_{N-1}$ Toda CFT}
\label{sec:fusion} 
 
We want to compute the relevant fusion and braiding matrices required for the computation of 't Hooft operators in ${\cal N}=2$ $SU(N)$ superconformal quiver gauge theories.
These are found by analytically continuing  conformal blocks.
For our purposes, it suffices to compute the  conformal blocks corresponding to the following correlation function on the four punctured sphere  
 \begin{equation}
 \label{correlafuse}
 \vev{V_{2Q-\alpha_2}(\infty)V_{{\rm \hat m}}(1)V_{\mu}(z,\bar{z})V_{\alpha_1}(0)}
 =
 \begin{matrix}
   \fmfframe(0,3)(0,3){
   \begin{fmfgraph*}(25,15)
     \fmfstraight
     \fmftop{tl,t1,t2,t3,tr}
     \fmfbottom{bl,b1,b3,br}
     \fmf{fermion,width=2}{t3,m3}
     \fmf{fermion,width=2}{t1,m1}
     \fmf{fermion,width=2}{br,b3}
     \fmf{fermion,width=2}{b1,bl}
     \fmfpoly{smooth,filled=shaded}{b3,m3,m1,b1}
     \fmfv{label=$\alpha_2$,label.angle=60,label.dist=6}{bl}
     \fmfv{label=$\alpha_1$,label.angle=120,label.dist=6}{br}
     \fmfv{label=$\mu$,label.angle=0,label.dist=4}{t3}
     \fmfv{label=${\rm \hat m}$,label.angle=180,label.dist=4}{t1}
   \end{fmfgraph*}}
 \end{matrix}
 \quad,
 \end{equation}
where $\mu=-bh_1$,  ${\rm \hat m}=-\kappa h_N$ and $\alpha_1,\alpha_2$ are non-degenerate. In the diagrammatic notation, flipping an arrow amounts to changing its label from $\alpha$ to $2Q-\alpha$, because of the two-point function \eqref{2pt}.  

The four-point function \eqref{correlafuse} involves two non-degenerate representations labeled by momenta $\alpha_1$ and $2Q-\alpha_2$, one semi-degenerate with momentum ${\rm \hat m}$ and one completely degenerate representation with momentun $\mu$. 
Using the null vectors of $V_{\mu}$ and $V_{-\kappa h_N}$, Fateev and Litvinov~\cite{Fateev:2007ab} prove that this specific four-point correlator obeys  an holomorphic and anti-holomorphic $N$-th order  differential equation. 
Writing
 \begin{equation}
 \label{correlafuseG}
 \vev{V_{2Q-\alpha_2}(\infty)V_{{\rm \hat m}}(1)V_{\mu}(z)V_{\alpha_1}(0)}
 =|z|^{-2b\langle \mu-Q, h_1\rangle} |1-z|^{2b\langle {\rm \hat m}, h_1\rangle} G(z,\bar z)\,,
 \end{equation}
it follows that $G(z,\bar{z})$ satisfies a hypergeometric differential equation\footnote{The choice of exponent for $|z|$ is different from \cite{Fateev:2007ab} and makes some equations more symmetric.}
 \begin{equation}
   \label{sd3}
   \left[z\prod_{j=1}^N \left(z\frac{d}{dz}+\sigma_j\right)
     - \prod_{j=1}^N\left(z\frac{d}{dz}-\tau_j\right)\right]
   G(z,\bar z) = 0\,,
 \end{equation}
where 
 \begin{align}
   \sigma_j &= b\langle Q-\alpha_2, h_j\rangle + b\langle{\rm \hat m}, h_1\rangle
   \nonumber
   \\
   \tau_j&=b\langle \alpha_1-Q, h_j\rangle + b\langle\mu,h_1\rangle\,.
   \label{parameters}
 \end{align}
Equation~\eqref{sd3} with $z \leftrightarrow \bar z$, but the same
$\sigma_j$ and $\tau_j$ parameter also holds. 

The hypergeometric differential equation \eqref{sd3} has singularities at $z=0$, $z=1$, and $z=\infty$. Its holomorphic solutions are thus analytic on the Riemann sphere, with three branch points at $0,1$ and $\infty$. We choose branch cuts along the real axis. The precise cuts depend on the solutions we consider. 
In the following, we show that the solutions with analytic expansions around $z=0,1$ and $\infty$ describe the correlator \eqref{correlafuseG} in the $s$, $t$ and $u$-channel respectively.

We will use hypergeometric functions of type $(N,N-1)$,  defined by the  series expansion
\begin{equation}
F\left(\left.\begin{smallmatrix} a_1 \cdots a_N \\ b_1\cdots b_{N-1}\end{smallmatrix}\right| z\right)
= \sum_{m=0}^\infty \frac{(a_1)_m \cdots (a_N)_m}
{m! (b_1)_m\cdots (b_{N-1})_m} z^m
= 1+ \frac{a_1\cdots a_N}{b_1\cdots b_{N-1}} z +\cdots
\label{series}
\end{equation}
where $(a)_m = a (a+1) \cdots (a+m-1)$. Generically,  the series converges for $|z|<1$.

\subsection{$s$- and $u$-channels, and braiding}

\subsubsection{Diagonal monodromy, and conformal blocks}

The $N$ solutions to \eqref{sd3} in the $s$-channel are given on the unit disc (minus $[-1,0]$) by
\begin{equation*}
  f_k^{(s)}(z)
  = z^{\tau_k} F\left(\left.\begin{smallmatrix}
    (\sigma_1 + \tau_k) \cdots (\sigma_N + \tau_k)\\
    (\tau_k - \tau_1 + 1)\,\widehat{\cdots}\, (\tau_k-\tau_N+1)
  \end{smallmatrix}\right| z \right)
  \quad k=1,\ldots,N,
\end{equation*}
where the hat denotes skipping the $k$-th term. These solutions $f_k^{(s)}$ can be analytically continued to the whole complex plane minus the cuts $(-\infty,0]\cup[1,+\infty)$ via the Mellin-Barnes representation (see Section~\ref{section:Mellin-Barnes}).

Under the  transformation   
\begin{equation*}
  w=1/z\qquad  \sigma_k\leftrightarrow \tau_k\,
\end{equation*}
the  differential equation \eqref{sd3} remains invariant. Therefore, the $N$ solutions to \eqref{sd3}  in the $u$-channel are  given for $|z|>1$, $z\not\in \bR^-$ by   
\begin{equation*}
  f^{(u)}_{k}(z)
  =z^{-\sigma_k}
  F\left(\begin{smallmatrix}
    (\tau_1 + \sigma_k) \cdots (\tau_N + \sigma_k)\\
    (\sigma_k - \sigma_1 + 1)\,\widehat{\cdots}\, (\sigma_k-\sigma_N+1)
  \end{smallmatrix}\left| \frac{1}{z} \right.\right)
  \quad k=1,\ldots,N.
\end{equation*}
Then, $f^{(u)}_{k}(z)$ is analytically continued to the complex plane minus the cut $(-\infty,1]$.

\paragraph{Conformal blocks in the $s$-channel.}

A general solution of \eqref{sd3} and its antiholomorphic counterpart can be expanded near $0$ as $\sum_{k,l} a_{kl} f_k^{(s)}(\bar z) f_l^{(s)}(z)$ with some coefficients $a_{kl}$. In order for the correlator \eqref{correlafuseG} to be single-valued, the coefficients $a_{kl}$ have to be diagonal: $G(z,\bar z) = \sum_k a_k f_k^{(s)}(\bar z) f_k^{(s)}(z)$. 

On the other hand, the OPE \eqref{fuserep} between $V_{\alpha_1}$ and $V_\mu$, which involves $N$ possible intermediate states, yields another expression of the correlator \eqref{correlafuseG} as a sum of $N$ series in powers of $z$ and $\bar z$. Comparing the leading powers of $z$, we obtain the decomposition in $s$-channel conformal blocks:
\begin{align*}
  &\vev{V_{2Q-\alpha_2}(\infty)V_{{\rm \hat m}}(1)V_{\mu}(z)V_{\alpha_1}(0)}
  \\
  \notag
  &= \sum_{k=1}^N C(2Q-\alpha_2,{\rm \hat m},\alpha_1- b h_k)
  C(2Q-\alpha_1 + b h_k,\mu,\alpha_1)
  \overline{\cF^{(s)}_{\alpha_1-bh_k}\!\left[\begin{matrix}
        {\rm \hat m} & \mu \\ \alpha_2 & \alpha_1
      \end{matrix}\right]}
  \cF^{(s)}_{\alpha_1-bh_k}\!\left[\begin{matrix}
      {\rm \hat m} & \mu \\ \alpha_2 & \alpha_1
    \end{matrix}\right].
\end{align*}
Here, we have absorbed the factors of \eqref{correlafuseG} in the definition of the conformal blocks
\begin{equation}
\label{zlks}
\cF^{(s)}_{\alpha_1-bh_k}\!\left[\begin{matrix}
{\rm \hat m} & \mu \\ \alpha_2 & \alpha_1
\end{matrix}\right]
=
 \begin{matrix}
   \begin{fmfgraph*}(30,15)
     \fmfstraight
     \fmftop{tl,t1,t2,t3,t4,tr}
     \fmfbottom{bl,b1,b2,b3,b4,br}
     \fmf{fermion,width=2}{br,b4,b1,bl}
     \fmf{fermion,width=2}{t1,b1}
     \fmf{fermion,width=0.5}{t4,b4}
     \fmf{phantom,tension=0,label=$\alpha_1-bh_k$,label.side=right}{b4,b1}
     \fmfv{label=$\alpha_1$,label.angle=120,label.dist=6}{br}
     \fmfv{label=$\mu$,label.angle=180,label.dist=4}{t4}
     \fmfv{label=${\rm \hat m}$,label.angle=0,label.dist=4}{t1}
     \fmfv{label=$\alpha_2$,label.angle=60,label.dist=6}{bl}
   \end{fmfgraph*}
 \end{matrix}
= 
\frac{(1-z)^{b\langle {\rm \hat m}, h_1\rangle}}{z^{b\langle \mu-Q, h_1\rangle - \tau_k}}
F\left(\left.\begin{smallmatrix}
(\sigma_1 + \tau_k) \cdots (\sigma_N + \tau_k)\\
(\tau_k - \tau_1 + 1)\,\widehat{\cdots}\, (\tau_k-\tau_N+1)
\end{smallmatrix}\right| z \right).
\end{equation}
The exponents of $z$ in this expression are $\tau_k - b\langle \mu-Q, h_1\rangle = \Delta(\alpha_1-b h_k)-\Delta(\alpha_1)-\Delta(\mu)$, plus integers, hence a diagonal monodromy around~0:
\begin{equation*}
M_{(0)} \cdot \cF^{(s)}_{\alpha_1-bh_k}\!\left[\begin{matrix}
{\rm \hat m} & \mu \\ \alpha_2 & \alpha_1
\end{matrix}\right]
 = e^{2\pi i (\Delta(\alpha_1-b h_k)-\Delta(\alpha_1)-\Delta(\mu))}\cF^{(s)}_{\alpha_1-bh_k}\!\left[\begin{matrix}
{\rm \hat m} & \mu \\ \alpha_2 & \alpha_1
\end{matrix}\right].
\end{equation*}

\paragraph{Conformal blocks in the $u$-channel.}
They are given by
\begin{equation}
\label{zlku}
\cF^{(u)}_{\alpha_2+bh_k}\!\left[\begin{matrix}
{\rm \hat m} & \mu \\ \alpha_2 & \alpha_1
\end{matrix}\right]
=
 \begin{matrix}
   \begin{fmfgraph*}(30,15)
     \fmfstraight
     \fmftop{tl,t1,t2,t3,t4,tr}
     \fmfbottom{bl,b1,b2,b3,b4,br}
     \fmf{fermion,width=2}{br,b4,b1,bl}
     \fmf{fermion,width=0.5,right=0.4}{t4,b1}
     \fmf{fermion,width=2,left=0.4,rubout}{t1,b4}
     \fmf{phantom,tension=0,label=$\alpha_2+bh_k$,label.side=right}{b4,b1}
     \fmfv{label=$\alpha_1$,label.angle=120,label.dist=6}{br}
     \fmfv{label=$\mu$,label.angle=0,label.dist=4}{t4}
     \fmfv{label=${\rm \hat m}$,label.angle=180,label.dist=4}{t1}
     \fmfv{label=$\alpha_2$,label.angle=60,label.dist=6}{bl}
   \end{fmfgraph*}
 \end{matrix}
= 
\frac{(z-1)^{b\langle {\rm \hat m}, h_1\rangle}}{z^{b\vev{\mu-Q,h_1}+\sigma_k}}
F\left(\begin{smallmatrix}
(\tau_1 + \sigma_k) \cdots (\tau_N + \sigma_k)\\
(\sigma_k - \sigma_1 + 1)\,\widehat{\cdots}\, (\sigma_k-\sigma_N+1)
\end{smallmatrix}\left| \frac{1}{z}\right. \right)\,.
\end{equation}
These conformal blocks have diagonal monodromy around~$\infty$,
\begin{equation*}
M_{(\infty)} \cdot \cF^{(u)}_{\alpha_2+bh_k}\!\left[\begin{matrix}
{\rm \hat m} & \mu \\ \alpha_2 & \alpha_1
\end{matrix}\right]
= e^{2\pi i (\Delta(\alpha_2+b h_k)+\Delta(\mu)-\Delta(\alpha_2))}
\cF^{(u)}_{\alpha_2+bh_k} \!\left[\begin{matrix}
{\rm \hat m} & \mu \\ \alpha_2 & \alpha_1
\end{matrix}\right]\,.
\end{equation*}
Here we used $- b\langle {\rm \hat m}, h_1\rangle + b\vev{\mu-Q,h_1} + \sigma_k = \Delta(\alpha_2+b h_k) + \Delta(\mu) - \Delta(\alpha_2)$.

They correspond to the $N$ possible intermediate states that appear in the fusion of $V_\mu$ with $V_{2Q-\alpha_2}$ \eqref{fuserep}: in this channel the correlator \eqref{correlafuseG} is
\begin{align*}
  &\vev{V_{2Q-\alpha_2}(\infty)V_{{\rm \hat m}}(1)V_{\mu}(z)V_{\alpha_1}(0)}
  \\
  \notag
  &=\sum_{k=1}^N C(2Q-\alpha_2,\mu, \alpha_2 + b h_k)
  C(2Q-\alpha_2 - b h_k,{\rm \hat m},\alpha_1)
  \overline{\cF^{(u)}_{\alpha_2+b h_k}\!\left[\begin{matrix}
        {\rm \hat m} & \mu \\ \alpha_2 & \alpha_1
      \end{matrix}\right]}
  \cF^{(u)}_{\alpha_2 + b h_k}\!\left[\begin{matrix}
      {\rm \hat m} & \mu \\ \alpha_2 & \alpha_1
    \end{matrix}\right].
\end{align*}

\subsubsection{Analytic continuation and Mellin-Barnes representation}
\label{section:Mellin-Barnes} 

In order to calculate the braiding matrix we need to find the way conformal blocks in the $s$-channel are written in terms of those in the $u$-channel. We first find the analytic continuation of the hypergeometric function. 
For $-\pi<\operatorname{arg} (- z)<\pi$ the hypergeometric function \eqref{series} can be expressed as the  Mellin-Barnes
integral
\begin{equation}
F\left(\left.\begin{smallmatrix}a_1\,\cdots\,a_{N-1}\,a_N\\
b_1\,\cdots\,b_{N-1}\end{smallmatrix}\right| z\right)=\frac{\Gamma(b_1)\cdots
  \Gamma(b_{N-1})}{\Gamma(a_1)\cdots \Gamma(a_N)}\frac{1}{2 \pi i}
\int_{-i\infty}^{i\infty} \frac{\Gamma(a_1+s)\cdots
  \Gamma(a_N+s)}{\Gamma(b_1+s)\cdots \Gamma(b_{N-1} +s)} \Gamma(-s)
(-z)^s ds\,,
\label{mellin}
\end{equation}
where the contour is chosen as follows: the poles $s=-a_k-m$,
$m\in\bN=\{0,1,\ldots\}$ of $\Gamma(a_k+s)$ lie to its left, and those of
$\Gamma(-s)$, namely $s=0,1,\ldots$ lie to its right. This integral
converges for any value of $z$ not on the positive real axis. 

We first close the contour towards the positive real axis. Each pole $m\in\bN$ is encircled clockwise, and the residue of $\Gamma(-s)$ at $s=m\geq 0$ is
$-(-1)^m / m!$. Summing over these poles, we reproduce the series expansion in \eqref{series}. 
Equation \eqref{mellin} is valid as long as both sides are defined, that is, for $|z|<1$, $z\not\in \bR^+$. Therefore, the Mellin-Barnes representation allows us to analytically continue the hypergeometric sum to the whole complex plane minus $[1,\infty)$.

Close instead the contour on the other side, encircling the poles of $\Gamma(a_k+s)$. Just as summing over the poles $0,1,\ldots$ of $\Gamma(s)$ gave rise to a hypergeometric function, the set of poles $\{-a_k,-a_k-1,\ldots\}$ of each $\Gamma(a_k+s)$ gives rise to a hypergeometric function, evaluated at $1/z$ instead of $z$ because of the change in the sign of $m$. With this new choice of closing the contour, the integral in \eqref{mellin} evaluates to
\begin{align*}
&\sum_{k=1}^N \sum_{m=0}^\infty 
\frac{\Gamma(a_1-a_k-m)\,\widehat{\cdots}\, \Gamma(a_N-a_k-m)}
     {\Gamma(b_1-a_k-m)\cdots \Gamma(b_{N-1}-a_k-m)} 
\Gamma(a_k+m) (-z)^{-a_k-m} \frac{(-1)^m}{m!}
\\
&=\sum_{k=1}^N \sum_{m=0}^\infty
\left[\prod_{1\leq j\leq N}^{j\neq k}\left[\frac{(-1)^m
      \Gamma(a_j-a_k)}{(1+a_k-a_j)_m}\right]
\prod_{1\leq j < N}\left[\frac{(1+a_k-b_j)_m}{(-1)^m
           \Gamma(b_j-a_k)}\right] 
(a_k)_m\Gamma(a_k) (-z)^{-a_k-m} \frac{(-1)^m}{m!}\right]
\end{align*}
where the hat denotes the omission of the $k$-th term from the product of
Gamma functions, and we used the identity $\Gamma(b-m) = (-1)^m \Gamma(b)/(1-b)_m$ to get the second line. This gives
\begin{align*}
&\sum_{k=1}^N \left[(-z)^{-a_k} 
\frac{\Gamma(a_1-a_k)\,\widehat\cdots\,\Gamma(a_N-a_k)}
     {\Gamma(b_1-a_k)\cdots \Gamma(b_{N-1}-a_k)}
\Gamma(a_k)
\sum_{m=0}^\infty
\frac{(1+a_k-b_1)_m\cdots (1+a_k-b_{N-1})_m}
     {(1+a_k-a_1)_m\,\widehat{\cdots}\, (1+a_k-a_N)_m}
\frac{(a_k)_m}{m!} z^{-m}
\right]
\\
&=\sum_{k=1}^N \left[(-z)^{-a_k} 
\frac{\Gamma(a_1-a_k)\,\widehat\cdots\,\Gamma(a_N-a_k)}
     {\Gamma(b_1-a_k)\cdots \Gamma(b_{N-1}-a_k)}
\Gamma(a_k)
F\left(\begin{smallmatrix}(1+a_k-b_1)\cdots (1+a_k-b_{N-1})\, a_k\\
(1+a_k-a_1) \,\widehat{\cdots} (1+a_k-a_N)\end{smallmatrix}\left|
 \frac{1}{z}\right.\right)
\right].
\end{align*}
This last expression only makes sense for $|z|>1$, $z\not\in\bR^+$. Therefore, the Mellin-Barnes representation is an analytic continuation of the hypergeometric sum to the whole complex plane minus the positive real axis.

From these expressions we deduce that   
\begin{align}
\label{qsf}
&F\left(\left.\begin{smallmatrix}a_1\,\cdots\,a_{N-1}\,a_N\\
b_1\,\cdots\,b_{N-1}\end{smallmatrix}\right| z\right)
\\
\notag
&=
\sum_{k=1}^N \left[(-z)^{-a_k} 
\frac{\Gamma(b_1)\cdots\Gamma(b_{N-1})}
     {\Gamma(a_1)\,\widehat{\cdots}\,\Gamma(a_N)}
\frac{\Gamma(a_1-a_k)\,\widehat{\cdots}\,\Gamma(a_N-a_k)}
     {\Gamma(b_1-a_k)\cdots\Gamma(b_{N-1}-a_k)}
F\left(\begin{smallmatrix}(1+a_k-b_1)\cdots (1+a_k-b_{N-1})\, a_k\\
(1+a_k-a_1) \,\widehat{\cdots} (1+a_k-a_N)\end{smallmatrix}
\left|\frac{1}{z}\right.\right) 
\right],
\end{align}
which has to be understood in terms of analytic
continuations to the complex plane minus the positive real axis.

\subsubsection{Braiding matrix}

Let us translate \eqref{qsf} into a relation between conformal blocks $\cF^{(s)}$ and $\cF^{(u)}$ defined in \eqref{zlks}, \eqref{zlku}. 
First write
\begin{align}
\label{spd}
\cF^{(s)}_{\alpha_1-bh_l}\!\left[\begin{matrix}
{\rm \hat m} & \mu \\ \alpha_2 & \alpha_1
\end{matrix}\right]
&= 
\frac{(1-z)^{b\langle {\rm \hat m}, h_1\rangle}}{z^{b\langle \mu-Q, h_1\rangle - \tau_l}}
F\left(\left.\begin{smallmatrix}
(\sigma_1 + \tau_l) \cdots (\sigma_N + \tau_l)\\
(\tau_l - \tau_1 + 1)\,\widehat{\cdots}\, (\tau_l-\tau_N+1)
\end{smallmatrix}\right| z \right)
\\
\notag
&=
\frac{(1-z)^{b\langle {\rm \hat m}, h_1\rangle}}{z^{b\langle \mu-Q, h_1\rangle - \tau_l}}
\sum_{k=1}^N \left[(-z)^{-\sigma_k-\tau_l} 
\, \boxed{\frac{\Gamma\cdots\Gamma}{\Gamma\cdots\Gamma}}\,
F\left(\begin{smallmatrix}(\sigma_k+\tau_1)\cdots (\sigma_k+\tau_N)\\
(1+\sigma_k-\sigma_1) \,\widehat{\cdots} (1+\sigma_k-\sigma_N)
\end{smallmatrix} \left|\frac{1}{z}\right.\right)
\right],
\end{align}
where the product of Gamma functions is
\begin{equation*}
\boxed{\frac{\Gamma\cdots\Gamma}{\Gamma\cdots\Gamma}}
=
\frac{\prod_{j\neq l}\Gamma(\tau_l-\tau_j+1)}
     {\prod_{j\neq k}\Gamma(\sigma_j+\tau_l)}
\frac{\prod_{j\neq k}\Gamma(\sigma_j-\sigma_k)}
     {\prod_{j\neq l}\Gamma(1-\sigma_k-\tau_j)}
=
\prod_{j\neq l} \frac{\Gamma(1+\tau_l - \tau_j)}{\Gamma(1-\sigma_k-\tau_j)}
\prod_{j\neq k} \frac{\Gamma(\sigma_j-\sigma_k)}{\Gamma(\sigma_j + \tau_l)}.
\end{equation*}
In \eqref{spd}, powers of $(-z)$ are computed with an argument $-\pi < \arg(-z) < \pi$, whereas for powers of $z$, we take $-\pi < \arg(z) < \pi$. With these conventions, $(-z)^{-\sigma_k-\tau_l} = e^{i\pi(\sigma_k+\tau_l)\epsilon} z^{-\sigma_k - \tau_l}$, where $\epsilon = \sign(\Im(z))$. Similarly, $(1-z)^{b\langle {\rm \hat m}, h_1\rangle} = e^{-i\pi b\langle {\rm \hat m}, h_1\rangle \epsilon} (z-1)^{b\langle {\rm \hat m}, h_1\rangle}$. All in all,
\begin{align}
\notag
\cF^{(s)}_{\alpha_1-bh_l}\!\left[\begin{matrix}
{\rm \hat m} & \mu \\ \alpha_2 & \alpha_1
\end{matrix}\right]
&= 
\sum_{k=1}^N \left[
\frac{e^{i\pi(\sigma_k+\tau_l-b\langle {\rm \hat m}, h_1\rangle)\epsilon}
(z-1)^{b\langle {\rm \hat m}, h_1\rangle}}{z^{b\langle \mu-Q, h_1\rangle +\sigma_k}}
\, \boxed{\frac{\Gamma\cdots\Gamma}{\Gamma\cdots\Gamma}}\,
F\left(\begin{smallmatrix}(\sigma_k+\tau_1)\cdots (\sigma_k+\tau_N)\\
(1+\sigma_k-\sigma_1) \,\widehat{\cdots} (1+\sigma_k-\sigma_N)
\end{smallmatrix} \left|\frac{1}{z}\right.\right)
\right]
\\
\notag
&= 
e^{-i\pi b \langle {\rm \hat m}, h_1\rangle \epsilon}
\sum_{k=1}^N \left[
e^{i\pi(\sigma_k+\tau_l)\epsilon}
\, \boxed{\frac{\Gamma\cdots\Gamma}{\Gamma\cdots\Gamma}}\,
\cF^{(u)}_{\alpha_2+bh_k}\!\left[\begin{matrix}
{\rm \hat m} & \mu \\ \alpha_2 & \alpha_1
\end{matrix}\right]
\right]
\\
\label{braid}
&=
\sum_{k=1}^N B_{lk}^{\epsilon} \!\left[\begin{matrix}
{\rm \hat m} & \mu \\ \alpha_2 & \alpha_1
\end{matrix}\right]
\cF^{(u)}_{\alpha_2+bh_k}\!\left[\begin{matrix}
{\rm \hat m} & \mu \\ \alpha_2 & \alpha_1
\end{matrix}\right].
\end{align}
Replacing $\sigma_i$ and $\tau_i$ by their explicit form~\eqref{parameters} gives
\begin{align}
\notag
&\begin{matrix}
  \begin{fmfgraph*}(35,15)
    \fmfstraight
    \fmftop{tl,t1,t2,t3,tr}
    \fmfbottom{bl,b1,b2,b3,br}
    \fmf{fermion,width=2}{br,b3,b1,bl}
    \fmf{fermion,width=0.5}{t3,b3}
    \fmf{fermion,width=2}{t1,b1}
    \fmfv{label=$\alpha_2$,label.angle=60,label.dist=6}{bl}
    \fmf{phantom,label=$\alpha_1-bh_l$,label.side=right}{b3,b1}
    \fmfv{label=$\alpha_1$,label.angle=120,label.dist=6}{br}
    \fmfv{label=${\rm \hat m}$,label.angle=180,label.dist=4}{t1}
    \fmfv{label=$\mu$,label.angle=0,label.dist=4}{t3}
  \end{fmfgraph*}
\end{matrix}
=
\sum_{k=1}^N 
e^{i\pi\epsilon \varphi} \prod_{j\neq l} 
\frac{\Gamma(1- b\langle \alpha_1-Q, h_j - h_l\rangle)}
{\Gamma(1+b\langle \alpha_2-Q,h_k\rangle 
        - b\langle \alpha_1-Q,h_j\rangle
        - b\langle \mu+{\rm \hat m}, h_1\rangle)}  
\cdot
\\  
\label{braidingRL}
&\cdot
 \prod_{j\neq k}
\frac{\Gamma(b\langle \alpha_2-Q, h_k- h_j\rangle)}
{\Gamma(b\langle \alpha_1-Q, h_l\rangle
        - b\langle \alpha_2-Q, h_j\rangle  
        + b\langle \mu+{\rm \hat m}, h_1\rangle )}
\begin{matrix}
  \begin{fmfgraph*}(35,15)
    \fmfstraight
    \fmftop{tl,t1,t2,t3,tr}
    \fmfbottom{bl,b1,b2,b3,br}
    \fmf{fermion,width=2}{br,b3,b1,bl}
    \fmffreeze
    \fmf{fermion,width=0.5}{t1,b1}
    \fmf{fermion,width=2}{t3,b3}
    \fmfv{label=$\alpha_2$,label.angle=60,label.dist=5}{bl}
    \fmf{phantom,label=$\alpha_2+b h_k$,label.side=left}{b1,b3}
    \fmfv{label=$\alpha_1$,label.angle=120,label.dist=5}{br}
    \fmfv{label=${\rm \hat m}$,label.angle=180,label.dist=4}{t3}
    \fmfv{label=$\mu$,label.angle=180,label.dist=4}{t1}
  \end{fmfgraph*}
\end{matrix}
\end{align}
with $\varphi= b\langle \alpha_1-Q, h_l\rangle - b \langle \alpha_2-Q, h_k\rangle + b\langle \mu, h_1\rangle$. To obtain from this the expression \eqref{braiding} used in the main text, flip the diagram around a vertical axis, changing the sign of $\epsilon$ in the process, and then apply the transformation $\alpha_1 - Q \leftrightarrow Q-\alpha_2$ in order to account for the relabeling and reversing of the arrows.

The braiding matrix $B^\epsilon_{lk}\!\left[\begin{smallmatrix} {\rm \hat m} & \mu \\ \alpha_2 & \alpha_1 \end{smallmatrix}\right]$ introduced in \eqref{braid} can be factored as the product
\begin{equation*}
B_{lk}^{\epsilon} \!\left[\begin{matrix}
{\rm \hat m} & \mu \\ \alpha_2 & \alpha_1
\end{matrix}\right] = e^{-i\pi b\langle {\rm \hat m}, h_1\rangle\epsilon}
D_l \check B_{lk}^\epsilon \wD_k
\end{equation*}
of diagonal matrices $D$ and $\widetilde D$, and a rather simple matrix $\check B$:
\begin{align*}
D_l &=
\dfrac{\Gamma(\tau_l-\tau_1+1) \cdots \Gamma(\tau_l - \tau_N +1)}
{\Gamma(\sigma_1+\tau_l) \cdots \Gamma(\sigma_N+\tau_l)},
\qquad
\wD_k =
\dfrac{\Gamma(\sigma_1 - \sigma_k)\,\widehat{\cdots}\, 
  \Gamma(\sigma_N-\sigma_k)}
{\Gamma(1-\sigma_k-\tau_1)\cdots \Gamma(1-\sigma_k-\tau_N)},
\\
\check B_{lk}^\epsilon &=
\Gamma(\sigma_k+\tau_l) \Gamma(-\sigma_k-\tau_l+1) 
e^{i\pi\epsilon(\sigma_k+\tau_l)}
= \dfrac{\pi e^{i\pi\epsilon(\sigma_k+\tau_l)}}{\sin \pi (\sigma_k+\tau_l)}.
\end{align*}

\subsection{$t$-channel, and fusion}

The $s$- and $u$-channel conformal blocks correspond to bases of solutions with diagonal monodromy around 0 and $\infty$ respectively. In the same way, we get $t$-channel conformal blocks from the monodromy around 1, expressed as the product of two braidings: $M_{(1)}=B^- (B^+)^{-1}$. 

In this paragraph, we denote $\phi = \pi b\langle {\rm \hat m}, h_1\rangle$. 

The operator $M_{(1)}- e^{2 i\phi}\id = (B^- -e^{2i\phi} B^+) (B^+)^{-1} = e^{i\phi} D_l (\check B_{lk}^- - \check B_{lk}^+)\wD_k (B^+)^{-1} $ is of rank~1: indeed, $\check B_{lk}^- - \check B_{lk}^+ = -2i\pi$ for all $l,k$. From this computation we deduce that $M_{(1)}$ has the eigenvalue $e^{2i\phi}$ with degeneracy $N-1$. The determinant of $M_{(1)}$ is easily seen to be 
\begin{equation*}
\det M_{(1)} = e^{2i\pi (N b\langle {\rm \hat m}, h_1\rangle - \Sigma)},
\end{equation*}
where $\Sigma=\sum_j (\sigma_j + \tau_j) = N b\langle \mu+{\rm \hat m},h_1\rangle$. Hence, the last eigenvalue is $e^{2i\phi - 2i\pi \Sigma}$.

Fateev and Litvinov \cite{Fateev:2005gs} prove that the OPE \eqref{semidegeneratefuserep} between the semi-degenerate $V_{{\rm \hat m}}$ and the degenerate $V_\mu$ has two terms. 
\begin{align}
  \label{t-decomposition}
  &\vev{V_{2Q-\alpha_2}(\infty)V_{{\rm \hat m}}(1)V_{\mu}(z)V_{\alpha_1}(0)}
  \\
  \notag
  &=
  C({\rm \hat m},\mu,2Q-{\rm \hat m} + b h_1)
  C({\rm \hat m} - b h_1, 2Q-\alpha_2,\alpha_1)
  |1-z|^{2[\Delta({\rm \hat m}-bh_1) - \Delta({\rm \hat m}) - \Delta(\mu)]} (1+\cdots)
  \\
  \notag
  &\quad +
  C({\rm \hat m},\mu,2Q - {\rm \hat m} + b h_N)
  C({\rm \hat m} - b h_N,2Q-\alpha_2,\alpha_1)
  |1-z|^{2[\Delta({\rm \hat m}-bh_N) - \Delta({\rm \hat m}) - \Delta(\mu)]} (1+\cdots)
\end{align}
where $(1+\cdots)$ are series in (positive) powers of $(1-z),\overline{(1-z)}$. Comparing the exponents of $(1-z)$ and the eigenvalues of $M_{(1)}$, one can write the second term
\begin{equation*}
  C({\rm \hat m},\mu,2Q - {\rm \hat m} + b h_N)
  C({\rm \hat m} - b h_N,2Q-\alpha_2,\alpha_1)
  \overline{\cF^{(t)}_{{\rm \hat m}-b h_N}\!\left[\begin{matrix}
        {\rm \hat m} & \mu \\ \alpha_2 & \alpha_1
      \end{matrix}\right]}
  \cF^{(t)}_{{\rm \hat m}- b h_N}\!\left[\begin{matrix}
      {\rm \hat m} & \mu \\ \alpha_2 & \alpha_1
    \end{matrix}\right],
\end{equation*}
in terms of the $t$-channel conformal block $\cF^{(t)}_{{\rm \hat m} - bh_N}$, the eigenfunction of $M_{(1)}$ with monodromy $e^{2i\phi-2i\pi\Sigma}$. It is given by
\begin{equation}
  \label{zlkt}
  \cF^{(t)}_{{\rm \hat m}- b h_N}\!\left[\begin{matrix}
      {\rm \hat m} & \mu \\ \alpha_2 & \alpha_1
    \end{matrix}\right]
  =
  \begin{matrix}
    \fmfframe(8,0)(0,0){\begin{fmfgraph*}(20,25)
      \fmftop{tl,tr}
      \fmfbottom{bl,br}
      \fmf{fermion,width=2}{br,b2,bl}
      \fmffreeze
      \fmf{fermion,width=0.5}{tr,v}
      \fmf{fermion,width=2}{tl,v}
      \fmf{fermion,width=2,label=${\rm \hat m}-bh_N$}{v,b2}
      \fmfv{label=$\alpha_2$,label.angle=-60,label.dist=6}{bl}
      \fmfv{label=$\alpha_1$,label.angle=-120,label.dist=6}{br}
      \fmfv{label=${\rm \hat m}$,label.angle=30,label.dist=4}{tl}
      \fmfv{label=$\mu$,label.angle=150,label.dist=4}{tr}
    \end{fmfgraph*}}
  \end{matrix}
  =
  \Gamma(N-\Sigma) \sum_{k=1}^N \wD_k^\leftrightarrow
  \cF^{(s)}_{\alpha_1-bh_k}\!\left[\begin{matrix}
      {\rm \hat m} & \mu \\ \alpha_2 & \alpha_1
    \end{matrix}\right],
\end{equation}
where the $\leftrightarrow$ denotes exchanging all $\sigma$ and $\tau$:
\begin{equation*}
\wD_k^{\leftrightarrow} 
= \dfrac{\Gamma(\tau_1 - \tau_k)\,\widehat{\cdots}\, \Gamma(\tau_N-\tau_k)}{\Gamma(-\tau_k-\sigma_1+1)\cdots \Gamma(-\tau_k-\sigma_N+1)}.
\end{equation*}
The overall normalization is chosen so that the leading order coefficient of $\cF^{(t)}_{{\rm \hat m}- b h_N}$ as a series in powers of $(1-z)$ is~1, as shown in~\cite{Norlund:1955}. 

The first term of \eqref{t-decomposition}, on the other hand, does not split as the product of a holomorphic times an antiholomorphic part, because the spaces of holomorphic/anti-holomorphic solutions of the hypergeometric differential equation \eqref{sd3} with monodromy $e^{2i\phi}$ around~1 are $(N-1)$-dimensional. This indicates that the holomorphic $W_N$ Ward identities have multiple solutions in this case, instead of a unique (normalized) solution $\cF^{(t)}_{{\rm \hat m}-bh_N}$ for a momentum ${\rm \hat m}-bh_N$. This is in striking contrast with the case of the Virasoro symmetry, for which the decomposition in holomorphic and anti-holomorphic conformal blocks always happens.

We obtain \eqref{fusion1} of the main text by setting ${\rm \hat m}=\mu^* = b h_N$ in \eqref{zlkt}. The internal momentum is then $\mu^*-bh_N=0$, corresponding to the identity operator $V_0 = \id$. In the construction of Verlinde loop operators, we also need to ``fuse $V_\mu$ and $V_{\mu^*}$ back together'': it is done using the projection onto the $e^{2i\phi-2i\pi\Sigma}$ eigenspace of $M_{(1)}$, given for general ${\rm \hat m}$ by
\begin{equation*}
  \pr\left[
  \begin{matrix}
    \begin{fmfgraph*}(30,15)
      \fmftop{tl,tr}
      \fmfbottom{bl,br}
      \fmf{fermion,width=2}{br,b1,b3,bl}
      \fmf{phantom}{br,b1} \fmf{phantom}{b3,bl}
      \fmf{phantom}{tr,t1,t2,t3,tl}
      \fmffreeze
      \fmf{fermion,width=0.5}{t1,b1}
      \fmf{fermion,width=2}{t3,b3}
      \fmfv{label=$\alpha_2$,label.angle=-60,label.dist=6}{bl}
      \fmf{phantom,label=$\alpha_1-bh_k$,label.side=left}{b1,b3}
      \fmfv{label=$\alpha_1$,label.angle=-120,label.dist=6}{br}
      \fmfv{label=$\mu$,label.angle=180,label.dist=4}{t1}
      \fmfv{label=${\rm \hat m}$,label.angle=0,label.dist=4}{t3}
    \end{fmfgraph*}
  \end{matrix}
  \right]
  = 
  \Gamma(1-(N-\Sigma)) 
  \left[\prod_j  \frac{\Gamma(\tau_k-\tau_j+1)}{\Gamma(\sigma_j+\tau_k)}\right]
  \begin{matrix}
    \fmfframe(0,2)(0,-2){\begin{fmfgraph*}(20,25)
      \fmftop{tl,tr}
      \fmfbottom{bl,br}
      \fmf{fermion,width=2}{br,b2,bl}
      \fmffreeze
      \fmf{fermion,width=0.5}{tr,v}
      \fmf{fermion,width=2}{tl,v}
      \fmf{phantom,tension=3}{v,v2,v3,b2}
      \fmffreeze
      \fmf{fermion,width=2}{v2,b2}
      \fmf{plain,width=2}{v,v2}
      \fmf{phantom,label=${\rm \hat m}-bh_N$,label.side=left}{v,v3} 
      \fmfv{label=$\alpha_2$,label.angle=60,label.dist=6}{bl}
      \fmfv{label=$\alpha_1$,label.angle=120,label.dist=6}{br}
      \fmfv{label=${\rm \hat m}$,label.angle=30,label.dist=4}{tl}
      \fmfv{label=$\mu$,label.angle=150,label.dist=4}{tr}
    \end{fmfgraph*}}
  \end{matrix}
\end{equation*}
In \eqref{fusion2} we set ${\rm \hat m}=\mu^*$.

%===================================================
\section{Some monodromies in Toda CFT}
\label{sec:SQCDmonodromies}

This appendix is devoted to the computation of monodromies of $V_\mu$ and $V_{\mu^*}$ along curves relevant for the computation of 't Hooft loops in conformal SQCD (Section~\ref{sec:SQCD}), that is, curves drawn on a sphere with two semi-degenerate punctures ${\rm \hat m}_2$ and ${\rm \hat m}_3^*$, and two full punctures labeled by momenta ${\rm m}_1$ and ${\rm m}_4^*$. 

In the process, some sums of products of sines are evaluated in terms of contour integrals. This method is also used in Section~\ref{sec:SQCD} and in Appendix~\ref{sec:Wilson}.

\subsection{Monodromies of $\mu$}

We first aim to compute the monodromies described by the second and third diagrams below:
\begin{align}
\begin{matrix}\fmfframe(0,5)(5,0){\begin{fmfgraph*}(30,20)
  \fmfcmd{style_def my_arrow expr p = 
    cdraw (subpath (length(p)/10,length(p)) of p);
    shrink(.7); cfill (marrow (p,.3)); endshrink; 
    enddef;}
  \fmfcmd{style_def my expr p = 
    cdraw (subpath (length(p)/10,length(p)) of p);
    enddef;}
  \fmfstraight
  \fmftop{00,10,20,30,40,50}
  \fmfbottom{08,18,28,38,48,58}
  \fmf{phantom}{00,02,04,06,08}
  \fmf{phantom}{10,11,12,13,14,15,16,17,18}
  \fmf{phantom}{20,22,24,26,28}
  \fmf{phantom}{30,32,34,36,38}
  \fmf{phantom}{40,41,42,43,44,45,46,47,48}
  \fmf{phantom}{50,52,54,56,58}
  \fmffreeze
  \fmf{fermion,width=.5}{10,16}
  \fmf{fermion,width=2,rubout}{30,36}
  \fmf{plain,right,width=.5}{45,44}
  \fmf{fermion,width=2}{56,36,16,06}
  \fmf{my,width=.5}{14,45}
  \fmf{my_arrow,width=.5}{13,44}
  \fmfv{label=$\mu$,label.angle=180}{10}
  \fmfv{label=${\rm \hat m}_2$,label.angle=0}{30}
  \fmfv{label=$\alpha'$,label.angle=-60}{06}
  \fmfv{label=${\rm m}_1$,label.angle=-90}{56}
  \fmf{phantom,label=$\alpha$,label.side=left}{36,16}
\end{fmfgraph*}}\end{matrix}
&
&
\begin{matrix}\fmfframe(0,5)(5,0){\begin{fmfgraph*}(30,20)
  \fmfcmd{style_def my_arrow expr p = 
    cdraw (subpath (length(p)/10,length(p)) of p);
    shrink(.7); cfill (marrow (p,.3)); endshrink; 
    enddef;}
  \fmfcmd{style_def my expr p = 
    cdraw (subpath (length(p)/10,length(p)) of p);
    enddef;}
  \fmfstraight
  \fmftop{00,10,20,30,40,50}
  \fmfbottom{08,18,28,38,48,58}
  \fmf{phantom}{00,02,04,06,08}
  \fmf{phantom}{10,11,12,13,14,15,16,17,18}
  \fmf{phantom}{20,22,24,26,28}
  \fmf{phantom}{30,32,34,36,38}
  \fmf{phantom}{40,41,42,43,44,45,46,47,48}
  \fmf{phantom}{50,52,54,56,58}
  \fmffreeze
  \fmf{fermion,width=.5}{10,16}
  \fmf{fermion,width=2,rubout}{30,36}
  \fmf{plain,right,width=.5}{45,44}
  \fmf{fermion,width=2}{56,36,16,06}
  \fmf{my,width=.5,rubout=5}{14,45}
  \fmf{my_arrow,width=.5}{13,44}
  \fmfv{label=$\mu$,label.angle=180}{10}
  \fmfv{label=${\rm \hat m}_2$,label.angle=0}{30}
  \fmfv{label=$\alpha'$,label.angle=-60}{06}
  \fmfv{label=${\rm m}_1$,label.angle=-90}{56}
  \fmf{phantom,label=$\alpha$,label.side=left}{36,16}
\end{fmfgraph*}}\end{matrix}
&
&
\begin{matrix}\fmfframe(0,5)(5,0){\begin{fmfgraph*}(30,20)
  \fmfcmd{style_def my_arrow expr p = 
    cdraw (subpath (length(p)/10,length(p)) of p);
    shrink(.7); cfill (marrow (p,.3)); endshrink; 
    enddef;}
  \fmfcmd{style_def my expr p = 
    cdraw (subpath (length(p)/10,length(p)) of p);
    enddef;}
  \fmfstraight
  \fmftop{00,10,20,30,40,50}
  \fmfbottom{08,18,28,38,48,58}
  \fmf{phantom}{00,02,04,06,08}
  \fmf{phantom}{10,11,12,13,14,15,16,17,18}
  \fmf{phantom}{20,22,24,26,28}
  \fmf{phantom}{30,32,34,36,38}
  \fmf{phantom}{40,41,42,43,44,45,46,47,48}
  \fmf{phantom}{50,52,54,56,58}
  \fmf{phantom}{47,47a}
  \fmf{phantom}{45,45a}
  \fmf{phantom}{45,45b}
  \fmf{phantom}{14,14b}
  \fmffreeze
  \fmf{fermion,width=.5}{10,16}
  \fmf{fermion,width=2,rubout}{30,36}
  \fmf{plain,right,width=.5}{47,44}
  \fmf{fermion,width=2,rubout}{56,36,16,06}
  \fmf{plain,left,width=.5,rubout=5}{47a,45a}
  \fmf{my,width=.5}{14b,45b}
  \fmf{my_arrow,width=.5}{13,44}
  \fmfv{label=$\mu$,label.angle=180}{10}
  \fmfv{label=${\rm \hat m}_2$,label.angle=0}{30}
  \fmfv{label=$\alpha'$,label.angle=-60}{06}
  \fmfv{label=${\rm m}_1$,label.angle=-90}{56}
  \fmf{phantom,label=$\alpha$,label.side=left}{36,16}
\end{fmfgraph*}}\end{matrix}
&
&
\begin{matrix}\fmfframe(0,5)(5,0){\begin{fmfgraph*}(30,20)
  \fmfcmd{style_def my_arrow expr p = 
    cdraw (subpath (length(p)/10,length(p)) of p);
    shrink(.7); cfill (marrow (p,.3)); endshrink; 
    enddef;}
  \fmfcmd{style_def my expr p = 
    cdraw (subpath (length(p)/10,length(p)) of p);
    enddef;}
  \fmfstraight
  \fmftop{00,10,20,30,40,50}
  \fmfbottom{08,18,28,38,48,58}
  \fmf{phantom}{00,02,04,06,08}
  \fmf{phantom}{10,11,12,13,14,15,16,17,18}
  \fmf{phantom}{20,22,24,26,28}
  \fmf{phantom}{30,32,34,36,38}
  \fmf{phantom}{40,41,42,43,44,45,46,47,48}
  \fmf{phantom}{50,52,54,56,58}
  \fmf{phantom}{47,47a}
  \fmf{phantom}{45,45a}
  \fmf{phantom}{45,45b}
  \fmf{phantom}{14,14b}
  \fmffreeze
  \fmf{fermion,width=.5}{10,16}
  \fmf{fermion,width=2,rubout}{30,36}
  \fmf{plain,right,width=.5}{47,44}
  \fmf{fermion,width=2,rubout}{56,36,16,06}
  \fmf{plain,left,width=.5,rubout=5}{47a,45a}
  \fmf{my,width=.5,rubout=5}{14b,45b}
  \fmf{my_arrow,width=.5}{13,44}
  \fmfv{label=$\mu$,label.angle=180}{10}
  \fmfv{label=${\rm \hat m}_2$,label.angle=0}{30}
  \fmfv{label=$\alpha'$,label.angle=-60}{06}
  \fmfv{label=${\rm m}_1$,label.angle=-90}{56}
  \fmf{phantom,label=$\alpha$,label.side=left}{36,16}
\end{fmfgraph*}}\end{matrix}
\label{SQCDdiagramlist}
\end{align}
The other two diagrams correspond to moves that will allow us to fix the dictionnary between some signs appearing in the formulae for braiding moves and the precise curve to which they correspond. For our purposes, $\alpha'=\alpha-bh_l$ for some $l=1,\cdots,N$.

Each one of these four moves consists of three steps. First braid $\mu$ around ${\rm \hat m}_2$ with some orientation $\epsilon_2=\pm 1$. Then take $\epsilon_1= 0,\pm 1$ times the monodromy of $\mu$ around ${\rm m}_1$. Finally, braid back $\mu$ around ${\rm \hat m}_2$, with an orientation controlled by $\epsilon_2'$. We denote by $M^{\epsilon_2,\epsilon_1,\epsilon_2'}$ the move consisting of these two braiding moves with $\epsilon_1$ rotations of $\mu$ around ${\rm m}_1$ in between: 
\[
M^{\epsilon_2,\epsilon_1,\epsilon_2'}\cdot
\left[
\begin{matrix}
  \begin{fmfgraph*}(35,15)
    \fmfstraight
    \fmftop{tl,t1,t2,t3,tr}
    \fmfbottom{bl,b1,b2,b3,br}
    \fmf{fermion,width=2}{br,b3,b1,bl}
    \fmf{fermion,width=2}{t3,b3}
    \fmf{fermion,width=0.5}{t1,b1}
    \fmfv{label=$\alpha'$,label.angle=60,label.dist=6}{bl}
    \fmf{phantom,label=$\alpha'+bh_l$,label.side=right}{b3,b1}
    \fmfv{label=${\rm m}_1$,label.angle=120,label.dist=6}{br}
    \fmfv{label=$\mu$,label.angle=180,label.dist=4}{t1}
    \fmfv{label=${\rm \hat m}_2$,label.angle=0,label.dist=4}{t3}
  \end{fmfgraph*}
\end{matrix}
\right]
=
\sum_r M_{lr}^{\epsilon_2,\epsilon_1,\epsilon_2'}
\left[
\begin{matrix}
  \begin{fmfgraph*}(35,15)
    \fmfstraight
    \fmftop{tl,t1,t2,t3,tr}
    \fmfbottom{bl,b1,b2,b3,br}
    \fmf{fermion,width=2}{br,b3,b1,bl}
    \fmf{fermion,width=2}{t3,b3}
    \fmf{fermion,width=0.5}{t1,b1}
    \fmfv{label=$\alpha'$,label.angle=60,label.dist=6}{bl}
    \fmf{phantom,label=$\alpha'+bh_r$,label.side=right}{b3,b1}
    \fmfv{label=${\rm m}_1$,label.angle=120,label.dist=6}{br}
    \fmfv{label=$\mu$,label.angle=180,label.dist=4}{t1}
    \fmfv{label=${\rm \hat m}_2$,label.angle=0,label.dist=4}{t3}
  \end{fmfgraph*}
\end{matrix}
\right].
\]
It is not easy to track down which sign of $\epsilon_2$ or $\epsilon_2'$ corresponds to which direction for the braiding. We will determine it by finding out that $M^{\epsilon_2,0,-\epsilon_2}$ is simply the identity matrix, corresponding to the monodromy along a trivial curve. Hence, this case corresponds to the first diagram in \eqref{SQCDdiagramlist}. We deduce immediately from this that the second and fourth diagrams have $\epsilon_2'=\epsilon_2$, while diagrams 1 and 3 have $\epsilon_2'=-\epsilon_2$. Of course, we know that $\epsilon_1=0$ for the two first diagrams, but we still have to determine the relative sign between $\epsilon_1$ and $\epsilon_2$ in the two last diagrams. This is done by proving that $M^{\epsilon,\epsilon,\epsilon}$ is diagonal, whereas $M^{\epsilon,-\epsilon,\epsilon}$ is not: therefore, diagram 4, which is simply a monodromy of $\mu$ around a tube in the pants decomposition, has $\epsilon_1=\epsilon_2$.

The matrices describing the braiding of $\mu$ around ${\rm \hat m}_2$ are~\eqref{braidingRL}, and its ``reversed'' counterpart,~\eqref{braiding}. In our current setting, the momenta $\alpha_1$ and $\alpha_2$ are replaced by ${\rm m}_1$ and $\alpha'=\alpha-bh_l$ (see Section~\ref{sec:SQCD}), and we can express braiding moves using $\langle\alpha-Q,h_j\rangle = ia_j$ and $\langle {\rm m}_1-Q,h_j\rangle = im_{1,j}$:
\begin{align*}
\notag
\begin{matrix}
  \begin{fmfgraph*}(35,15)
    \fmfstraight
    \fmftop{tl,t1,t2,t3,tr}
    \fmfbottom{bl,b1,b2,b3,br}
    \fmf{fermion,width=2}{br,b3,b1,bl}
    \fmf{fermion,width=2}{t3,b3}
    \fmf{fermion,width=0.5}{t1,b1}
    \fmfv{label=$\alpha'$,label.angle=60,label.dist=6}{bl}
    \fmf{phantom,label=$\alpha$,label.side=right}{b3,b1}
    \fmfv{label=${\rm m}_1$,label.angle=120,label.dist=6}{br}
    \fmfv{label=$\mu$,label.angle=180,label.dist=4}{t1}
    \fmfv{label=${\rm \hat m}_2$,label.angle=0,label.dist=4}{t3}
  \end{fmfgraph*}
\end{matrix}
=
\sum_{k=1}^N 
&e^{\pi\epsilon_2 b(m_{1,k} - a_l)} 
\prod_{j\neq l} \frac{\Gamma(bq+ ib(a_j-a_l))}
{\Gamma(bq +iba_j - ibm_{1,k} - b\langle {\rm \hat m}_2, h_1\rangle)}  
\cdot
\\  
&\cdot
 \prod_{j\neq k}
\frac{\Gamma(ib(m_{1,j}-m_{1,k}))}
{\Gamma(-iba_l + ibm_{1,j} + b\langle {\rm \hat m}_2, h_1\rangle )}
\begin{matrix}
  \begin{fmfgraph*}(35,15)
    \fmfstraight
    \fmftop{tl,t1,t2,t3,tr}
    \fmfbottom{bl,b1,b2,b3,br}
    \fmf{fermion,width=2}{br,b3,b1,bl}
    \fmffreeze
    \fmf{fermion,width=2}{t1,b1}
    \fmf{fermion,width=0.5}{t3,b3}
    \fmfv{label=$\alpha'$,label.angle=60,label.dist=5}{bl}
    \fmf{phantom,label=${\rm m}_1-b h_k$,label.side=left}{b1,b3}
    \fmfv{label=${\rm m}_1$,label.angle=120,label.dist=5}{br}
    \fmfv{label=$\mu$,label.angle=180,label.dist=4}{t3}
    \fmfv{label=${\rm \hat m}_2$,label.angle=180,label.dist=4}{t1}
  \end{fmfgraph*}
\end{matrix}
\end{align*}
and
\begin{align*}
\notag
\begin{matrix}
  \begin{fmfgraph*}(35,15)
    \fmfstraight
    \fmftop{tl,t1,t2,t3,tr}
    \fmfbottom{bl,b1,b2,b3,br}
    \fmf{fermion,width=2}{br,b3,b1,bl}
    \fmffreeze
    \fmf{fermion,width=2}{t1,b1}
    \fmf{fermion,width=0.5}{t3,b3}
    \fmfv{label=$\alpha'$,label.angle=60,label.dist=5}{bl}
    \fmf{phantom,label=${\rm m}_1-b h_k$,label.side=left}{b1,b3}
    \fmfv{label=${\rm m}_1$,label.angle=120,label.dist=5}{br}
    \fmfv{label=$\mu$,label.angle=180,label.dist=4}{t3}
    \fmfv{label=${\rm \hat m}_2$,label.angle=180,label.dist=4}{t1}
  \end{fmfgraph*}
\end{matrix}
=
\sum_{r=1}^N 
&e^{\pi b\epsilon'_2 (m_{1,k} - a_r  + ib(1-\delta_{lr}))} 
\prod_{j\neq k} \frac{\Gamma(1- ib(m_{1,j}-m_{1,k}))}
{\Gamma(bq-b^2\delta_{lr} + iba_r - ibm_{1,j} - b\langle {\rm \hat m}_2, h_1\rangle)}  
\cdot
\\  
&\cdot
 \prod_{j\neq r}
\frac{\Gamma(b^2(\delta_{lj}-\delta_{lr}) + ib(a_r-a_j) )}
{\Gamma(b^2(\delta_{lj}-1) - iba_j + ibm_{1,k} + b\langle {\rm \hat m}_2, h_1\rangle )}
\begin{matrix}
  \begin{fmfgraph*}(35,15)
    \fmfstraight
    \fmftop{tl,t1,t2,t3,tr}
    \fmfbottom{bl,b1,b2,b3,br}
    \fmf{fermion,width=2}{br,b3,b1,bl}
    \fmf{fermion,width=2}{t3,b3}
    \fmf{fermion,width=0.5}{t1,b1}
    \fmfv{label=$\alpha'$,label.angle=60,label.dist=6}{bl}
    \fmf{phantom,label=$\alpha'+bh_r$,label.side=right}{b3,b1}
    \fmfv{label=${\rm m}_1$,label.angle=120,label.dist=6}{br}
    \fmfv{label=$\mu$,label.angle=180,label.dist=4}{t1}
    \fmfv{label=${\rm \hat m}_2$,label.angle=0,label.dist=4}{t3}
  \end{fmfgraph*}
\end{matrix}
\end{align*}

The braiding of $\mu$ with ${\rm m}_1$ contributes $e^{2i\pi \epsilon_1 [\Delta({\rm m}_1 - bh_k) - \Delta({\rm m}_1) - \Delta(\mu)]} = e^{2i\pi \epsilon_1 b(im_{1,k} + \langle Q,h_1\rangle)}$. Therefore,
\begin{align*}
M_{lr}^{\epsilon_2,\epsilon_1,\epsilon_2'}
=\sum_k e^{\pi (\phi_{lr} - 2n b m_{1,k})} 
\prod_{j\neq l} \frac{\Gamma(bq+ ib(a_j - a_l))}{\Gamma(bq+iba_j - ibm_{1,k} - b\langle {\rm \hat m}_2, h_1\rangle)}  
\prod_{j\neq k} \frac{\Gamma(ib(m_{1,j}-m_{1,k}))}{\Gamma(-iba_l + ibm_{1,j} + b\langle {\rm \hat m}_2, h_1\rangle)}
\\
\cdot
\prod_{j\neq k} \frac{\Gamma(1- ib(m_{1,j}-m_{1,k}))} {\Gamma(bq-b^2\delta_{lr} + iba_r - ibm_{1,j} - b\langle {\rm \hat m}_2, h_1\rangle)} 
\prod_{j\neq r} \frac{\Gamma(b^2(\delta_{lj} - \delta_{lr})+ib(a_r-a_j))}{\Gamma(b^2(\delta_{lj}-1) - iba_j + ibm_{1,k} + b\langle {\rm \hat m}_2, h_1\rangle )}
\end{align*}
where $n=\epsilon_1 - (\epsilon_2+\epsilon_2')/2$ is an integer, and
\[
\phi_{lr} = -\epsilon_2 ba_l - \epsilon_2' ba_r
   + \epsilon_2' i b^2 (1-\delta_{lr})
   + \epsilon_1 i (N-1) bq.
\]

We factor out the part that is independent of $k$ in $M_{lr}^{\epsilon_2,\epsilon_1,\epsilon_2'}$, and use Euler's reflection relation $\Gamma(x)\Gamma(1-x) = \pi/\sin\pi x$ to get
\begin{align}
\notag
M_{lr}^{\epsilon_2,\epsilon_1,\epsilon_2'}
&=\frac{\prod_{j\neq l} \Gamma(bq+ ib(a_j-a_l)) \prod_{j\neq r}\Gamma(b^2(\delta_{lj}-\delta_{lr})+ib(a_r-a_j))}{\prod_j \Gamma(- iba_l + ibm_{1,j} + b\langle {\rm \hat m}_2, h_1\rangle ) \Gamma(bq-b^2\delta_{lr} + iba_r -ib m_{1,j} - b\langle {\rm \hat m}_2, h_1\rangle)}
\\
\label{Mlr}
&
\cdot
e^{\pi \phi_{lr}} \left[\pi \sum_k e^{-2\pi n b m_{1,k}} 
\frac{\prod_{j\neq l,r}' \sin\pi(b^2(\delta_{jl}-1) - iba_j + ibm_{1,k} + b\langle {\rm \hat m}_2, h_1\rangle )}
{\prod_{j\neq k} \sin\pi(ib(m_{1,j}-m_{1,k}))}\right]
\end{align}
where we use the notation $\prod_{j\neq l,r}' A_j = \big(\prod_j A_j\big)/(A_lA_r)$. 
We denote by $S_{lr}^{\epsilon_2,\epsilon_1,\epsilon_2'}$ the sum between brackets, and write it in terms of parameters slightly shifted compared to those in~\eqref{parameters}
\begin{align*}
  \sigma_j &= b\langle Q-\alpha', h_j\rangle + b\langle{\rm \hat m}_2+\mu, h_1\rangle
  = b^2 (\delta_{lj}-1) - iba_j + b\langle {\rm \hat m}_2,h_1\rangle
  \\
  \tau_j &=b\langle {\rm m}_1-Q, h_j\rangle = ibm_{1,j}
\end{align*}
Then,
\[
S_{lr}^{\epsilon_2,\epsilon_1,\epsilon_2'} = 
\left[\sum_k  e^{2i\pi n \tau_k} 
\frac{\prod_{j\neq l,r}' \sin\pi(\tau_k + \sigma_j)}
{\prod_{j\neq k} \sin\pi(\tau_j - \tau_k)}\right].
\]

Consider the integral
\[
I = \frac{(-1)^{N-1}}{2i} \oint dz e^{2i\pi n z} 
\frac{\prod_{j\neq l,r}' \sin\pi(z + \sigma_j)}
{\prod_j \sin\pi(z - \tau_j)}
\] 
along the contour enclosing the rectangle $\{x<\Re(z)<x+1, -y<\Im (z)<y\}$, with $x\in\mathbb{R}$ generic, and $y\to \infty$. 
 Note that the integrand has period 1.  It has poles at each $\tau_j$, and, if $l=r$, also at $\sigma_l$. The residue theorem implies that
\[
I = S_{lr}^{\epsilon_2,\epsilon_1,\epsilon_2'} 
+(-1)^{N-1} \delta_{lr} e^{-2i\pi n \sigma_l} \frac{\prod_{j\neq l} \sin\pi(\sigma_j - \sigma_l)}{\prod_j \sin\pi( - \sigma_l - \tau_j)},
\]
hence
\[
S_{lr}^{\epsilon_2,\epsilon_1,\epsilon_2'} 
=  \left[\delta_{lr} e^{-2i\pi n \sigma_l} \frac{\prod_{j\neq l} \sin\pi(\sigma_j - \sigma_l)}{\prod_j \sin\pi( \sigma_l + \tau_j)} + I\right].
\]

On the other hand, $I$ can be computed directly. The integrals along the ``vertical'' sides $\Re(z) = x,x+1$ cancel. Therefore,
\begin{equation}
\label{integral7}
I = \frac{(-1)^{N-1}}{2i} \int_x^{x+1} dx' 
\left[e^{2i\pi n z} \frac{\prod_{j\neq l,r}' \sin\pi(z + \sigma_j)}{\prod_j \sin\pi(z - \tau_j)}\right]^{z=x'-iy}_{z=x'+iy}
\end{equation}
where we use the notation $[f(z)]^{z=x'-iy}_{z=x'+iy} = f(x'-iy)-f(x'+iy)$. The $x'$ integral picks up the $n$-th Fourier coefficient of the product of sines. 

To obtain the relevant Fourier coefficient, we write $\sin\pi(z+\sigma_j) = \frac{1}{2i} (e^{i\pi(z+\sigma_j)}-e^{-i\pi(z+\sigma_j)})$, and similarly for $\sin\pi(z-\tau_j)$. Then, depending on whether $z\to - i\infty$ or $i\infty$, we expand the fraction as a converging series in positive or negative powers of $e^{i\pi z}$ respectively. For $z=x-i\epsilon y$, $\Re(i\epsilon z) \to +\infty$, and the product of sines in \eqref{integral7} is written 
\[
(2i\epsilon)^2 \frac{\prod_{j\neq l,r}' (e^{i\epsilon\pi(z + \sigma_j)}-e^{-i\epsilon\pi(z+\sigma_j)})}{\prod_j (e^{i\epsilon\pi(z - \tau_j)}-e^{-i\epsilon\pi(z-\tau_j)})}
= (2i)^2 e^{-2i\epsilon\pi z}e^{i\pi\epsilon (\sum_j \tau_j + \sum_j \sigma_j - \sigma_l - \sigma_r)} \frac{\prod_{j\neq l,r}' (1-e^{-2i\epsilon\pi(z+\sigma_j)})}{\prod_j (1-e^{-2i\epsilon\pi(z-\tau_j)})}.
\]
Furthermore, $\sum_j \tau_j + \sum_j \sigma_j - \sigma_l - \sigma_r = -b^2(N-2+\delta_{lr}) + ib(a_l+a_r) + (N-2) b\langle {\rm \hat m}_2, h_1 \rangle$. Expanding the remaining fraction of products yields further negative powers of $e^{2i\epsilon\pi z}$, which only contribute to higher Fourier coefficients, $|n|\geq 2$. 
\begin{itemize}
\item If $n=0$, $I=0$.
\item If $n= \pm 1$, $I= n (-1)^{N-1} 2i e^{i\pi n [-b^2(N-2+\delta_{lr}) + ib(a_l+a_r) + (N-2) b\langle {\rm \hat m}_2, h_1 \rangle]}$.
\item If $|n|\geq 2$, $I$ is some non-zero sum of exponentials, which we do not need to explicit for our purposes.
\end{itemize}

We group the results for $n = 0,\pm 1$ in an ad hoc but useful expression:
\[
I = n\cdot 2i e^{n i\pi [b^2(1-\delta_{lr}) -  N bq/2]}e^{- nb\pi [a_l + a_r + (N-2)\hat m_2]]}.
\]
Hence, as long as $n=\epsilon_1 - (\epsilon_2 + \epsilon_2')/2$ is $-1$, 0, or 1,
\[
S_{lr}^{\epsilon_2,\epsilon_1,\epsilon_2'} =
\left[\delta_{lr} e^{-2i\pi n \sigma_l} \frac{\prod_{j\neq l} \sin\pi(\sigma_j - \sigma_l)}{\prod_j \sin\pi(\sigma_l + \tau_j)}
+  n\cdot 2i e^{n i\pi [b^2(1-\delta_{lr})-N bq/2]} e^{- n\pi b[a_l+a_r +(N-2) \hat m_2]}\right].
\]

We now use this expression of the sum in \eqref{Mlr} to get a general expression for $M_{lr}^{\epsilon_2,\epsilon_1,-\epsilon_2'}$, valid if $n=\epsilon_1 - (\epsilon_2 + \epsilon_2')/2$ is $-1$, 0, or 1. This results in
\begin{align}
\label{Mresult}
&M_{lr}^{\epsilon_2,\epsilon_1,\epsilon_2'}
=\frac{\prod_{j\neq l} \Gamma(bq+ ib(a_j-a_l)) \prod_{j\neq r}\Gamma(b^2(\delta_{lj}-\delta_{lr})+ib(a_r-a_j))}{\prod_j \Gamma(bq/2- iba_l + ibm_{1,j} + ib\hat m_2) \Gamma(bq/2-b^2\delta_{lr} + iba_r -ib m_{1,j} - ib\hat m_2)}\cdot
\\
\notag
&\cdot
e^{\pi\phi_{lr}} 
\left[\delta_{lr} \pi e^{-2i\pi n \sigma_l} \frac{\prod_{j\neq l} \sin\pi(\sigma_j - \sigma_l)}{\prod_j \sin\pi(\sigma_l + \tau_j)}
+  n\cdot 2i\pi e^{n i\pi [b^2(1-\delta_{lr})-N bq/2]} e^{- n\pi b[a_l+a_r +(N-2) \hat m_2]}\right].
\end{align}

It is convenient to express separately diagonal and off-diagonal coefficients. For $l\neq r$,
\begin{align}
\notag
&M_{lr}^{\epsilon_2,\epsilon_1,\epsilon_2'}
= n\cdot 2i\pi 
e^{i\pi b^2 (\epsilon_2'+n) + i\pi bq [(N-1)\epsilon_1-Nn/2]}
e^{-(n+\epsilon_2) \pi ba_l - (n+\epsilon_2') \pi ba_r - n\pi b(N-2) \hat m_2}
\cdot
\\
\label{Moffdiagonal}
&\cdot
\frac{\prod_{j\neq l} \Gamma(bq+ ib(a_j-a_l)) \prod_{j\neq r}\Gamma(b^2(\delta_{lj}-\delta_{lr})+ib(a_r-a_j))}{\prod_j \Gamma(bq/2- iba_l + ibm_{1,j} + ib\hat m_2) \Gamma(bq/2-b^2\delta_{lr} + iba_r -ib m_{1,j} - ib\hat m_2)}
\end{align}
Note that for $n=0$, $M$ is diagonal, and for $n=\pm 1$, the first exponential becomes
\[
(-1)^{\epsilon_2'+n} e^{i\pi bq [\epsilon_2'+n + (N-1)\epsilon_1-Nn/2]}
= e^{i\pi [N\epsilon_1+(N/2)(\epsilon_2+\epsilon_2')+ (\epsilon_2'-\epsilon_2)]bq/2}.
\]

For $l=r$, the Gamma functions combine as $[\prod_j \sin\pi(\sigma_l+\tau_j)]/[\pi\cdot \prod_{j\neq l} \sin\pi(\sigma_j - \sigma_l)]$ (using Euler's reflection formula), and this product cancels the similar product multiplying $\delta_{lr}$. Pulling out $e^{-2i\pi n \sigma_l}$ from the bracket, one gets
\begin{equation}
\label{Mdiagonal}
M_{ll}^{\epsilon_2,\epsilon_1,\epsilon_2'}
=e^{\pi\phi_{ll}-2i\pi n\sigma_l} 
\left[1
+   n\cdot 2i e^{-n i\pi N bq/2} e^{2i\pi n \sigma_l} e^{- n\pi b[2a_l +(N-2) \hat m_2]}\frac{\prod_j \sin\pi(\sigma_l + \tau_j)}{\prod_{j\neq l} \sin\pi(\sigma_j - \sigma_l)} \right]
\end{equation}
We then note that $\phi_{ll}-2in\sigma_l = -2\epsilon_1 ba_l + i bq [(N-1)\epsilon_1-n]  + 2bn \hat m_2$. The exponentials in the second term also simplify: 
$e^{-n i\pi N bq/2} e^{2i\pi n \sigma_l} e^{- n\pi b[2a_l +(N-2) \hat m_2]} 
= e^{-n i\pi (N-2) bq/2} e^{- n\pi b N\hat m_2}$.

\begin{figure}[ht]
\begin{center}
\begin{align*}
\begin{matrix}\fmfframe(0,5)(5,0){\begin{fmfgraph*}(30,20)
  \fmfcmd{style_def my_arrow expr p = 
    cdraw (subpath (length(p)/10,length(p)) of p);
    shrink(.7); cfill (marrow (p,.3)); endshrink; 
    enddef;}
  \fmfcmd{style_def my expr p = 
    cdraw (subpath (length(p)/10,length(p)) of p);
    enddef;}
  \fmfstraight
  \fmftop{00,10,20,30,40,50}
  \fmfbottom{08,18,28,38,48,58}
  \fmf{phantom}{00,02,04,06,08}
  \fmf{phantom}{10,11,12,13,14,15,16,17,18}
  \fmf{phantom}{20,22,24,26,28}
  \fmf{phantom}{30,32,34,36,38}
  \fmf{phantom}{40,41,42,43,44,45,46,47,48}
  \fmf{phantom}{50,52,54,56,58}
  \fmffreeze
  \fmf{fermion,width=.5}{10,16}
  \fmf{fermion,width=2,rubout}{30,36}
  \fmf{plain,right,width=.5}{45,44}
  \fmf{fermion,width=2}{56,36,16,06}
  \fmf{my,width=.5}{14,45}
  \fmf{my_arrow,width=.5}{13,44}
  \fmfv{label=$\mu$,label.angle=180}{10}
  \fmfv{label=${\rm \hat m}_2$,label.angle=0}{30}
  \fmfv{label=$\alpha'$,label.angle=-60}{06}
  \fmfv{label=${\rm m}_1$,label.angle=-90}{56}
  \fmf{phantom,label=$\alpha$,label.side=left}{36,16}
\end{fmfgraph*}}\end{matrix}
&
\begin{cases}
\epsilon_1=0\\
\epsilon_2'=-\epsilon_2
\end{cases}
&
\begin{matrix}\fmfframe(0,5)(5,0){\begin{fmfgraph*}(30,20)
  \fmfcmd{style_def my_arrow expr p = 
    cdraw (subpath (length(p)/10,length(p)) of p);
    shrink(.7); cfill (marrow (p,.3)); endshrink; 
    enddef;}
  \fmfcmd{style_def my expr p = 
    cdraw (subpath (length(p)/10,length(p)) of p);
    enddef;}
  \fmfstraight
  \fmftop{00,10,20,30,40,50}
  \fmfbottom{08,18,28,38,48,58}
  \fmf{phantom}{00,02,04,06,08}
  \fmf{phantom}{10,11,12,13,14,15,16,17,18}
  \fmf{phantom}{20,22,24,26,28}
  \fmf{phantom}{30,32,34,36,38}
  \fmf{phantom}{40,41,42,43,44,45,46,47,48}
  \fmf{phantom}{50,52,54,56,58}
  \fmffreeze
  \fmf{fermion,width=.5}{10,16}
  \fmf{fermion,width=2,rubout}{30,36}
  \fmf{plain,right,width=.5}{45,44}
  \fmf{fermion,width=2}{56,36,16,06}
  \fmf{my,width=.5,rubout=5}{14,45}
  \fmf{my_arrow,width=.5}{13,44}
  \fmfv{label=$\mu$,label.angle=180}{10}
  \fmfv{label=${\rm \hat m}_2$,label.angle=0}{30}
  \fmfv{label=$\alpha'$,label.angle=-60}{06}
  \fmfv{label=${\rm m}_1$,label.angle=-90}{56}
  \fmf{phantom,label=$\alpha$,label.side=left}{36,16}
\end{fmfgraph*}}\end{matrix}
&
\begin{cases}
\epsilon_1=0\\
\epsilon_2'=\epsilon_2
\end{cases}
\\
\text{(a)}&
&
\text{(b)}&
\\
\\
\begin{matrix}\fmfframe(0,5)(5,0){\begin{fmfgraph*}(30,20)
  \fmfcmd{style_def my_arrow expr p = 
    cdraw (subpath (length(p)/10,length(p)) of p);
    shrink(.7); cfill (marrow (p,.3)); endshrink; 
    enddef;}
  \fmfcmd{style_def my expr p = 
    cdraw (subpath (length(p)/10,length(p)) of p);
    enddef;}
  \fmfstraight
  \fmftop{00,10,20,30,40,50}
  \fmfbottom{08,18,28,38,48,58}
  \fmf{phantom}{00,02,04,06,08}
  \fmf{phantom}{10,11,12,13,14,15,16,17,18}
  \fmf{phantom}{20,22,24,26,28}
  \fmf{phantom}{30,32,34,36,38}
  \fmf{phantom}{40,41,42,43,44,45,46,47,48}
  \fmf{phantom}{50,52,54,56,58}
  \fmf{phantom}{47,47a}
  \fmf{phantom}{45,45a}
  \fmf{phantom}{45,45b}
  \fmf{phantom}{14,14b}
  \fmffreeze
  \fmf{fermion,width=.5}{10,16}
  \fmf{fermion,width=2,rubout}{30,36}
  \fmf{plain,right,width=.5}{47,44}
  \fmf{fermion,width=2,rubout}{56,36,16,06}
  \fmf{plain,left,width=.5,rubout=5}{47a,45a}
  \fmf{my,width=.5}{14b,45b}
  \fmf{my_arrow,width=.5}{13,44}
  \fmfv{label=$\mu$,label.angle=180}{10}
  \fmfv{label=${\rm \hat m}_2$,label.angle=0}{30}
  \fmfv{label=$\alpha'$,label.angle=-60}{06}
  \fmfv{label=${\rm m}_1$,label.angle=-90}{56}
  \fmf{phantom,label=$\alpha$,label.side=left}{36,16}
\end{fmfgraph*}}\end{matrix}
&
\begin{cases}
\epsilon_1=\pm 1\\
\epsilon_2'=-\epsilon_2=\pm 1
\end{cases}
&
\begin{matrix}\fmfframe(0,5)(5,0){\begin{fmfgraph*}(30,20)
  \fmfcmd{style_def my_arrow expr p = 
    cdraw (subpath (length(p)/10,length(p)) of p);
    shrink(.7); cfill (marrow (p,.3)); endshrink; 
    enddef;}
  \fmfcmd{style_def my expr p = 
    cdraw (subpath (length(p)/10,length(p)) of p);
    enddef;}
  \fmfstraight
  \fmftop{00,10,20,30,40,50}
  \fmfbottom{08,18,28,38,48,58}
  \fmf{phantom}{00,02,04,06,08}
  \fmf{phantom}{10,11,12,13,14,15,16,17,18}
  \fmf{phantom}{20,22,24,26,28}
  \fmf{phantom}{30,32,34,36,38}
  \fmf{phantom}{40,41,42,43,44,45,46,47,48}
  \fmf{phantom}{50,52,54,56,58}
  \fmf{phantom}{47,47a}
  \fmf{phantom}{45,45a}
  \fmf{phantom}{45,45b}
  \fmf{phantom}{14,14b}
  \fmffreeze
  \fmf{fermion,width=.5}{10,16}
  \fmf{fermion,width=2,rubout}{30,36}
  \fmf{plain,right,width=.5}{47,44}
  \fmf{fermion,width=2,rubout}{56,36,16,06}
  \fmf{plain,left,width=.5,rubout=5}{47a,45a}
  \fmf{my,width=.5,rubout=5}{14b,45b}
  \fmf{my_arrow,width=.5}{13,44}
  \fmfv{label=$\mu$,label.angle=180}{10}
  \fmfv{label=${\rm \hat m}_2$,label.angle=0}{30}
  \fmfv{label=$\alpha'$,label.angle=-60}{06}
  \fmfv{label=${\rm m}_1$,label.angle=-90}{56}
  \fmf{phantom,label=$\alpha$,label.side=left}{36,16}
\end{fmfgraph*}}\end{matrix}
&
\begin{cases}
\epsilon_1=\pm 1\\
\epsilon_2'=\epsilon_2=\epsilon_1
\end{cases}
\\
\text{(c)}&
&
\text{(d)}&
\end{align*}
\parbox{5in}{
\caption{Monodromies that can appear in the computation of Verlinde loop operators encircling two punctures on the four-punctured sphere associated to $\Ncal=2$ conformal SQCD. The signs $\epsilon_2$ and $\epsilon_2'$ encode the direction of the braiding, and $\epsilon_1=0,\pm 1$ depending on how many times $\mu$ rotates around ${\rm m}_1$ after the first braiding.
\label{fig:SQCDmonodromies}
}}
\end{center}
\end{figure}
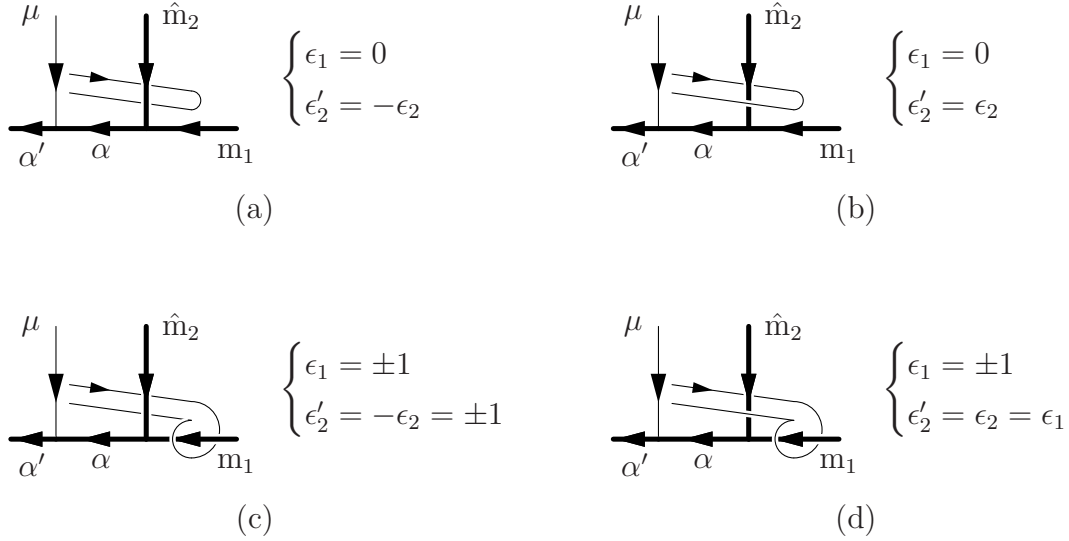

We now apply \eqref{Mdiagonal} and \eqref{Moffdiagonal} to four sets of $\epsilon$'s, and match the different cases to the monodromies depicted in Figure~\ref{fig:SQCDmonodromies}. We can fix as a convention that every case depicted has $\epsilon_2=+1$.

\paragraph{Case (a)} $\epsilon_1=0$, $\epsilon_2'=-\epsilon_2$.
Then $n=\epsilon_1-(\epsilon_2+\epsilon_2')/2=0$, so  $M_{lr}$ is diagonal, and the diagonal terms are very simple:
\[
M_{ll}^{\epsilon_2,0,-\epsilon_2}
=e^{\pi[ -2\epsilon_1 ba_l + i bq [(N-1)\epsilon_1-n]  + 2bn \hat m_2]}
= 1.
\]
This monodromy is the identity. This fixes the dictionary from relative signs of $\epsilon_2$ and $\epsilon_2'$ to directions of the braiding: opposite signs lead to braidings inverse of each other (Figure~\ref{fig:SQCDmonodromies}~(a,c)), whereas same signs correspond to braiding on two different sides of ${\rm \hat m}_2$, as in cases (b,d).

\paragraph{Case (d)} $\epsilon_1=\pm 1$, $\epsilon_2'=\epsilon_2$. Then $n = \epsilon_1 - (\epsilon_2+\epsilon_2')/2 = \epsilon_1-\epsilon_2$ is 0 or $\pm 2$ depending on the relative sign of $\epsilon_2$ and $\epsilon_1$. If $|n|=2$, the integral \eqref{integral7} is non-zero, and $M_{lr}$ is not diagonal. But the case depicted in Figure~\ref{fig:SQCDmonodromies}~(d) is homotopic to the curve wrapping around $\alpha'$, which simply corresponds to a phase, and thus to a diagonal $M$: the correct signs have to obey $n=0$. 

All in all, we want to compute $M_{lr}^{\epsilon,\epsilon,\epsilon}$, which has $n=0$. Again, the answer is very simple:
\begin{align*}
M_{lr}^{\epsilon,\epsilon,\epsilon}
=\delta_{lr} e^{\pi[-2\epsilon_1 ba_l + i bq [(N-1)\epsilon_1-n]  + 2bn \hat m_2]}
= \delta_{lr} e^{\pi\epsilon[-2 ba_l + i bq (N-1)]}.
\end{align*}
This result is compatible with the argument we expect for the exponential:
\[
\Delta(\alpha') - \Delta(\alpha) -\Delta(\mu)
= iba_l +(N-1)bq/2
\]
Understanding cases (a) and (d) has allowed us to fix the dictionary between the curves and the $\epsilon$'s. All the diagrams of Figure~\ref{fig:SQCDmonodromies} are drawn for $\epsilon_2= +1$ and $\epsilon_1 = 0,+1$.

\paragraph{Case (b)} $\epsilon_1=0$, $\epsilon_2'=\epsilon_2$. 
Then $n = - \epsilon_2$, so $M_{lr}$ is not diagonal. Diagonal terms are
\begin{align}
\label{monollb}
M_{ll}^{\epsilon_2,0,\epsilon_2}
=e^{\pi \epsilon_2 [i bq - 2b \hat m_2]}
\left[1
- 2i \epsilon_2 e^{\epsilon_2 i\pi (N-2) bq/2} e^{\epsilon_2 \pi b N\hat m_2} \frac{\prod_j \sin\pi(bq/2 - iba_l + ib\hat m_2 + ibm_{1,j})}{\prod_{j\neq l} \sin\pi(bq+ib(a_j-a_l))} \right]
\end{align}
For $l\neq r$, one gets
\begin{align}
\label{monolrb}
M_{lr}^{\epsilon_2,0,\epsilon_2}
&= - 2i\pi \epsilon_2 
e^{\epsilon_2 i\pi Nbq/2}
e^{\epsilon_2 \pi b(N-2) \hat m_2}
\cdot
\\
\notag
&\cdot
\frac{\prod_{j\neq l} \Gamma(bq+ ib(a_j-a_l)) \prod_{j\neq r}\Gamma(b^2\delta_{lj}+ib(a_r-a_j))}{\prod_j \Gamma(bq/2- iba_l + ibm_{1,j} + ib\hat m_2) \Gamma(bq/2 + iba_r -ib m_{1,j} - ib\hat m_2)}
\end{align}

\paragraph{Case (c)} $\epsilon_1=\pm 1$, $\epsilon_2'=-\epsilon_2=\pm 1$. Now, $n = \epsilon_1 - (\epsilon_2+\epsilon_2')/2 = \epsilon_1$. 
Diagonal coefficients are
\begin{align}
\label{monollc}
M_{ll}^{\epsilon_2,\epsilon_1,-\epsilon_2}
&=e^{\epsilon_1 \pi[i bq (N-2)-2ba_l  + 2b\hat m_2]} \cdot
\\
\notag
&\cdot \left[1
+   2i\epsilon_1 e^{-\epsilon_1 i\pi (N-2) bq/2} e^{- \epsilon_1\pi b N\hat m_2}
\frac{\prod_j \sin\pi(bq/2 - iba_l + ib\hat m_2 + ibm_{1,j})}{\prod_{j\neq l} \sin\pi(bq+ib(a_j-a_l))} \right]
\end{align}

Off diagonal matrix coefficients ($l\neq r$) are
\begin{align}
\label{monolrc}
M_{lr}^{\epsilon_2,\epsilon_1,-\epsilon_2}
&=  2i\pi\epsilon_1 
e^{i\pi [N\epsilon_1 -2\epsilon_2]bq/2-(\epsilon_1+\epsilon_2) \pi ba_l - (\epsilon_1-\epsilon_2') \pi ba_r}
e^{- \epsilon_1\pi b(N-2) \hat m_2}
\cdot
\\
\notag
&\cdot
\frac{\prod_{j\neq l} \Gamma(bq+ ib(a_j-a_l)) \prod_{j\neq r}\Gamma(b^2\delta_{lj}+ib(a_r-a_j))}{\prod_j \Gamma(bq/2- iba_l + ibm_{1,j} + ib\hat m_2) \Gamma(bq/2 + iba_r -ib m_{1,j} - ib\hat m_2)}.
\end{align}
The first exponential is:
\begin{align*}
\begin{cases}
e^{\pi \epsilon_1 [i(N-2)bq/2 -2ba_l]}
&\text{if $\epsilon_2=\epsilon_1$}
\\
e^{\pi \epsilon_1 [i(N+2)bq/2 -2ba_r]}
&\text{if $\epsilon_2=-\epsilon_1$}
\end{cases}
&&\text{hence}&&
\begin{cases}
M^{\epsilon,\epsilon,-\epsilon}
= M^{\epsilon,\epsilon,\epsilon} M^{-\epsilon,0,-\epsilon}
\\
M^{-\epsilon,\epsilon,\epsilon}
=  M^{-\epsilon,0,-\epsilon} M^{\epsilon,\epsilon,\epsilon}.
\end{cases}
\end{align*}
These relations, obtained by an explicit computation of the matrix coefficients, are consistent with the fact that the monodromy described in Figure~\ref{fig:SQCDmonodromies}~(c) is the product of those described in cases (d) and (b), with an order depending on the relative signs of $\epsilon_2$ and $\epsilon_1$.

\subsection{Monodromies of $\mu^*$}

Recall the conjugation map $(\sum \lambda_j h_j)^* := - \sum \lambda_j h_{N+1-j}$ acting on momenta. It is useful to denote $l^*=N+1-l$, such that $h_l^*= - h_{N+1-l} = - h_{l^*}$.

We replace ${\rm \hat m}_2\to {\rm \hat m}_3$, ${\rm m}_1\to {\rm m}_4$, $\alpha\to 2Q-\alpha^*$ in the monodromy (b) of Figure~\ref{fig:SQCDmonodromies},\footnote{The real number $a_j$ defined by $\langle \alpha-Q,h_j\rangle = ia_j$ is then mapped to $a_{j^*}$.} and reverse the arrows carrying momenta $\alpha$ and $\alpha'=\alpha-bh_l$, so that they now carry momenta $\alpha^*$ and $\alpha^*+bh_l$. 
\[
\begin{matrix}\fmfframe(0,5)(5,0){\begin{fmfgraph*}(30,20)
  \fmfcmd{style_def my_arrow expr p = 
    cdraw (subpath (length(p)/10,length(p)) of p);
    shrink(.7); cfill (marrow (p,.3)); endshrink; 
    enddef;}
  \fmfcmd{style_def my expr p = 
    cdraw (subpath (length(p)/10,length(p)) of p);
    enddef;}
  \fmfstraight
  \fmftop{00,10,20,30,40,50}
  \fmfbottom{08,18,28,38,48,58}
  \fmf{phantom}{00,02,04,06,08}
  \fmf{phantom}{10,11,12,13,14,15,16,17,18}
  \fmf{phantom}{20,22,24,26,28}
  \fmf{phantom}{30,32,34,36,38}
  \fmf{phantom}{40,41,42,43,44,45,46,47,48}
  \fmf{phantom}{50,52,54,56,58}
  \fmf{phantom}{47,47a}
  \fmf{phantom}{45,45a}
  \fmf{phantom}{45,45b}
  \fmf{phantom}{14,14b}
  \fmffreeze
  \fmf{fermion,width=.5}{10,16}
  \fmf{fermion,width=2,rubout}{30,36}
%  \fmf{plain,right,width=.5}{47,44}
  \fmf{fermion,width=2,rubout}{56,36}
  \fmf{fermion,width=2}{06,16,36}
%  \fmf{plain,left,width=.5,rubout=5}{47a,45a}
  \fmf{plain,left,width=.5}{44,45}
  \fmf{my,width=.5,rubout=5}{14b,45b}
  \fmf{my_arrow,width=.5}{13,44}
  \fmfv{label=$\mu$,label.angle=180}{10}
  \fmfv{label=${\rm \hat m}_3$,label.angle=0}{30}
  \fmfv{label=$\alpha^*+bh_l$,label.angle=180}{06}
  \fmfv{label=${\rm m}_4$,label.angle=-90}{56}
  \fmf{phantom,label=$\alpha^*$,label.side=left}{36,16}
\end{fmfgraph*}}\end{matrix}
\begin{cases}
\epsilon_1=0\\
\epsilon_2'=\epsilon_2=\pm 1
\end{cases}
\]

The $(\alpha^*, \alpha^*+bh_l-bh_r)$ matrix coefficient of the resulting operator is 
\[
M_{lr}^{\epsilon_2,0,\epsilon_2}[{\rm \hat m}_2\to {\rm \hat m}_3, {\rm m}_1\to {\rm m}_4, a_j\to a_{j^*}],
\]
where the square brackets indicate which substitutions to make to get the result. 

We then use the invariance of the theory under conjugation of all momenta: conjugating all momenta does not change the resulting matrix elements. Hence, the $(\alpha, \alpha-bh_{l^*}+bh_{r^*})$ matrix element of
\[
\begin{matrix}\fmfframe(0,5)(5,0){\begin{fmfgraph*}(30,20)
  \fmfcmd{style_def my_arrow expr p = 
    cdraw (subpath (length(p)/10,length(p)) of p);
    shrink(.7); cfill (marrow (p,.3)); endshrink; 
    enddef;}
  \fmfcmd{style_def my expr p = 
    cdraw (subpath (length(p)/10,length(p)) of p);
    enddef;}
  \fmfstraight
  \fmftop{00,10,20,30,40,50}
  \fmfbottom{08,18,28,38,48,58}
  \fmf{phantom}{00,02,04,06,08}
  \fmf{phantom}{10,11,12,13,14,15,16,17,18}
  \fmf{phantom}{20,22,24,26,28}
  \fmf{phantom}{30,32,34,36,38}
  \fmf{phantom}{40,41,42,43,44,45,46,47,48}
  \fmf{phantom}{50,52,54,56,58}
  \fmf{phantom}{47,47a}
  \fmf{phantom}{45,45a}
  \fmf{phantom}{45,45b}
  \fmf{phantom}{14,14b}
  \fmffreeze
  \fmf{fermion,width=.5}{10,16}
  \fmf{fermion,width=2,rubout}{30,36}
%  \fmf{plain,right,width=.5}{47,44}
  \fmf{fermion,width=2,rubout}{56,36}
  \fmf{fermion,width=2}{06,16,36}
%  \fmf{plain,left,width=.5,rubout=5}{47a,45a}
  \fmf{plain,left,width=.5}{44,45}
  \fmf{my,width=.5,rubout=5}{14b,45b}
  \fmf{my_arrow,width=.5}{13,44}
  \fmfv{label=$\mu^*$,label.angle=180}{10}
  \fmfv{label=${\rm \hat m}_3^*$,label.angle=0}{30}
  \fmfv{label=$\alpha-bh_{l^*}$,label.angle=180}{06}
  \fmfv{label=${\rm m}_4^*$,label.angle=-90}{56}
  \fmf{phantom,label=$\alpha$,label.side=left}{36,16}
\end{fmfgraph*}}\end{matrix}
\begin{cases}
\epsilon_1=0\\
\epsilon_2'=\epsilon_2=\pm 1
\end{cases}
\]
is $M_{lr}^{\epsilon_2,0,\epsilon_2}[{\rm \hat m}_2\to {\rm \hat m}_3, {\rm m}_1\to {\rm m}_4, a_j\to a_{j^*}]$ as well.

The last step is then to flip the diagram along a vertical axis. The only thing that can be changed by such a reflexion is an overall sign in the matching between $\epsilon_2,\epsilon_1,\epsilon_2'$, and the orientations in diagrams. We will understand whether there is a change in signs by applying the list of transformations just described for (b) to the monodromy (d), and combining this new monodromy ``(d*)'' with (d). In order to combine these two monodromies, we do the change of indices $l\to l^*$ and $r\to r^*$ and rename all signs $\epsilon\to \epsilon^*$ in the monodromy of $\mu^*$.

After the moves that we describe, (d) becomes some new monodromy (d*):
\[
\begin{matrix}\fmfframe(0,5)(5,0){\begin{fmfgraph*}(30,20)
  \fmfcmd{style_def my_arrow expr p = 
    cdraw (subpath (length(p)/10,length(p)) of p);
    shrink(.7); cfill (marrow (p,.3)); endshrink; 
    enddef;}
  \fmfcmd{style_def my expr p = 
    cdraw (subpath (length(p)/10,length(p)) of p);
    enddef;}
  \fmfstraight
  \fmftop{50,40,30,20,10,00}
  \fmfbottom{58,48,38,28,18,08}
  \fmf{phantom}{00,02,04,06,08}
  \fmf{phantom}{10,11,12,13,14,15,16,17,18}
  \fmf{phantom}{20,22,24,26,28}
  \fmf{phantom}{30,32,34,36,38}
  \fmf{phantom}{40,41,42,43,44,45,46,47,48}
  \fmf{phantom}{50,52,54,56,58}
  \fmf{phantom}{47,47a}
  \fmf{phantom}{45,45a}
  \fmf{phantom}{45,45b}
  \fmf{phantom}{14,14b}
  \fmffreeze
  \fmf{fermion,width=.5}{10,16}
  \fmf{fermion,width=2,rubout}{30,36}
  \fmf{plain,left,width=.5}{47,44}
  \fmf{fermion,width=2,rubout}{56,36}
  \fmf{fermion,width=2}{06,16,36}
  \fmf{plain,right,width=.5,rubout=5}{47a,45a}
  \fmf{my,width=.5,rubout=5}{14b,45b}
  \fmf{my_arrow,width=.5}{13,44}
  \fmfv{label=$\mu^*$,label.angle=0}{10}
  \fmfv{label=${\rm \hat m}_3^*$,label.angle=180}{30}
  \fmfv{label=$\alpha-bh_l$,label.angle=-90}{06}
  \fmfv{label=${\rm m}_4^*$,label.angle=-90}{56}
  \fmf{phantom,label=$\alpha$,label.side=right}{36,16}
\end{fmfgraph*}}\end{matrix}
\quad
{\epsilon_2'}^*=\epsilon_2^*=\epsilon_1^*=\epsilon
\]
The $(\alpha, \alpha-bh_l+bh_r)$ matrix coefficient of the resulting operator is then $\delta_{lr} e^{\pi\epsilon [-2 ba_l + i bq (N-1)]}$. To determine the correspondence between orientations in the diagrams and signs of $\epsilon$'s, we consider the combination of (d*) and (d), as in the diagram below. It only yields a phase, which depends on the orientations $\epsilon_1$ and $\epsilon_1^*$ of the two curves:
\[
e^{\pi\epsilon_1^* [-2 ba_l + i bq (N-1)]}
e^{\pi\epsilon_1 [-2 ba_l + i bq (N-1)]}
= e^{\pi (\epsilon_1^*+\epsilon_1)[-2ba_l + ibq(N-1)]}.
\]
If the orientations of both sides are mirrors of each other, as in
\[
\begin{matrix}\fmfframe(0,5)(5,0){\begin{fmfgraph*}(60,20)
  \fmfcmd{style_def my_arrow expr p = 
    cdraw (subpath (length(p)/10,length(p)) of p);
    shrink(.7); cfill (marrow (p,.3)); endshrink; 
    enddef;}
  \fmfcmd{style_def my expr p = 
    cdraw (subpath (length(p)/10,length(p)) of p);
    enddef;}
  \fmfstraight
  \fmftop{s50,s40,s30,s20,s10,00,10,20,30,40,50}
  \fmfbottom{s58,s48,s38,s28,s18,08,18,28,38,48,58}
  \fmf{phantom}{00,02,04,06,08}
  \fmf{phantom}{s10,s11,s12,s13,s14,s15,s16,s17,s18}
  \fmf{phantom}{s20,s22,s24,s26,s28}
  \fmf{phantom}{s30,s32,s34,s36,s38}
  \fmf{phantom}{s40,s41,s42,s43,s44,s45,s46,s47,s48}
  \fmf{phantom}{s50,s52,s54,s56,s58}
  \fmf{phantom}{10,11,12,13,14,15,16,17,18}
  \fmf{phantom}{20,22,24,26,28}
  \fmf{phantom}{30,32,34,36,38}
  \fmf{phantom}{40,41,42,43,44,45,46,47,48}
  \fmf{phantom}{50,52,54,56,58}
  \fmf{phantom}{s47,s47a}
  \fmf{phantom}{s45,s45a}
  \fmf{phantom}{s45,s45b}
  \fmf{phantom}{s14,s14b}
  \fmf{phantom}{47,47a}
  \fmf{phantom}{45,45a}
  \fmf{phantom}{45,45b}
  \fmf{phantom}{14,14b}
  \fmffreeze
  \fmf{fermion,width=.5}{10,16}
  \fmf{fermion,width=2,rubout}{30,36}
  \fmf{plain,right,width=.5}{47,44}
  \fmf{fermion,width=2,rubout}{56,36,16,s16,s36}
  \fmf{plain,left,width=.5,rubout=5}{47a,45a}
  \fmf{my,width=.5,rubout=5}{14b,45b}
  \fmf{my_arrow,width=.5}{13,44}
  \fmf{fermion,width=.5}{s10,s16}
  \fmf{fermion,width=2,rubout}{s30,s36}
  \fmf{plain,left,width=.5}{s47,s44}
  \fmf{fermion,width=2,rubout}{s56,s36}
  \fmf{plain,right,width=.5,rubout=5}{s47a,s45a}
  \fmf{my,width=.5,rubout=5}{s14b,s45b}
  \fmf{my_arrow,width=.5}{s13,s44}
  \fmfv{label=$\mu^*$,label.angle=0}{s10}
  \fmfv{label=${\rm \hat m}_3^*$,label.angle=180}{s30}
  \fmfv{label=${\rm m}_4^*$,label.angle=-90}{s56}
  \fmfv{label=$\mu$,label.angle=0}{10}
  \fmfv{label=${\rm \hat m}_2$,label.angle=0}{30}
  \fmfv{label=${\rm m}_1$,label.angle=-90}{56}
  \fmf{phantom,label=$\alpha$,label.side=left}{36,16}
  \fmf{phantom,label=$\alpha-bh_l$,label.side=right}{s16,16}
  \fmf{phantom,label=$\alpha$,label.side=right}{s36,s16}
\end{fmfgraph*}}\end{matrix}
\]
the diagram should give the identity, and therefore, the signs have to be opposite of each other: $\epsilon_1^*=-\epsilon_1$. 

This implies that if we keep the convention that diagrams in Figure~\ref{fig:SQCDmonodromies} are drawn with $\epsilon_2=+1$ and $\epsilon_1=0,1$, then the mirror image of the non-starred diagram (b):
\[
\begin{matrix}\fmfframe(0,5)(5,0){\begin{fmfgraph*}(30,20)
  \fmfcmd{style_def my_arrow expr p = 
    cdraw (subpath (length(p)/10,length(p)) of p);
    shrink(.7); cfill (marrow (p,.3)); endshrink; 
    enddef;}
  \fmfcmd{style_def my expr p = 
    cdraw (subpath (length(p)/10,length(p)) of p);
    enddef;}
  \fmfstraight
  \fmftop{50,40,30,20,10,00}
  \fmfbottom{58,48,38,28,18,08}
  \fmf{phantom}{00,02,04,06,08}
  \fmf{phantom}{10,11,12,13,14,15,16,17,18}
  \fmf{phantom}{20,22,24,26,28}
  \fmf{phantom}{30,32,34,36,38}
  \fmf{phantom}{40,41,42,43,44,45,46,47,48}
  \fmf{phantom}{50,52,54,56,58}
  \fmffreeze
  \fmf{fermion,width=.5}{10,16}
  \fmf{fermion,width=2,rubout}{30,36}
  \fmf{plain,left,width=.5}{45,44}
  \fmf{fermion,width=2}{56,36}
  \fmf{fermion,width=2}{06,16,36}
  \fmf{my,width=.5,rubout=5}{14,45}
  \fmf{my_arrow,width=.5}{13,44}
  \fmfv{label=$\mu^*$,label.angle=0}{10}
  \fmfv{label=${\rm \hat m}_3^*$,label.angle=180}{30}
  \fmfv{label=$\alpha'$,label.angle=-120}{06}
  \fmfv{label=${\rm m}_4^*$,label.angle=-90}{56}
  \fmf{phantom,label=$\alpha$,label.side=right}{36,16}
\end{fmfgraph*}}\end{matrix}
\quad
\begin{cases}
\epsilon_1^*=0\\
{\epsilon_2'}^*=\epsilon_2^*
\end{cases}
\]
is drawn for $\epsilon_2^*=-1$. The diagonal and off-diagonal terms ($l\neq r$) are
\begin{align}
\label{monollb*}
M_{ll}^{*,\epsilon,0,\epsilon} 
&=
e^{\pi\epsilon (ibq-2b\hat m_3)} \left[1  
- 2i \epsilon e^{\pi \epsilon [i(N-2)bq/2+Nb\hat m_3)]}
\frac{\prod_j \sin\pi(bq/2 - iba_l + ib\hat m_3+ibm_{4,j}))}{\prod_{j\neq l} \sin\pi(bq + i(a_j-a_l))}\right]
\\
\notag
M_{lr}^{*,\epsilon_2^*,0,\epsilon_2^*} 
&= -2i\pi \epsilon_2^* e^{\epsilon_2^* i \pi Nbq/2} e^{\epsilon_2^* \pi b(N-2)\hat m_3} \cdot
\\
\label{monolrb*}
&
\cdot
\frac{\prod_{j\neq l} \Gamma(bq+ ib(a_j-a_l)) \prod_{j\neq r}\Gamma(b^2\delta_{lj} +ib(a_r-a_j))}{\prod_j \Gamma(bq/2 - iba_l  + ib\hat m_3 + ibm_{4,j}) \Gamma(bq/2 + iba_r - ib\hat m_3 -ibm_{4,j})}.
\end{align}

%===================================================
\section{Wilson loops in $\Ncal=2$ gauge theories}
\label{sec:Wilson} 

In this appendix  we consider a general Riemann surface $C_{g,n}$ with genus $g$, and $n$ punctures labeled by (non-degenerate or semidegenerate) momenta, in some pants decomposition $\sigma$. We construct the Verlinde loop operator $\Ocal_\mu(p^r)$ corresponding to the monodromy of the chiral vertex operator $V_\mu(z) = V_{-bh_1}(z)$ along the curve $p^r$ winding $r$ times around a tube of the decomposition $\sigma$. As we make explicit at the end of this appendix, $\Ocal_\mu(p^r)$ captures the expectation value of a Wilson loop operator in the ${\cal N}=2$ corresponding gauge theory. The cases $r=\pm 1$ reproduce the result of \cite{Passerini:2010pr}.

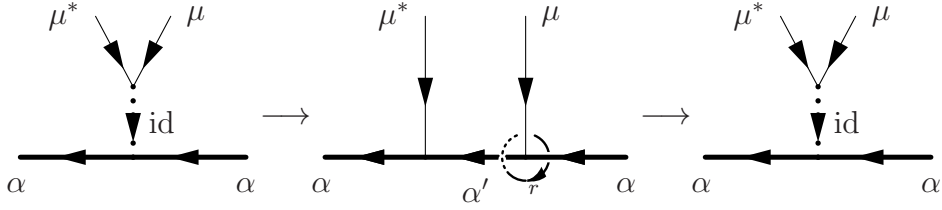
\begin{figure}[t]
\begin{center}
\[
  \begin{array}{c}
    \begin{fmfgraph*}(30,25)
      \fmfstraight
      \fmftop{tl,t1,t3,tr}
      \fmfbottom{bbl,bb2,bbr}
      \fmf{phantom}{t1,t2,t3}
      \fmffreeze
      \fmf{phantom}{bbl,bl,mbl,mtl,tl}
      \fmf{phantom}{bbr,br,mbr,mtr,tr}
      \fmf{phantom}{bb2,b2,mb2,mt2,t2}
      \fmffreeze
      \fmf{fermion,width=2}{br,b2,bl}
      \fmffreeze
      \fmf{fermion,width=0.5}{t1,v}
      \fmf{fermion,width=0.5}{t3,v}
      \fmf{dots_arrow,width=2,tension=2,label=$\id$}{v,b2}
      \fmfv{label=$\alpha$,label.angle=-100,label.dist=8}{bl}
      \fmfv{label=$\alpha$,label.angle=-90,label.dist=8}{br}
      \fmfv{label=$\mu^*$,label.angle=180}{t1}
      \fmfv{label=$\mu$,label.angle=0}{t3}
    \end{fmfgraph*}
  \end{array}
\longrightarrow
  \begin{array}{c}
    \begin{fmfgraph*}(40,25)
      \fmfcmd{style_def tipped_arrow expr p = 
        cdraw (subpath (length(p)/8,length(p)) of p);
        shrink(.7); cfill (harrow (p,1)); endshrink; 
        enddef;}
      \fmfcmd{style_def short expr p = 
        cdraw (subpath (0,.28length(p)) of p);
        cdraw (subpath (.35length(p),.38length(p)) of p);
        cdraw (subpath (.45length(p),.48length(p)) of p);
        cdraw (subpath (.55length(p),.58length(p)) of p);
        cdraw (subpath (.65length(p),.68length(p)) of p);
        cdraw (subpath (.75length(p),.78length(p)) of p);
        cdraw (subpath (.85length(p),.88length(p)) of p);
        enddef;}
      \fmfstraight
      \fmftop{00,02,04,06}
      \fmfbottom{80,82,84,86}
      \fmf{phantom}{00,20,40,60,80}
      \fmf{phantom}{02,22,42,62,82}
      \fmf{phantom}{04,24,44,64,84}
      \fmf{phantom}{06,26,46,66,86}
      \fmffreeze
      \fmf{phantom}{62,63,64}
      \fmf{phantom}{84,74,64}
      \fmf{phantom}{64,54,44}
      \fmffreeze
      \fmf{tipped_arrow,left}{54,74}
      \fmf{fermion,width=2,rubout}{66,64,62,60}
      \fmffreeze
      \fmf{short,left,rubout=3}{74,54}
      \fmf{fermion,width=0.5,tension=0}{02,62}
      \fmf{fermion,width=0.5,tension=0}{04,64}
      \fmfv{label=$\alpha$,label.angle=-100,label.dist=8}{60}
      \fmfv{label=$\alpha'$,label.angle=-90,label.dist=8}{63}
      \fmfv{label=$\alpha$,label.angle=-90,label.dist=8}{66}
      \fmfv{label=$\mu^*$,label.angle=180}{02}
      \fmfv{label=$\mu$,label.angle=0}{04}
      \fmfv{label={\scriptsize $r$},label.angle=-60,label.dist=2}{74}
    \end{fmfgraph*}
  \end{array}
\longrightarrow
  \begin{array}{c}
    \begin{fmfgraph*}(30,25)
      \fmfstraight
      \fmftop{tl,t1,t3,tr}
      \fmfbottom{bbl,bb2,bbr}
      \fmf{phantom}{t1,t2,t3}
      \fmffreeze
      \fmf{phantom}{bbl,bl,mbl,mtl,tl}
      \fmf{phantom}{bbr,br,mbr,mtr,tr}
      \fmf{phantom}{bb2,b2,mb2,mt2,t2}
      \fmffreeze
      \fmf{fermion,width=2}{br,b2,bl}
      \fmffreeze
      \fmf{fermion,width=0.5}{t1,v}
      \fmf{fermion,width=0.5}{t3,v}
      \fmf{dots_arrow,width=2,tension=2,label=$\id$}{v,b2}
      \fmfv{label=$\alpha$,label.angle=-100,label.dist=8}{bl}
      \fmfv{label=$\alpha$,label.angle=-90,label.dist=8}{br}
      \fmfv{label=$\mu^*$,label.angle=180}{t1}
      \fmfv{label=$\mu$,label.angle=0}{t3}
    \end{fmfgraph*}
  \end{array}
\]
\vskip-5mm
\parbox{5in}{
\caption{Calculating the Wilson loop as the monodromy associated to
  moving a degenerate field along a loop $r$ times. Here,
  $\mu=-bh_1$, and $\alpha' = \alpha - b h_k$ for some $k$.
\label{fig:Wilson}
}}
\end{center}
\end{figure}

We denote by $\alpha$ the momentum flowing through the tube encircled by $p$. The operator $\Ocal_\mu (p^r)$ consists in a fusion move \eqref{fusion1}, the monodromy of $\mu$ around the tube $\alpha$, which only contributes a phase, and finally \eqref{fusion2}, as in Figure~\ref{fig:Wilson}. None of these moves affect $\alpha$. In other words, the Verlinde loop operator acts  diagonally:
\begin{equation*}
{\cal O}_\mu (p^r)\cdot {\cal F}^{(\sigma)}_{\alpha,E}={\cal O}_\alpha^{(r)} {\cal F}^{(\sigma)}_{\alpha,E}\,,
\end{equation*}
where $E$ denotes the external momenta, as well as all internal momenta other than $\alpha$. 

Combine \eqref{fusion1} and the monodromy to get
\begin{align*}
\left[
\begin{array}{c}
  \fmfframe(0,3)(0,3){\begin{fmfgraph*}(20,20)
    \fmftop{tl,tr}
    \fmfbottom{bl,br}
    \fmf{fermion,width=2}{br,b2,bl}
    \fmffreeze
    \fmf{fermion,width=0.5}{tl,v}
    \fmf{fermion,width=0.5}{tr,v}
    \fmf{dots_arrow,width=2,label=id,tension=1.5}{v,b2} 
    \fmfv{label=$\alpha$,label.angle=-60,label.dist=6}{bl}
    \fmfv{label=$\alpha$,label.angle=-120,label.dist=6}{br}
    \fmfv{label=$\mu^*$,label.angle=30,label.dist=4}{tl}
    \fmfv{label=$\mu$,label.angle=150,label.dist=4}{tr}
  \end{fmfgraph*}}
\end{array}
\right]
&=
\frac{\Gamma(Nbq)}{\Gamma(bq)}
\sum_{k=1}^N \left[
\left(\prod_{i\neq k}
  \frac{\Gamma(b\langle \alpha-Q,h_i - h_k\rangle)}
  {\Gamma(bq + b\langle \alpha-Q,h_i-h_k\rangle)}
\right)
\begin{array}{c}
  \begin{fmfgraph*}(35,15)
    \fmftop{tl,tr}
    \fmfbottom{bl,br}
    \fmf{fermion,width=2}{br,b1,b3,bl}
    \fmf{phantom}{br,b1} \fmf{phantom}{b3,bl}
    \fmf{phantom}{tr,t1,t2,t3,tl}
    \fmffreeze
    \fmf{fermion,width=0.5}{t1,b1}
    \fmf{fermion,width=0.5}{t3,b3}
    \fmfv{label=$\alpha$,label.angle=-60,label.dist=6}{bl}
    \fmf{phantom,label=$\alpha-bh_k$,label.side=right}{b1,b3}
    \fmfv{label=$\alpha$,label.angle=-120,label.dist=6}{br}
    \fmfv{label=$\mu$,label.angle=180,label.dist=4}{t1}
    \fmfv{label=$\mu^*$,label.angle=0,label.dist=4}{t3}
  \end{fmfgraph*}
\end{array}
\right]
\\
&\mathop{\mapsto}^{r}
\frac{\Gamma(Nbq)}{\Gamma(bq)}
\sum_{k=1}^N \left[
\left(\prod_{i\neq k}
  \frac{\Gamma(b\langle \alpha-Q,h_i - h_k\rangle)}
  {\Gamma(bq+b\langle \alpha-Q,h_i-h_k\rangle)}
\right)
e^{2i\pi r\varphi}
\begin{array}{c}
  \begin{fmfgraph*}(35,15)
    \fmftop{tl,tr}
    \fmfbottom{bl,br}
    \fmf{fermion,width=2}{br,b1,b3,bl}
    \fmf{phantom}{br,b1} \fmf{phantom}{b3,bl}
    \fmf{phantom}{tr,t1,t2,t3,tl}
    \fmffreeze
    \fmf{fermion,width=0.5}{t1,b1}
    \fmf{fermion,width=0.5}{t3,b3}
    \fmfv{label=$\alpha$,label.angle=-60,label.dist=6}{bl}
    \fmf{phantom,label=$\alpha-bh_k$,label.side=right}{b1,b3}
    \fmfv{label=$\alpha$,label.angle=-120,label.dist=6}{br}
    \fmfv{label=$\mu$,label.angle=-150,label.dist=4}{t1}
    \fmfv{label=$\mu^*$,label.angle=-30,label.dist=4}{t3}
  \end{fmfgraph*}
\end{array}
\right],
\end{align*}
where the phase $\varphi$ is the power of $z$ in the expansion of the conformal block $\Fcal^{(s)}_{\alpha-bh_k}\left[\begin{smallmatrix} \mu^*&\mu\\ 2Q-\alpha&\alpha\end{smallmatrix}\right]$:
\[
\varphi = \Delta(\alpha- b h_k) - \Delta(\alpha) - \Delta(\mu)
= b\langle \alpha -Q, h_k\rangle + b\langle Q,h_1\rangle.
\]

We then project onto the identity internal state using~\eqref{fusion2}. Together with Euler's reflection formula $\Gamma(x)\Gamma(1-x) = \pi/\sin\pi x$, this yields
\begin{equation*}
\frac{\sin \pi bq}{\sin \pi N bq}
\sum_{k=1}^N \left[
\left(\prod_{i\neq k}
  \frac{\sin \pi (bq + b\langle \alpha-Q,h_i-h_k\rangle)}
       {\sin \pi  b\langle \alpha-Q,h_i-h_k\rangle}
\right)
e^{2i\pi r(b\langle \alpha -Q, h_k\rangle + b\langle
  Q,h_1\rangle)}
\right]
\left[
\begin{array}{c}
  \fmfframe(0,3)(0,3){\begin{fmfgraph*}(20,20)
    \fmftop{tl,tr}
    \fmfbottom{bl,br}
    \fmf{fermion,width=2}{br,b2,bl}
    \fmffreeze
    \fmf{fermion,width=0.5}{tl,v}
    \fmf{fermion,width=0.5}{tr,v}
    \fmf{dots_arrow,width=2,label=id}{v,b2} 
    \fmfv{label=$\alpha$,label.angle=-60,label.dist=6}{bl}
    \fmfv{label=$\alpha$,label.angle=-120,label.dist=6}{br}
    \fmfv{label=$\mu^*$,label.angle=30,label.dist=4}{tl}
    \fmfv{label=$\mu$,label.angle=150,label.dist=4}{tr}
  \end{fmfgraph*}}
\end{array}
\right].
\end{equation*}

Thus, using the residue theorem,
\begin{align*}
\Ocal_\alpha^{(r)}
&=\frac{\sin \pi bq}{\sin \pi N bq}
\sum_{k=1}^N \left[
\left(\prod_{i\neq k}
  \frac{\sin \pi (bq + b\langle \alpha-Q,h_i-h_k\rangle)}
       {\sin \pi  b\langle \alpha-Q,h_i-h_k\rangle}
\right)
e^{2i\pi r(b\langle \alpha -Q, h_k\rangle + b\langle
  Q,h_1\rangle)}
\right]
\\
&=\frac{e^{2i\pi r b \langle Q, h_1\rangle}}{\sin \pi N bq}\frac{1}{2i}
\oint dz \prod_i \frac{\sin\pi (bq + b\langle \alpha-Q, h_i\rangle + z)}
{\sin\pi (b\langle \alpha-Q, h_i\rangle+z)} e^{- 2i\pi r z},
\end{align*}
where the contour is the rectangle enclosing the region $\{x<\Re(z)<x+1, -y<\Im(z)<y\}$, with $x\in\mathbb{R}$ generic, and $y\to \infty$.\footnote{More precisely, we need $x+\Re(b\langle\alpha-Q, h_k\rangle)\not\in \mathbb{Z}$, and $y>|\Im(b\langle\alpha-Q,h_k\rangle)|$.} Because the sines are periodic, the integrals along the vertical sides cancel each other. For the two other sides of the rectangle, we expand the product of sines using
\[
\sin \pi (b \langle \alpha-Q, h_i\rangle +z) 
= \frac{1}{2i} (e^{i\pi (b \langle \alpha-Q, h_i\rangle +z) } - e^{-i\pi (b \langle \alpha-Q, h_i\rangle +z) }),
\]
and group terms with the same power of $e^{2i\pi z}$. Integrating the product of sines with the factor $e^{-2i\pi r z}$ picks up the $r$-th coefficient, which we can express as a sum of exponentials.

We get
\[
\Ocal_\alpha^{(r)} = \left\{
\begin{aligned}
  \frac{e^{i\pi r bq (N-1)}}{1-e^{2i\pi N bq}} 
  &\sum_{\substack{m_1+\cdots+m_N = r\\ m_j \geq 0}} e^{2i\pi b\langle \alpha-Q,\sum_i m_i h_i\rangle} 
  \left(1- e^{2i\pi b q}\right)^{c[m]} &\quad \text{if $r>0$}
  \\
  \frac{e^{i\pi r bq (N-1)}}{1-e^{-2i\pi N bq}} 
  &\sum_{\substack{m_1+\cdots+m_N = -r\\ m_j \geq 0}} e^{-2i\pi b\langle \alpha-Q,\sum_i m_i h_i\rangle} 
  \left(1- e^{-2i\pi b q}\right)^{c[m]} &\quad \text{if $r<0$},
\end{aligned}\right.
\]
where $c[m]$ is the number of non-zero $m_i$. 

In the $b=1$ limit, this formula simplifies drastically: only the terms with $c[m]=1$ remain, giving
\begin{equation}
\label{result-Wilson}
\Ocal_\alpha^{(r)}|_{b=1}
= \frac{1}{N} \sum_{k=1}^N e^{2i\pi r \langle \alpha-Q, h_k\rangle}
= \frac{1}{N} \Tr e^{-2\pi r a},
\end{equation}
where $a$ is a hermitian generator of $\mathfrak{su}(N)$ with eigenvalues $a_k = -i\langle \alpha-Q, h_k\rangle$, and the trace is taken in the fundamental representation of $SU(N)$.

In the construction described in \cite{Gaiotto:2009we}, each tube of the pants decomposition corresponds to a vector multiplet in the Lagrangian description of the gauge theory constructed from the data of $C_{g,n}$, $\sigma$, and labels of the punctures. Our result \eqref{result-Wilson} yields the expectation value of a multiply-wound Wilson  loop in  the fundamental or antifundamental representation  of the gauge group for the corresponding vector multiplet, depending on the orientation of the curve. The computation \eqref{result-Wilson} indicates that the Verlinde loop operator $\Ocal_\mu(p^{-r})$ encapsulates the expectation value of the Wilson loop in the fundamental representation,\footnote{Note that the direction of $p$, and hence the sign of $r$ is a matter of conventions.} thus   reproducing    Pestun's  result  \cite{Pestun:2007rz}
\begin{equation*}
\vev{W_r}_{{\cal N}=2}=\int [da] \frac{1}{N} \Tr (e^{2\pi r a}) \overline{Z_\text{Nekrasov}(a,m,\tau)}\, Z_\text{Nekrasov}(a,m,\tau)\,.
\end{equation*}

\clearpage
\end{fmffile}
\bibliography{loopstoda}
\end{document}